\documentclass{elsart}

\usepackage{amsfonts}
\usepackage{amssymb}
\usepackage{latexsym}
\usepackage[dvips]{graphicx}

\input epsf

\newcommand{\vecPsi}{\mathbf{\Psi}}

\newcommand{\vecF}{\mathbf{F}}

\newcommand{\vecK}{\mathbf{K}}

\newcommand{\vecU}{\mathbf{U}}

\newcommand{\vecX}{\mathbf{X}}

\newcommand{\veca}{\mathbf{a}}

\newcommand{\vece}{\mathbf{e}}

\newcommand{\veck}{\mathbf{k}}

\newcommand{\vecn}{\mathbf{n}}

\newcommand{\vecp}{\mathbf{p}}
\newcommand{\vecq}{\mathbf{q}}

\newcommand{\vecu}{\mathbf{u}}

\newcommand{\vecx}{\mathbf{x}}

\newcommand{\vecz}{\mathbf{z}}
\newcommand{\veczero}{\mathbf{0}}
\newcommand{\commentOut}[1]{}
\newcommand{\D}{\displaystyle}
\newcommand{\eps}{\varepsilon}

\begin{document}

\maketitle

\begin{frontmatter}
\title{Stability of Rossby waves in the $\beta$-plane approximation}
\author[YL]{Youngsuk Lee},
\corauth[cor]{Corresponding author.}
\ead{youngsuk@math.wisc.edu}
\author[LMS]{Leslie M. Smith\corauthref{cor}}
\ead{lsmith@math.wisc.edu}

\address[YL]{Department of Mathematics, University of Wisconsin - Madison,
53706, USA.}
\address[LMS]{Department of Mathematics and Mechanical Engineering,
University of Wisconsin - Madison, 53706, USA.}

\begin{abstract}

Floquet theory is used to describe the unstable spectrum at large scales of the
$\beta$-plane equation linearized about Rossby waves.
Base flows consisting of one to three Rossby wave are
considered analytically using continued fractions and
the method of multiple scales, while base flow with more than
three Rossby waves are studied numerically.  It is demonstrated
that the mechanism for instability changes from inflectional to
triad resonance at an $O(1)$ transition Rhines number $Rh$, independent
of the Reynolds number.    For a single Rossby wave base flow,
the critical Reynolds number $Re^c$ for instability is found in various
limits.  In the limits $Rh\rightarrow \infty$ and $k\rightarrow 0$,
the classical value $Re^c = \sqrt{2}$ is recovered.  For
$Rh \rightarrow 0$ and all orientations of the Rossby wave
except zonal and meridional, the base flow is unstable for all
Reynolds numbers; a zonal Rossby wave is stable, while a
meridional Rossby wave has critical Reynolds number $Re^c = \sqrt{2}$.
For more isotropic base flows consisting of many Rossby waves
(up to forty), the most unstable mode is purely zonal for
$2 \leq Rh < \infty$ and is nearly zonal for $Rh = 1/2$, where
the transition Rhines number is again $O(1)$, independent of
the Reynolds number and consistent with a change in the mechanism
for instability from inflectional to triad resonance.
%
%
%
\end{abstract}


\end{frontmatter}

\section{Introduction and the main results}
\label{sec_blah}

%
%
%

The $\beta$-plane equation is a model for large-scale
geophysical flows, in which the effect of the earth's sphericity
is modeled by a linear variation $\beta$ of the Coriolis parameter.
Due to the $\beta$ term, solutions of the linearized $\beta$-plane
equation are dispersive waves, called Rossby waves, with
anisotropic dispersion relation given by $\omega({\bf k}) = -\beta k_x/k^2$,
where $\omega({\bf k})$ denotes the wave frequency.  Zero-frequency
linear eigenmodes are zonal flows and form a slow manifold.
In the absence of the $\beta$ term, the model reduces to the
two-dimensional (2D), incompressible Navier Stokes equations.

The present work is motivated in part by numerical simulations of
$\beta$-plane flow driven by stochastic forcing at small scales
(Chekhlov, Orszag, Galperin and Staroselsky \cite{CH}, 
Smith and Waleffe \cite{SW1},
Marcus, Kundu and Lee \cite{ML}, Huang, Galperin and Sukoriansky \cite{HGS}).  
These authors show that
the presence of the $\beta$ term leads to anisotropic transfer of
energy to large scales and accumulation of energy on the slow
manifold corresponding to zonal flows. 
The generation of slow, large scales is also characteristic
of other models of geophysical flows, including three-dimensional (3D) rotating
flow \cite{SW1}, and the 2D and 3D Boussinesq equations for rotating stably 
stratified flows  \cite{SW2}, \cite{S}.  The accumulation of energy
in slow modes is interesting given that resonant triad interactions
cannot transfer energy from fast waves directly to slow modes with zero
frequency as shown, for example, in Longuet-Higgins and Gill \cite{LG}
for the $\beta$-plane equation. 
Newell \cite{N} showed that higher-order interactions, such as resonant
quartets, can lead to energy transfer from Rossby waves to zonal flows.
Towards a complete understanding
of the simulations \cite{CH}, \cite{SW1}, \cite{ML}, \cite{HGS}, 
this paper describes the large-scale
instability of base flows maintained by a single or
several deterministic force(s), for values of $\beta$ ranging from
$0 < \beta < \infty$, and for finite Reynolds numbers.  

The results
herein extend previous stability analyses of the 2D Navier
Stokes equations and the $\beta$-plane equation.
Using Floquet theory and a continued fraction formulation,
Meshalkin and Sinai \cite{MS} studied the linear stability
of the 2D Navier Stokes equations with a Kolmogorov base flow, 
and found the critical Reynolds number
$Re^c=\sqrt{2}$ for large-scale instability with wavenumbers $k \ll 1$.
For Reynolds numbers slightly above the critical value, the 
instability occurs for modes with wavevector ${\bf k}$ perpendicular
to the wavevector of the Kolmogorov base flow.
Friedlander and Howard \cite{FR1},
Belenkaya, Friedlander and Yudovich \cite{FR2} and Li \cite{LI} used 
similar techniques to study the stability of the
2D inviscid Navier Stokes equations (the 2D Euler equations).
Sivashinsky \cite{SI} used Multiple Scales
analysis on the perturbation equation for the 2D Navier Stokes
equations with a Kolmogorov base flow, and obtained
a nonlinear amplitude equation describing 
large-scale flow with Reynolds number slightly
above the critical value $Re^c = \sqrt{2}$.  Generalizing the
work of Sivashinsky \cite{SI} to the $\beta$-plane, 
nonlinear amplitude equations for super-critical, large-scale flow
were derived by Frisch {\it et al.}\ \cite{FLV} for
a zonal Kolmogorov base flow and small $\beta \ll 1$, and
by Manfroi and Young \cite{MY2} for a nearly
meridional Rossby wave base flow and fixed $\beta = O(1)$.
Instability of resonant triad interactions is absent from
\cite{FLV} because the base flow is zonal and $\beta$ is
small, and from \cite{MY2} partly because the Reynolds number is
only slightly supercritical (slightly larger than $\sqrt{2}$,
see \ref{sec_large_beta}).

In all of the above stability analyses, the large-scale instabilities have
a preferred direction because the base flows are anisotropic 
(consisting of a single mode).
More isotropic base flows for the 2D Navier Stokes equations
were considered by Sivashinsky and Yakhot \cite{SY}, who
found that large-scale instability is not 
observed for any finite Reynolds number
if the base flow is sufficiently isotropic.
They used Multiple Scales analysis for the linearized
perturbation equation, with isotropic scaling in 
space and periodic base flows consisting of several modes. 
Recently, Novikov and Papanicolau \cite{NP} have extended the 
linear analysis of \cite{SY}
to nonlinear analysis for a rectangular base flow.

For inviscid $\beta$-plane flow with large Rhines number,
Lorenz \cite{L} found growth rates of linearly unstable modes
from a low-order truncation of the infinite dimensional eigenvalue
problem associated with a base flow consisting of a Rossby wave and a
constant zonal flow.
Gill \cite{G} considered a Rossby wave base flow, and using the continued
fraction formulation, noted the transition from inflectional instability
for $Rh\to \infty$ to resonant triad instability for $Rh\to 0$.

In this paper, 
using the continued fraction formulation and 
a Rossby wave base flow with wavevector $\vecp$, 
we obtain the equation for the linear growth rate at the response 
wavevector $\veck$.
We study the equation at various Rhines numbers $Rh \propto 1/\beta$ to find
the growth rate analytically or numerically, and obtain
critical Reynolds numbers in different limits. 
For $Rh \to \infty$ and $k\to 0$, the critical Reynolds number
approaches the value $\sqrt{2}$ for the 2D Navier Stokes equations.
Manfroi and Young \cite{MY2} have also recently studied linear
stability of the $\beta$-plane equation in the $Rh \to \infty$ limit, 
but with Kolmogorov base flows instead of 
Rossby wave base flows.  Using Multiple Scales analysis
in the limits $Rh\to \infty$ and $k\to 0$, 
they find a critical Reynolds number smaller than the value 
$\sqrt{2}$ for the 2D Navier Stokes equations.
Our analysis is compared to that of Manfroi and Young \cite{MY}
in Section 5.  For small Rhines numbers (large $\beta$), 
we discuss the instability of resonant triad interactions.
We also generalize the work of Sivashinsky and Yakhot \cite{SY}
for more isotropic base flows to the $\beta$-plane equation.


The organization of the paper and the main results are as follows.
In Section \ref{sec_intro}, the linear stability problem is
formulated.  For forcing of a single mode, Floquet theory and 
continued fractions lead to
an infinite dimensional eigenvalue problem.  For finite Reynolds numbers, 
convergence of the continued fractions is guaranteed, and the convergence
rate is increased for large-scale instability with $k \ll 1$.
(Convergence of the continued fractions is discussed in Appendix
\ref{app_CF}.)
In Section \ref{sec_small_beta}, we consider the base flow wavevector
$\vecp = (p_x,p_y) = (\cos(\alpha_\vecp),\sin(\alpha_\vecp))$ and
response wavevector $\veck$, with  $k \ll 1$ and fixed $Rh$.  In this case
the stability problem can be well approximated by a 
low dimensional ($3 \times 3$) 
eigenvalue problem that may be solved analytically.  
This truncated system is expanded in powers of $k$ to find the
growth rate of the perturbation with $k\ll 1$.
For fixed $Re$ and $Rh$, this growth rate is given by
\begin{eqnarray*}
\Re(\sigma(\veck;\alpha_\vecp,Re,Rh)) &=& -\frac{k^2}{Re}
\left[
1-\frac{Re^2}{2}\frac{\sin^2{\alpha}}{(1+Re^2 {\omega}^2)^2}
\times\right. \\
\\
& &\left. \left( 1-8\cos^2{\alpha} + Re^2 {\omega}^2\right)
\right] + O(k^3)
\end{eqnarray*}
where $\Re(z)$ denotes the real part of a complex number
$z$, $\omega(\veck)$ is the wave frequency
given by the dispersion relation 
$\D \omega(\veck) = -Rh^{-1}k_x/k^2$, and $\alpha$ is the angle between
the base flow wavevector $\vecp$ and the perturbation wavevector
$\veck$.  The marginal Reynolds number for long-wave instability 
$Re^0(\veck;\alpha_\vecp,Rh)$ corresponds to zero growth rate, and
is found by setting to zero the lowest-order term in
the above expression for the growth rate 
$\Re(\sigma(\veck;\alpha_\vecp,Re,Rh))$.
For fixed $\alpha_\vecp$ and $Rh$, $Re^0(\veck;\alpha_\vecp,Rh)$ for
$k\rightarrow 0$ is the same as the critical Reynolds number
for $k\rightarrow 0$.  We show that $Re^0(\veck;\alpha_\vecp,Rh)$ 
is always greater than the critical Reynolds number
$Re^c =  \sqrt{2}$ for isotropic, 2D, Kolmogorov flow
with $\veck \perp \vecp$ \cite{MS}, \cite{SY},
and \cite{WAL} (see also \cite{FLV} and \cite{MY2}).  
Indeed,  for $Rh\to \infty$ we find
\begin{eqnarray*}
\lim_{Rh\rightarrow \infty}
Re^0(\veck; \alpha_\vecp, Rh) = \sqrt{2},\quad \veck \perp \vecp.
\end{eqnarray*}
\noindent 
For fixed $Rh$ and $|\veck|\rightarrow 0$, we find
\begin{eqnarray*}
&&\lim_{k_0\rightarrow 0} \min_{|\veck|<k_0}
Re^0(\veck;\alpha_\vecp,Rh)
\\
\\
&&
\quad =
\left\{
\begin{array}{ll}
\D{
\sqrt{\frac{2}{\cos^2\alpha_\vecp(1-8\sin^2\alpha_\vecp)}}
},
& \ 0\le \alpha_\vecp\le \sin^{-1}\frac{1}{4} \\
\\
8|\tan{\alpha_\vecp}|,
& \ \sin^{-1}\frac{1}{4} \le \alpha_\vecp \le \pi/2
\end{array}
\right. 
\end{eqnarray*}
which is a generalization of the result for the
critical Reynolds number $Re^c = \sqrt{2}$ for
2D isotropic flow in the limit $k \rightarrow 0$ \cite{SY}.

In the next Section \ref{sec_mid_beta}, we consider $|\veck| < 1$ and
$0.1 \leq Rh \leq \infty$ by numerical computation of growth
rates for a less severe truncation (31 by 31) of the 
infinite dimensional eigenvalue problem.  An important observation
is that the nature of the instability to large scales changes
from an inflectional instability to a triad resonance as
the Rhines number is decreased from $Rh = \infty$ (Figure~\ref{pic_CN_0}), 
where the transition Rhines number is $O(1)$
(See Gill \cite{G}).
For forcing wavevector with $\alpha_\vecp \neq \pi/2$ and $Rh < 1$, there is
a band of unstable wavevectors near $\veck$ resonant with
$\pm \vecp$, and the width of this band decreases as $Rh$
decreases.  As shown in reference \cite{LG}, resonant triad interactions 
cannot transfer energy to modes with $k_x=0$, and indeed the 
numerical computations show the most unstable wavevector $\veck$
near the resonant trace has $k_x \neq 0$.
For large Rhines numbers and Reynolds numbers above
critical ({\it e.g.}, $Re = 10$), the inflectional instability is
strongest for $\alpha_\vecp = 0$ and perturbation wavevector
with $k_x=0$.  

The limit $Rh \rightarrow 0$ is studied in Section \ref{sec_large_beta}.
The main result is that for $Rh\rightarrow 0$, the critical Reynolds
number is given by
\begin{eqnarray*}
\lim_{Rh \rightarrow 0} 
Re^{c}(\alpha_\vecp,Rh) = 
\left\{
\begin{array}{ll}
\sqrt{2}, &\quad \alpha_\vecp = 0 \\
0,       &\quad 0 < \alpha_\vecp  < \pi/2 \\
\infty,   &\quad \alpha_\vecp = \pi/2.
\end{array}
\right. 
\end{eqnarray*}
The critical Reynolds number is $\infty$ for $\alpha_\vecp = \pi/2$
in the sense that for any Reynolds number, the base flow
is stable if $Rh$ is small enough. Thus, resonant triad interactions
reduce $Re^c$ below the value $\sqrt{2}$ for small $Rh$ and 
$0<\alpha_\vecp <\pi/2$.

In Section \ref{sec_comp_my}, we compare our results with the recent
work by Manfroi and Young \cite{MY2} using Multiple Scales analysis.
In \cite{MY2}, Manfroi and Young consider the $\beta$-plane with
Kolmogorov base flows, rather than the Rossby wave base flows considered
herein.  Because their choice of base flow is different from ours,
the critical Reynolds number for $k\rightarrow 0$ and 
$Rh \rightarrow \infty$ is different from $\sqrt{2}$.    
We confirm their results, and reproduce 
our results for Rossby wave base flows using Multiple Scales
analysis.

In Section \ref{sec_manyf}, base flows consisting of 
several Rossby waves are considered.
The growth rate of large-scale instability is found analytically
for large $Rh$ following \cite{WAL}.   
In Appendix \ref{sec_app_MS}, the method of \cite{WAL} in
Fourier space is shown to be equivalent to the Multiple Scales
analysis following \cite{SY}.  An important result is that an
``isotropic'' base flow consisting of three Rossby waves is
unstable above $Re > \sqrt{32/3}$, 
in contrast to the case of pure 2D flow, which
is stable for a similar base flow \cite{SY}.  We also 
numerically studied the
growth rate of large-scale instability for base flows of many
Rossby waves (up to forty), as a function of $Re$ and $Rh$.
For $2<Rh \ll \infty$ and large enough $Re$, the most unstable
large-scale mode is purely zonal with $k_x=0$.
For small $Rh < 1$ and large enough $Re$, the most unstable
large-scale mode has wavevector $\veck$ with $k_x$ small but nonzero
and $k_x < k_y$, again indicating a tendency towards large-scale
zonal flows.   
A summary is given in Section \ref{sec_disc}.


\section{Linear stability of the $\beta$-plane equation}\label{sec_intro}

\subsection{Base flows and perturbations}\label{sec_basic_setting}

The governing equation for the stream function $\Psi$
($\nabla\times\Psi\hat{\vecz}=\vecu$) in the $\beta$-plane approximation
is given by
\begin{eqnarray}
(\nabla^2\Psi)_t + J(\nabla^2\Psi,\Psi)+\beta\Psi_x
=\nu \nabla^4\Psi + (\nabla\times\vecF)\cdot \hat{\vecz}
\label{beta_eqn_orig}
\end{eqnarray}
where $J(A,B)=A_xB_y-A_yB_x$ is the Jacobian,
$\beta$ is the linear variation of the Coriolis parameter,
$\nu$ is the viscosity and $\vecF$ is a force.
For any nonzero vector $\veca$, 
$\hat{\veca}=\veca/|\veca|$ is the unit vector in the direction of
$\veca$.
With the length scale $p^{-1}$ and the velocity scale $U^{o}$,
the corresponding nondimensional form of (\ref{beta_eqn_orig})
is 
\begin{eqnarray}
(\nabla^2\Psi)_t + J(\nabla^2\Psi,\Psi)+\frac{1}{Rh}\Psi_x
=\frac{1}{Re} \nabla^4\Psi + \frac{1}{Re} (\nabla\times\vecF)\cdot \hat{\vecz}
\label{beta_eqn}
\end{eqnarray}
where the Reynolds number $Re$ and the Rhines number $Rh$
are given by
\begin{eqnarray*}
Re = \frac{U^0}{\nu p}\quad 
\textrm{and} 
\quad
Rh = \frac{p^2 U^0}{\beta}.
\end{eqnarray*}
The inviscid limit of (\ref{beta_eqn}) with
$\vecF=\veczero$ yields wave solutions,
called Rossby waves,
\begin{eqnarray*}
\vecU(\vecx,t;\veck) = (\hat{\veck}\times\hat{\vecz})\hat{U}(\veck)
\exp(i\theta(\vecx,t;\veck)) + c.c.
\end{eqnarray*}
where $c.c$ is the complex conjugate and the phase $\theta(\vecx,t;\veck)$
is given by
\begin{eqnarray}
\theta(\vecx,t;\veck) = \veck\cdot \vecx - \omega(\veck) t
\label{theta}
\end{eqnarray}
with the dispersion relation $\omega(\veck)$
\begin{eqnarray}
\omega(\veck) = -\frac{1}{Rh} \frac{k_x}{k^2}.
\label{dispersion}
\end{eqnarray}
\noindent
In this paper each wavevector $\veck=(k_x,k_y)$
has the polar coordinates representation $(k,\alpha_\veck)$ so that
$\veck=k(\cos\alpha_\veck,\sin\alpha_\veck)$.

Consider a base flow $\vecU^{0}(\vecx,t)$ for 
(\ref{beta_eqn}) given
by a sum of $m$ Rossby waves of unit wavevectors
$\vecp_j=(\cos\alpha_{\vecp_j},\sin\alpha_{\vecp_j})$,
\begin{eqnarray}
\vecU^{0}(\vecx,t;\{\vecp_j\}_{j=1}^{m})
=\sum_{j=1}^{m} (\vecp_j\times \hat{\vecz})\sin\theta(\vecx,t;\vecp_j).
\label{base_flow}
\end{eqnarray}
Throughout this paper, we always assume that wavevectors $\vecp_j$
are unit vectors unless otherwise specified.
This flow has a stream function
\begin{eqnarray}
\Psi^{0}(\vecx,t;\{\vecp_j\}_{j=1}^{m}) 
= - \sum_{j=1}^{m} \cos\theta(\vecx,t;\vecp_j)
\label{psi_general}
\end{eqnarray}
and can be maintained by forcings at $m$ wavevectors 
$\vecp_j$ (and their conjugates)
with the following force:
\begin{eqnarray*}
\vecF(\vecx,t;\{\vecp_j\}_{j=1}^{m})=\sum_{j=1}^{m}(\vecp_j\times \hat{\vecz})
\sin\theta(\vecx,t;\vecp_j).
\end{eqnarray*}
Note that $\Psi^{0}$ satisfies $\nabla^2\Psi^{0} + \Psi^{0} = 0$.
In particular, for $m=1$ and $Rh=\infty$ ($\beta=0$), this flow
is a sinusoidal Kolmogorov flow for the 2D
Navier Stokes equations.

A small perturbation $\psi(\vecx,t)=\Psi(\vecx,t)-\Psi^{0}(\vecx,t)$
satisfies the following linearized perturbation equation,
\begin{eqnarray}
(\nabla^{2}\psi)_t 
+\Psi^{0}_x(\psi+\nabla^2\psi)_y 
-\Psi^{0}_y(\psi+\nabla^2\psi)_x
+Rh^{-1}\psi_{x} - Re^{-1}\nabla^{4}\psi
=0,
\label{beta_eqn_pert}
\end{eqnarray}
where the quadratic nonlinear terms in $\psi$ have been dropped
for the linear stability analysis.
To study the linear stability at wavevector $\veck$, we use
Floquet theory to seek a solution $\psi$ of the form
\begin{eqnarray}
\psi(\vecx,t;\{\vecp_j\}_{j=1}^{m}) = e^{\lambda t}e^{i\theta(\vecx,t;\veck)}
\sum_{\vecn} \psi_{\vecn}\exp\left(
i\sum_{j=1}^{m} n_j\theta(\vecx,t;\vecp_j)
\right),
\label{pert_form}
\end{eqnarray}
where the first sum is over $\vecn=(n_1,\cdots,n_m)$.
Inserting this form of $\psi$ into (\ref{beta_eqn_pert}),
the following infinite dimensional linear eigenvalue problem is obtained
\begin{eqnarray}
&&\D{
\left[\lambda+Re^{-1}q^2(\vecn)
+i\omega_{\vecn}\right]\psi_{\vecn}
}
\nonumber\\
\nonumber\\
&&\D{
-\frac{1}{2q^2(\vecn)}
\sum_{j=1}^{m}(\vecp_j\times\vecq(\vecn-\hat{\vece}_j))
(q^2(\vecn-\hat{\vece}_j)-1)\psi_{\vecn-\hat{\vece}_j}
}
\nonumber\\
\nonumber\\
&&\D{
+\frac{1}{2q^2(\vecn)}
\sum_{j=1}^{m} (\vecp_j\times\vecq(\vecn+\hat{\vece}_j))
(q^2(\vecn+\hat{\vece}_j)-1)\psi_{\vecn+\hat{\vece}_j}
} = 0
\label{ILS_m}
\end{eqnarray}
where $\vecq(\vecn) = \veck+\sum_{j=1}^{m}n_j\vecp_j$,
$q(\vecn)=|\vecq(\vecn)|$ and 
$\hat{\vece}_j$ is the $j$-th unit vector ($\hat{\vece}_j=
(0,\cdots,1,\cdots,0)$).
Also, $\omega_\vecn$
is given by
\begin{eqnarray}
\omega_{\vecn}=-\omega(\veck)-\left(\sum_{j=1}^{m}n_j\omega(\vecp_j)\right)
+\omega(\vecq(\vecn)).
\label{triad_phses_m}
\end{eqnarray}

Our goal is to find 
the eigenvalue $\lambda$ of (\ref{ILS_m}) with the largest
real part for each $\veck$ with fixed $Re$, $Rh$ and $\vecp_j$,
and this eigenvalue is denoted by
$\sigma(\veck;\{\alpha_{\vecp_j}\}_{j=1}^{m},Re,Rh)$.
Thus, the linear growth rate is given by its real part
$\Re(\sigma(\veck;\{\alpha_{\vecp_j}\}_{j=1}^{m},Re,Rh))$
where $\Re(z)$ is the real part of any complex number $z$.
If the growth rate is positive for $\veck$, the base flow is
unstable to a perturbation with wavevector $\veck$.

Although it is not easy to analyze (\ref{ILS_m}) for general $m$, 
the case of $m=1$ can be studied using the technique of
continued fractions following Meshalkin and Sinai \cite{MS}. For $m=1$ with
$\vecp = \vecp_1$, the base flow (\ref{psi_general}) is
\begin{equation}\label{base_flow_one}
\Psi^{0}(\vecx,t;\vecp) = -\cos\theta(\vecx,t;\vecp),
\end{equation}
and (\ref{ILS_m}) reduces to
\begin{eqnarray}
[\lambda + Re^{-1} q^2(n) + i\omega_n]\psi_n
-\frac{\vecp\times\veck\cdot\hat{\vecz}}{2q^2(n)}(q^2(n-1) - 1)\psi_{n-1}
\nonumber\\
\nonumber\\
+\frac{\vecp\times\veck\cdot\hat{\vecz}}{2q^2(n)}(q^2(n+1) - 1)\psi_{n+1} =
0.\label{ILS}
\end{eqnarray}
Here, $\vecq(n) = \veck+n\vecp$, $q(n) = |\vecq(n)|$ and 
\begin{equation}\label{triad_phase}
\omega_n = -\omega(\veck) -n \omega(\vecp) + \omega(\veck+\vecp).
\end{equation}
We define $a_n(\lambda, \veck, \alpha_\vecp, Re,Rh)$ and 
$d_n(\veck,\alpha_\vecp,\psi_n)$ by
\begin{eqnarray}
\begin{array}{l}
a_n(\lambda, \veck,\alpha_\vecp,Re,Rh) = 
\D{
-\frac{2q^2(n)(\lambda + Re^{-1} q^2(n) + i\omega_n)}
{(\vecp\times\veck\cdot\hat{\vecz})(q^2(n) -1 )}
},\\
\\
d_n(\veck,\alpha_\vecp,\psi_n) = (q^2(n)-1)\psi_n.
\end{array}
\label{an_dn}
\end{eqnarray}
Then, (\ref{ILS}) can be written as
\begin{eqnarray*}
a_n d_n + d_{n-1} - d_{n+1}=0.
\end{eqnarray*}
Formally, one can show that $a_n$ satisfies
\begin{eqnarray}
\displaystyle{
a_0
+\frac{1}{a_1+\frac{1}{a_2+\frac{1}{a_3+\cdots}}}
+\frac{1}{a_{-1}+\frac{1}{a_{-2}+\frac{1}{a_{-3}+\cdots}}}
=0.
}
\label{CF}
\end{eqnarray}

This idea was first explored by Meshalkin and Sinai \cite{MS} for
the 2D Navier Stokes equations 
in the special case when $\veck$ and $\vecp$ are perpendicular
to each other. They obtained the classical critical Reynolds number
$\sqrt{2}$ for the first instability at large scales. Later, 
Friedlander {\it et al.} \cite{FR1}, Belenkaya {\it et al.} 
\cite{FR2} and Li \cite{LI} 
used similar techniques to study
the stability of a Kolmogorov flow for the inviscid ($Re=\infty$)
Navier Stokes equations (Euler equations). 
In this paper, we consider only the viscous ($Re<\infty$) case 
to have fast convergence
of the continued fractions in equation (\ref{CF}). 
The convergence is
straightforward since for fixed $\lambda$, $\veck$, $\alpha_\veck$, $Re$ and
$Rh$, we have
\begin{eqnarray*}
|a_n|\rightarrow\infty \textrm{ as } |n|\rightarrow\infty
\end{eqnarray*}
due to the viscous term $Re^{-1} q(n)^2$ in the numerator of $a_n$.
In Appendix \ref{app_CF}, 
we discuss 0 the convergence of the continued fractions
in equation (\ref{CF}). For each $\lambda$
satisfying (\ref{CF}), one can construct $\{\psi_n\}$
so that (\ref{pert_form}) is well-defined. For this discussion,
we refer to \cite{FR1}, \cite{FR2} and \cite{LI}.
Unfortunately, the technique of continued fractions can not be extended
for $m>1$. 

\subsection{Numerical computation of linear growth rates}
\label{sec_numeric}

For all figures of numerically computed linear growth rates, 
contours of the real part of
$\sigma(\veck;\{\alpha_{\vecp_j}\}_{j=1}^{m},Re,Rh)$
are plotted. To indicate the unstable wavevectors, the shaded
regions consist of $\veck$ with 
$\Re(\sigma(\veck;\{\alpha_{\vecp_j}\}_{j=1}^{m},Re,Rh))>0$ 
and the darker regions
consist of $\veck$ with
\begin{eqnarray*}
\Re(\sigma(\veck;\{\alpha_{\vecp_j}\}_{j=1}^{m},Re,Rh)) > 0.9\ 
\max_{\tilde{\veck}}\
\Re(\sigma(\tilde{\veck};\{\alpha_{\vecp_j}\}_{j=1}^{m},Re,Rh))
\end{eqnarray*} 
where the maximum is taken over all 
$\tilde{\veck}$ under consideration in each figure. Thus, the wavevector
with the strongest instability is inside the darker regions.

For base flow (\ref{base_flow_one}) with $m=1$, the linear growth rate 
$\Re(\sigma(\veck;\alpha_\vecp,Re,Rh))$ is the real part of
the solution to (\ref{CF}) with the largest real part.  
Since we consider the viscous
case ($Re<\infty$), the continued fractions in (\ref{CF}) have fast
convergence. Thus, we can truncate the continued fractions in (\ref{CF})
for $|n|>N$ for some large $N$. This is equivalent to truncating the
infinite dimensional eigenvalue problem (\ref{ILS}) for $|n|>N$ and
considering the $(2N+1)\times (2N+1)$ finite dimensional eigenvalue
problem.
We find all eigenvalues for the truncated eigenvalue problem using the
MATLAB function 'eig'. Then, $\sigma(\veck;\alpha_\vecp,Re,Rh)$
is the eigenvalue with the largest real part. For most numerical results,
we use $N=15$. We check the results with $N=16$ for some cases and
find no significant difference between the results for $N=15$ and
$N=16$.

\begin{figure}
\begin{center}
\begin{tabular}{c}
\includegraphics[width=8cm]{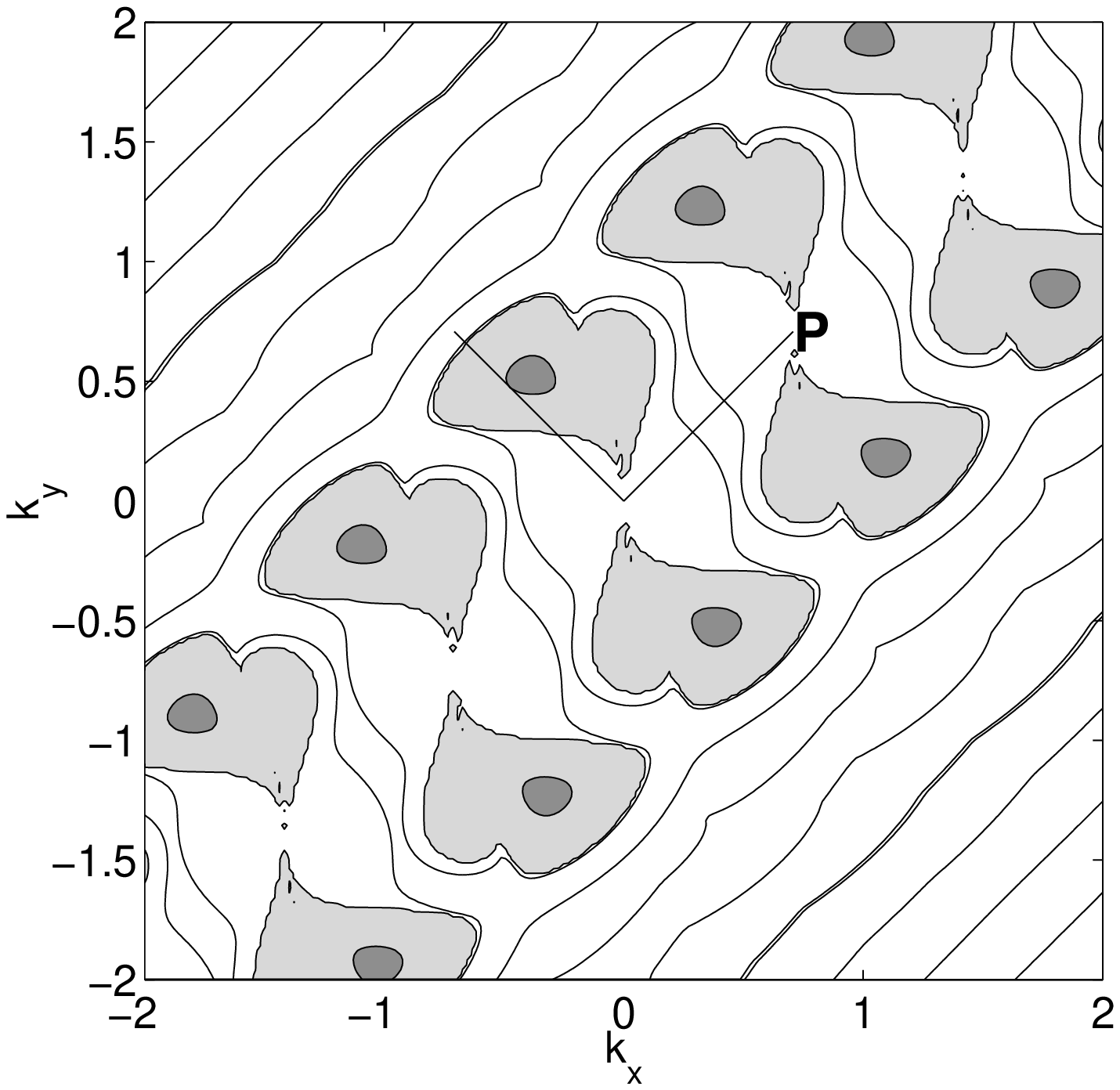}
\end{tabular}
\caption{Contour plot of $\Re(\sigma(\veck;\alpha_\vecp,Re,Rh))$ for
for $\alpha_\vecp=\pi/4$ at $Re = 10$ and $Rh=2$.
$\Re(\sigma(\veck;\alpha_\vecp,Re,Rh))$ is positive in the shaded regions.
}
\label{pic_basic}
\end{center}
\end{figure}

For the real part of $\sigma(\veck;\alpha_\vecp,Re,Rh)$, 
one can check 
the symmetry
\begin{eqnarray*}
\Re(\sigma(\veck;\alpha_\vecp,Re,Rh)) = 
\Re(\sigma(-\veck;\alpha_\vecp,Re,Rh))
\end{eqnarray*}
from the form of our perturbation (\ref{pert_form}) with $m=1$. 
This enables us to consider wavevectors in the half-plane.
Another interesting
observation is that 
$\Re(\sigma(\veck;\alpha_\vecp,Re,Rh))$ is invariant under the shift by
$j\vecp$ for $j=0,\pm 1, \pm 2, \cdots$. By the form of our perturbation
(\ref{pert_form}), one can also show that if $\lambda$ is an eigenvalue
of (\ref{ILS}) for $\veck=\veck_0$, then
\begin{eqnarray*}
\lambda + i(-\omega(\veck_0)-j\omega(\vecp)+\omega(\veck_0 + j\vecp))
\end{eqnarray*}
is an eigenvalue of (\ref{ILS}) for $\veck=\veck_0+j\vecp$.
Thus, $\Re(\sigma(\veck;\alpha_\vecp,Re,Rh)) = 
\Re(\sigma(\veck+j\vecp;\alpha_\vecp,Re,Rh))$ for any 
$j=0,\pm 1, \pm 2,\cdots$.
As an example to illustrate these properties, 
Figure~\ref{pic_basic} shows contours of 
$\Re(\sigma(\veck;\alpha_\vecp,Re, Rh))$ for $\alpha_\vecp=\pi/4$,
$Re =  10$ and $Rh = 2$ in 
$-2<k_x<2$ and $-2<k_y<2$ with a grid of size $200\times 200$. 
From Figure~\ref{pic_basic}, we observe the symmetry and the shift
invariance.
Thus, it is sufficient to consider wavevectors $\veck$ with
$|\veck\cdot\vecp|\le 1/2$ and $(\vecp\times\veck)\cdot\hat{\vecz}\ge 0$.
For special cases $\alpha_\vecp=0$ and $\alpha_\vecp=\pi/2$, 
we have the additional symmetry
$\Re(\sigma((k_x,k_y);\alpha_\vecp,Re,Rh)) = 
\Re(\sigma((\pm k_x,\pm k_y);\alpha_\vecp,Re,Rh))$ which is
due to the symmetry of $\omega(\veck) = -Rh^{-1}k_x/k^2$.  
This allows us to consider only wavectors $\veck$ with $0\le k_x\le 1/2$
and $k_y \ge 0$ for $\alpha_\vecp = 0$,
and $k_x\ge 0$ and $0\le k_y\le 1/2$ for $\alpha_\vecp = \pi/2$.
It is not easy to find the growth rate numerically for $m>1$
since the construction of
the matrix for the eigenvalue problem of (\ref{ILS_m}) is nontrivial.
Thus, later in Section \ref{sec_manyf}, we use ``severe'' truncation
of (\ref{ILS_m}) to find approximate growth rates for $m>1$.

\section{Large Scale Analysis for fixed $Rh$}\label{sec_small_beta}

In this section, we derive the growth rate for small $k \ll 1$
with base flow (\ref{base_flow_one})
at fixed $\alpha_\vecp$, $Re$ and $Rh$ 
by asymptotic expansion of equation (\ref{CF})
of continued fractions.
From the growth rate, we also find the critical Reynolds number
in the limit $|\veck|\to 0$.
From (\ref{CF}), we show that
the linear growth rate 
$\Re(\sigma(\veck;\alpha_\vecp,Re,Rh))$
is given by 
\begin{eqnarray}
\Re(\sigma(\veck;\alpha_\vecp,Re,Rh))
&=& 
-\frac{k^2}{Re}
\left[
1-\frac{Re^2}{2}\frac{\sin^2{\alpha}}{(1+Re^2\omega^2)^2}
\times \right.\nonumber\\
\nonumber\\
&&\left.\left( 1-8\cos^2{\alpha}+Re^2\omega^2\right)
\right] + H_3 + H_4
\label{growth_rate}
\end{eqnarray}
where $\omega = \omega(\veck)$ is the dispersion relation
(\ref{dispersion})
and $\alpha=\alpha_\veck-\alpha_\vecp$ is the angle between 
the response wavevector $\veck$ and the base flow wavevector $\vecp$.
The higher order terms $H_3$ and $H_4$ satisfy
$$ |H_3| \le C_3 k^3,\qquad  |H_4| \le C_4 k^4 $$
for some nonnegative constants $C_3$ and $C_4$ depending on $\alpha_\vecp$, 
$Re$ and $Rh$ but not on $\veck$. 
The explicit expression for $H_3$ is given below
(see (\ref{higher_order})).
In particular, $C_3$ is proportional to $Rh^{-1}$ and
is zero when $Rh=\infty$. 
Thus, if $Rh=\infty$, then
the growth rate (\ref{growth_rate}) is reduced to
\begin{eqnarray*}
\Re(\sigma(\veck;\alpha_\vecp,Re,Rh=\infty))
=-\frac{k^2}{Re}\left[
1-\frac{Re^2}{2}\sin^{2}{\alpha}(1-8\cos^2{\alpha})
\right] + O(k^4)
\end{eqnarray*}
which is the large-scale growth rate obtained by Sivashinsky and
Yakhot \cite{SY} and by Waleffe \cite{WAL}.
Note that no assumption is made on $Rh$ for (\ref{growth_rate}).
However, since the higher order term $H_3$ is $O(k^3/Rh)$,
the leading order term in (\ref{growth_rate}) is a good approximation
of the actual growth rate if $Rh$ is large.
The leading order term
can also be derived using Multiple Scales analysis by assuming 
that the Rhines number is large as shown in Appendix \ref{sec_app_MS}. 
See \cite{S}, \cite{SY}, \cite{FLV}, \cite{MY2} and 
\cite{MY} for more details. 
Note that one advantage of using equation (\ref{CF}) of continued fractions is
that we can obtain the growth rate (\ref{growth_rate}) up to 
third order in $k$ explicitly.

\subsection{Derivation of the growth rate}\label{sec_small_beta_der}

This section is devoted to deriving the growth rate formula
(\ref{growth_rate}) from (\ref{CF}) for $k\ll 1$. 
We will assume that 
\begin{eqnarray}
\Re(\sigma(\veck;\alpha_\vecp,Re,Rh))\ge -\frac{1}{2Re}.
\label{derive_assumption}
\end{eqnarray}
This assumption is required as follows.
In general, the infinite dimensional eigenvalue problem (\ref{ILS})
has infinitely many eigenvalues for each wavevector $\veck$.
In particular for $\veck = \veczero$, (\ref{ILS}) has eigenvalues
\begin{eqnarray*}
0,-1^2/Re,-2^2/Re,\cdots,
\end{eqnarray*} 
and thus 
$\Re(\sigma(\veczero;\alpha_\vecp,Re,Rh))=0$.
Therefore, for $k \ll 1$ we can expect that (\ref{ILS})
has eigenvalues
\begin{eqnarray*}
O(k),  -1^2/Re + O(k), -2^2/Re + O(k),\cdots. 
\end{eqnarray*}
Assumption (\ref{derive_assumption}) guarantees that 
we obtain the first eigenvalue which has the largest real part
for $k\ll 1$.
In fact, with assumption (\ref{derive_assumption}) one finds that 
$\Re(\sigma(\veck;\alpha_\vecp,Re,Rh))=O(k^2)$ as
shown below (see (\ref{sigma_small_k})).

For $k\ll 1$, we show that (\ref{CF}) can be reduced to
\begin{eqnarray}
a_0 = - \left(\frac{1}{a_1} + \frac{1}{a_{-1}}\right) + R_5.
\label{TCF}
\end{eqnarray}
The remainder $R_5$ satisfies
\begin{eqnarray*}
|R_5| \le C_5 k^5,
\end{eqnarray*}
where 
$C_5$ is a constant depending on $\alpha_\vecp$ and $Re$ but not
on $Rh$ and $\veck$, and 
\begin{eqnarray*}
\lim_{Re\to 0} \frac{C_5}{Re} = 0.
\end{eqnarray*}
To this end, we have 
\begin{eqnarray}
\left|\sigma(\veck;\alpha_\vecp,Re,Rh) + 
(1/Re)q^2(n) + i\omega_n\right| > \frac{1}{3Re} 
\quad n =\pm 1, \pm 2,\ldots
\label{part_top}
\end{eqnarray}
by assumption (\ref{derive_assumption}). 
The expression inside the absolute value is a part of
the numerator of $a_n$ in (\ref{an_dn}).
Other expressions in (\ref{an_dn}) satisfy
\begin{eqnarray*}
\vecp\times\veck\cdot z = O(k),
\quad
q^2(\pm 1)-1 = O(k),
\quad
q^2(\pm n) - 1 = O(1),\quad n=2,3,\ldots,
\end{eqnarray*}
for $k\ll 1$.
Using these and (\ref{part_top}),
one can show that
$a_{n}$ in (\ref{an_dn}) satisfy
\begin{eqnarray}
\begin{array}{l}
\D{\frac{1}{|a_{\pm 1}|}} \le C Re\, k^2 = O(Re\, k^2)\\
\\
\D{\frac{1}{|a_n|}} \le C Re\, k = O(Re\, k),
\quad \textrm{for} \quad n=\pm 2, \pm 3, \ldots
\end{array}
\label{small_k_prop0}
\end{eqnarray}
for some constant $C$ independent of $Rh$.
Using (\ref{small_k_prop0}), one finds that 
\begin{eqnarray}
\frac{1}{a_{\pm 2}+\frac{1}{a_{\pm 3} +\frac{1}{a_{\pm 4}+\cdots}}} = O(Re\, k),
\label{small_k_prop1}
\end{eqnarray}
so that we have
\begin{eqnarray}
\frac{1}{a_{\pm 1}+\frac{1}{a_{\pm 2} +\frac{1}{a_{\pm 3}+\cdots}}} 
= \frac{1}{a_{\pm 1}} + O(Re^3 k^5).
\label{small_k_prop2}
\end{eqnarray}
The details of the derivation of 
(\ref{small_k_prop1}) and (\ref{small_k_prop2})
are given in Appendix \ref{app_CF}.
Equation (\ref{TCF}) follows from
(\ref{CF}) using (\ref{small_k_prop1}) and (\ref{small_k_prop2}).

The growth rate (\ref{growth_rate}) is obtained 
from (\ref{TCF}) by asymptotic expansions
of $a_{\pm 1}$ in $k$, as will now be shown.
A Taylor expansion of (\ref{triad_phase}) gives 
\begin{eqnarray}
\omega_{\pm 1} 
&=& -\omega(\veck) \mp \omega(\vecp) \pm \omega(\vecp\pm\veck)\nonumber\\
&=& -\omega(\veck) +D\omega(\vecp)\cdot\veck 
\pm \D{\frac{1}{2}}\veck^{T}D^2\omega(\vecp)\cdot\veck
+ O(k^3/Rh)\label{omega_taylor}
\end{eqnarray}
where $D\omega(\vecp)$ and $D^2\omega(\vecp)$ are the first and
second derivatives of $\omega$ at $\vecp$ respectively and
$\veck^{T}$ is the transpose of $\veck$.
From (\ref{omega_taylor}) one can see that
$\omega_1$ and $\omega_{-1}$ are the same up to 
$O(k^2/Rh)$ for $k\ll 1$. For simplicity, let 
$\Omega_1 = D\omega(\vecp)\cdot\veck$, 
$\Omega_2 = \D{\frac{1}{2}} \veck^T D^2\omega(\vecp)\cdot\veck$
and $\sigma = \sigma(\veck;\alpha_\vecp,Re,Rh)$.
Note that $\Omega_1=O(k/Rh)$ and $\Omega_2=O(k^2/Rh)$.
From (\ref{an_dn}), $a_0$ and $a_{\pm 1}$ can be rewritten as
\begin{eqnarray}
\begin{array}{l}
\D{
a_0 = \frac{2k(\sigma + k^2/Re)}{\sin{\alpha}(1-k^2)}
}\\
\\
\D{
a_{\pm 1} =
-\frac{2(1\pm 2k\cos{\alpha}+k^2)(\sigma + (1\pm 2k\cos{\alpha}+k^2)/Re
+i\omega_{\pm 1})}
{k^2\sin{\alpha}(\pm 2\cos{\alpha} + k)}
}.
\end{array}
\label{an_expression}
\end{eqnarray}
For simplicity, we set
\begin{eqnarray}
\begin{array}{ll}
q_{\pm } = \sqrt{1 \pm 2k\cos\alpha+k^2},
& 
\quad
d_o = \sigma + Re^{-1} -i\omega,
\\
d_{\pm 1} = \pm 2k\cos\alpha/Re + i\Omega_1,
&
\quad
d_{\pm 2} = k^2/Re \mp i\Omega_2
\end{array}
\label{def_q_di}
\end{eqnarray}
Expanding $1/a_{\pm 1}$ in $k$ and using (\ref{omega_taylor})
and \ref{def_q_di},
we find that
\begin{eqnarray}
-\frac{1}{a_{\pm 1}}
&=&
\frac{k^2 \sin{\alpha}(\pm 2\cos{\alpha}+k)}
{2q_{\pm 1}^2
\left\{
d_0 + d_1
+ d_2 + O(k^3/Rh) 
\right\}
}\nonumber \\
\nonumber \\
&=&
\frac{k^2\sin\alpha(2\cos{\alpha}\pm k)}
{ 2q_{\pm 1}^2 d_0
\D{
\left(
1+\frac{d_{\pm 1}}{d_0}
+\frac{d_{\pm 2}}{d_0}
+O(k^3/Rh) 
\right)}
}\nonumber \\
\nonumber \\
&=&
\frac{k^2\sin\alpha}{2 d_0}
(\pm 2\cos\alpha+k)\Big( 1\mp 2k\cos\alpha+k^2(4\cos^2{\alpha}-1)+
\nonumber\\
\nonumber \\
& &
O(k^3)\Big)\times
\D{
\left(1-\frac{d_{\pm 1}}{d_0}
-\frac{d_{\pm 2}}{d_0}
+\left( \frac{d_{\pm 1}}{d_0}\right)^2
+O(k^3)
\right)
}\nonumber\\
\nonumber\\
&=&
\frac{k^2\sin\alpha}{2 d_0}
\left[
\pm 2\cos\alpha + 
k\left\{
1 - 4\cos^2\alpha\left(1+\frac{Re^{-1}}{d_0}\right)
\right\}
\right.
\nonumber\\
\nonumber\\
& & \left.
\mp \frac{2i\Omega_1\cos\alpha}{d_0}
\pm G_1 + i G_2
\right] + O(k^5),\label{an_expansion}
\end{eqnarray}
where $G_1$ and $G_2$ are given by
\begin{eqnarray*}
G_1 &=&  \frac{2\cos\alpha}{d_0}
\left( \Omega_2
+2k\cos\alpha\Omega_1
+\frac{2k\cos\alpha\Omega_1/Re}{d_0}
\right)
-\frac{k\Omega_1}{d_0}, \\
\\
G_2 & = & \frac{1}{d_0}
\left(
2\cos\alpha(\Omega_2+2k\cos\alpha\Omega_1)-k\Omega_1
+\frac{8k\cos^2\alpha\Omega_1/Re}{d_0}
\right).
\end{eqnarray*}
Assumption (\ref{derive_assumption})
guarantees that
the denominator $d_0= \sigma + Re^{-1} -i\omega$ is not zero.

Using (\ref{an_expansion}) and (\ref{def_q_di}), (\ref{TCF}) leads to
\begin{eqnarray}
\sigma
&=&
\D{
-\frac{k^2}{Re}
\left[
1 - \frac{\sin^2{\alpha}}{2(\sigma+Re^{-1}-i\omega)}
\left\{ 1 - 4\cos^2{\alpha}\left( 1+\frac{Re^{-1}}{\sigma + Re^{-1}-i\omega}
\right) \right\}
\right]
} \nonumber\\
\nonumber \\
& &
+ i\frac{k\sin^2\alpha}{2(\sigma+Re^{-1}-i\omega)}G_2 + O(k^4).
\label{sigma_small_k}
\end{eqnarray}
From expression (\ref{sigma_small_k}), we see that 
$|\sigma(\veck;\alpha_\vecp,Re,Rh)| = O(k^2/Re)$ 
as mentioned before.
Expanding the denominator $(\sigma+Re^{-1}-i\omega)$, we obtain 
\begin{eqnarray}
\sigma(\veck) &= &
\D{
-\frac{k^2}{Re}
\left[
1 - \frac{\sin^2{\alpha}}{2(Re^{-1}-i\omega)}
\left\{ 1 - 4\cos^2{\alpha}\left( 1+\frac{Re^{-1}}{Re^{-1}-i\omega}
\right) \right\}
\right]
}\nonumber \\
\nonumber\\
& & + i\frac{k\sin^{2}\alpha}{2(Re^{-1}-i\omega)}\tilde{G}_2
 + O(k^4)\label{sigma_small_k_2}
\end{eqnarray}
where $\tilde{G}_2$ is given by
$$
\tilde{G}_2 = \frac{1}{Re^{-1}-i\omega}
\left(2\cos\alpha(\Omega_2+2k\cos\alpha\Omega_1) - k\Omega_1
+\frac{8k\cos^2\alpha\Omega_1/Re}{Re^{-1}-i\omega}\right).
$$
\noindent
Note that $\tilde{G}_2$ is $O(k^2/Rh)$ since 
$\Omega_1 = O(k/Rh)$ and $\Omega_2 = O(k^2/Rh)$. In particular,
$\tilde{G}_2 = 0$ if $Rh=\infty$.
Finally, the growth rate (\ref{growth_rate})
is obtained by taking the real part of $\sigma$ 
in (\ref{sigma_small_k_2}). 
The third order term $H_3$ in (\ref{growth_rate}) is given by
\begin{eqnarray}
H_3=\Re\left( i\frac{k\sin^{2}\alpha}{2(Re^{-1}-i\omega)}\tilde{G}_2\right)
\label{higher_order}
\end{eqnarray}
and $H_4$ is the fourth order remainder.
Note that $H_3$ is $O(k^3/Rh)$ since $\tilde{G}_2=O(k^2/Rh)$.
In particular $H_3=0$ if $Rh=\infty$.

\subsection{Marginal Reynolds numbers $Re^{0}(\veck;\alpha_\vecp,Rh)$}
\label{sec_cre_smallB}

The critical Reynolds number is the smallest nonnegative Reynolds above
which a base flow base becomes unstable. With base flow
(\ref{base_flow_one}), the critical Reynolds number depends on
$\alpha_\vecp$ and $Rh$, and is denoted by $Re^{c}(\alpha_\vecp,Rh)$.
Similarly, the critical Reynolds number 
$Re^{c}(\veck;\alpha_\vecp,Rh)$ 
at a fixed wavevector $\veck$
is defined by the smallest nonnegative Reynolds number above which 
$\Re(\sigma(\veck;\alpha_\vecp,Re,Rh))$ is positive, and can be found by
setting the growth rate equal to zero.
In other words, if the Reynolds number is larger than
$Re^{c}(\veck;\alpha_\vecp,Rh)$,
then the wavevector $\veck$ becomes unstable.
If such a Reynolds number does not
exist at $\veck$, we define $Re^{c}(\veck;\alpha_\vecp,Rh)=\infty$
in the sense that the wavevector $\veck$ is stable for all finite
Reynolds numbers. 
Note that $Re^{c}(\veck;\alpha_\vecp,Rh)=\infty$ 
does not imply that the wavevector $\veck$
is unstable for the inviscid ($Re = \infty$) $\beta$-plane equation.
In terms of $Re^{c}(\veck;\alpha_\vecp,Rh)$, $Re^{c}(\alpha_\vecp;Rh)$
satisfies
\begin{eqnarray*}
Re^{c}(\alpha_\vecp,Rh) = \min_{\veck}Re^{c}(\veck;\alpha_\vecp,Rh).
\end{eqnarray*}

The critical Reynolds number $Re^{c}(\veck;\alpha_\vecp,Rh)$
for $k\ll 1$
may be found by setting the
growth rate (\ref{growth_rate}) to be zero. 
However, it is nontrivial to find $Re^{c}(\veck;\alpha_\vecp,Rh)$
from the growth rate (\ref{growth_rate}) since the explicit expression for
the higher-order term $H_4$ in (\ref{growth_rate}) is not known. 
Instead, we study the smallest nonnegative 
Reynolds number which make the leading order term of
(\ref{growth_rate}) equal to 
zero by solving 
\begin{eqnarray}
Re^2 = \frac{2(1+Re^2\omega^2)^2}{\sin^2{\alpha}
\left\{1-8\cos^2\alpha + Re^2\omega^2\right\}}
,\ \textrm{and}\
1-8\cos^2{\alpha} + Re^2\omega^2 > 0.
\label{re0_def}
\end{eqnarray}
It is called the marginal Reynolds number and
denoted by $Re^{0}(\veck;\alpha_\vecp,Rh)$. 
Similar to the critical
Reynolds number, we define $Re^{0}(\veck;\alpha_\vecp,Rh)=\infty$
if no Reynolds number satisfying (\ref{re0_def}) exists for $\veck$. 
\noindent
In particular, for $Rh=\infty$, the marginal Reynolds number is
given by
\begin{eqnarray*}
Re^{0}(\veck;\alpha_\vecp,Rh=\infty)
=
\sqrt{\frac{2}{\sin^2{\alpha}(1-8\cos^2{\alpha})}}
\quad \textrm{ for }1-8\cos^2{\alpha} > 0,
\end{eqnarray*}
and its minimum  is equal to $\sqrt{2}$ for $\alpha=\pi/2$
as shown by Sivashinsky and Yakhot \cite{SY} and by Waleffe \cite{WAL}. 
In general,  it is clear that
$Re^{0}(\veck;\alpha_\vecp,Rh)$
is different from the critical Reynolds number
$Re^{c}(\veck;\alpha_\vecp,Rh)$ but
can be understood as the critical Reynolds number
in the limit $k\to 0$.\footnote{
$Re^{0}(\veck;\alpha_\vecp,Rh)$ is the critical Reynolds number 
in the limit $k\to 0$ in the sense that
\begin{eqnarray*}
\lim_{k_0\to 0}\min_{|\veck|<k_0} Re^{0}(\veck;\alpha_\vecp,Rh) =
\lim_{k_0\to 0}\min_{|\veck|<k_0} Re^{c}(\veck;\alpha_\vecp,Rh).
\end{eqnarray*}
This can be shown by an argument similar to the one in 
Section \ref{sec_re_k0}.
}
Since the higher order term $H_3$ in the growth rate
(\ref{growth_rate}) is proportional to $1/Rh$, 
$Re^{0}$ is meaningful for
large Rhines numbers.

We show that the marginal Reynolds number 
satisfies
\begin{eqnarray}
Re^{0}(\veck;\alpha_\vecp,Rh) \ge \sqrt{2},
\label{re0gesqrt2}
\end{eqnarray}
where the value $\sqrt{2}$ is the classical critical Reynolds number 
for the 2D Navier Stokes equations
\cite{MS}, \cite{SY}, and \cite{WAL} (see also \cite{FLV} and \cite{MY2}).
This inequality can be easily shown as follows:
if $(1-8\cos^2{\alpha} + Re^2\omega^2)>0$, then from (\ref{re0_def})
\begin{eqnarray*}
&&\left[Re^0(\veck;\alpha_\vecp,Rh)\right]^2 - 2\\
\\
&&
\quad
=\D{
\frac{Re^4\omega^4 + 2(2-\sin^2{\alpha})Re^2\omega^2 
+ 2(1-\sin^2{\alpha}(1-8\cos^2{\alpha}))}
{\sin^2{\alpha}((1-8\cos^2{\alpha})+Re^2\omega^2)}
}
\ge 0
\end{eqnarray*}
since $(2-\sin^2{\alpha})$ and 
$\left(1-\sin^2{\alpha}(1-8\cos^2{\alpha})\right)$ are
nonnegative for all $\alpha$. 
In particular, when $\alpha=\alpha_\veck-\alpha_\vecp = \pi/2$, we have
\begin{eqnarray}
\D{\lim_{Rh\rightarrow\infty}\left[Re^0(\veck;\alpha_\vecp,Rh)\right]^2=2}.
\label{cre_cont_one}
\end{eqnarray}
Then, (\ref{re0gesqrt2}) and (\ref{cre_cont_one}) imply that
for any $\alpha_\vecp$,
\begin{eqnarray}
\lim_{Rh\to \infty} \min_{\veck} Re^{0}(\veck;\alpha_\vecp,Rh) = \sqrt{2},
\label{contsqrt2}
\end{eqnarray}
recovering the classical critical Reynolds number $\sqrt{2}$
for the 2D Navier Stokes equation.
This is different from the recent result 
by Manfroi and Young \cite{MY}
for the $\beta$-plane equation with a steady Kolmogorov base flow.
They observe that the critical Reynolds number is less than $\sqrt{2}$
in the limit $Rh\to\infty$. We compare their results with ours
in Section \ref{sec_comp_my}.

\subsection{Critical Reynolds numbers in the limit $k\to 0$}
\label{sec_re_k0}

\begin{figure}
\begin{center}
\begin{tabular}{cc}
(a) $Re=1.67$
&
(b) $Re=8\tan{\pi/4}=8$
\\
\includegraphics[height= 6cm]{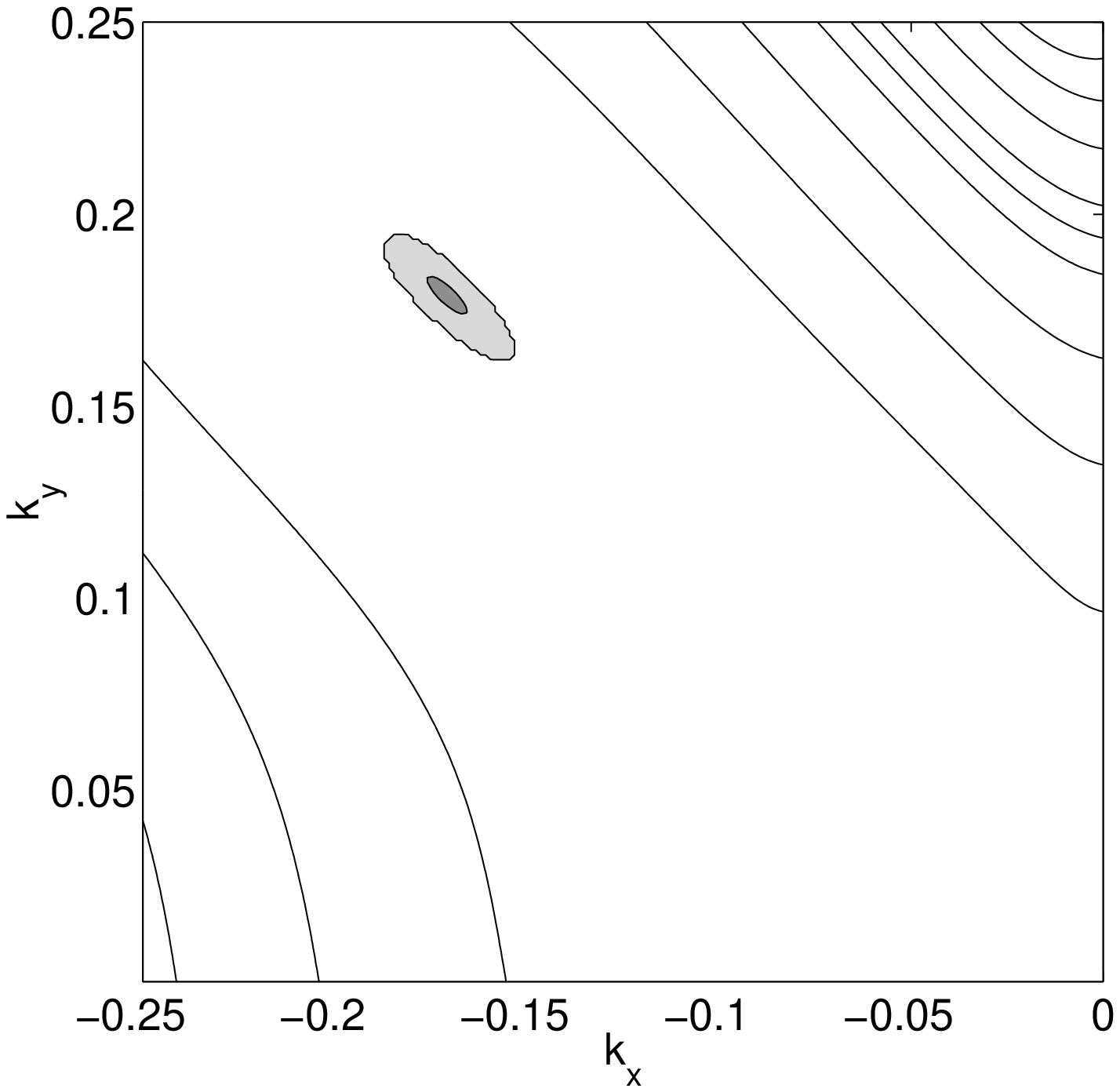}
&
\includegraphics[height = 6cm]{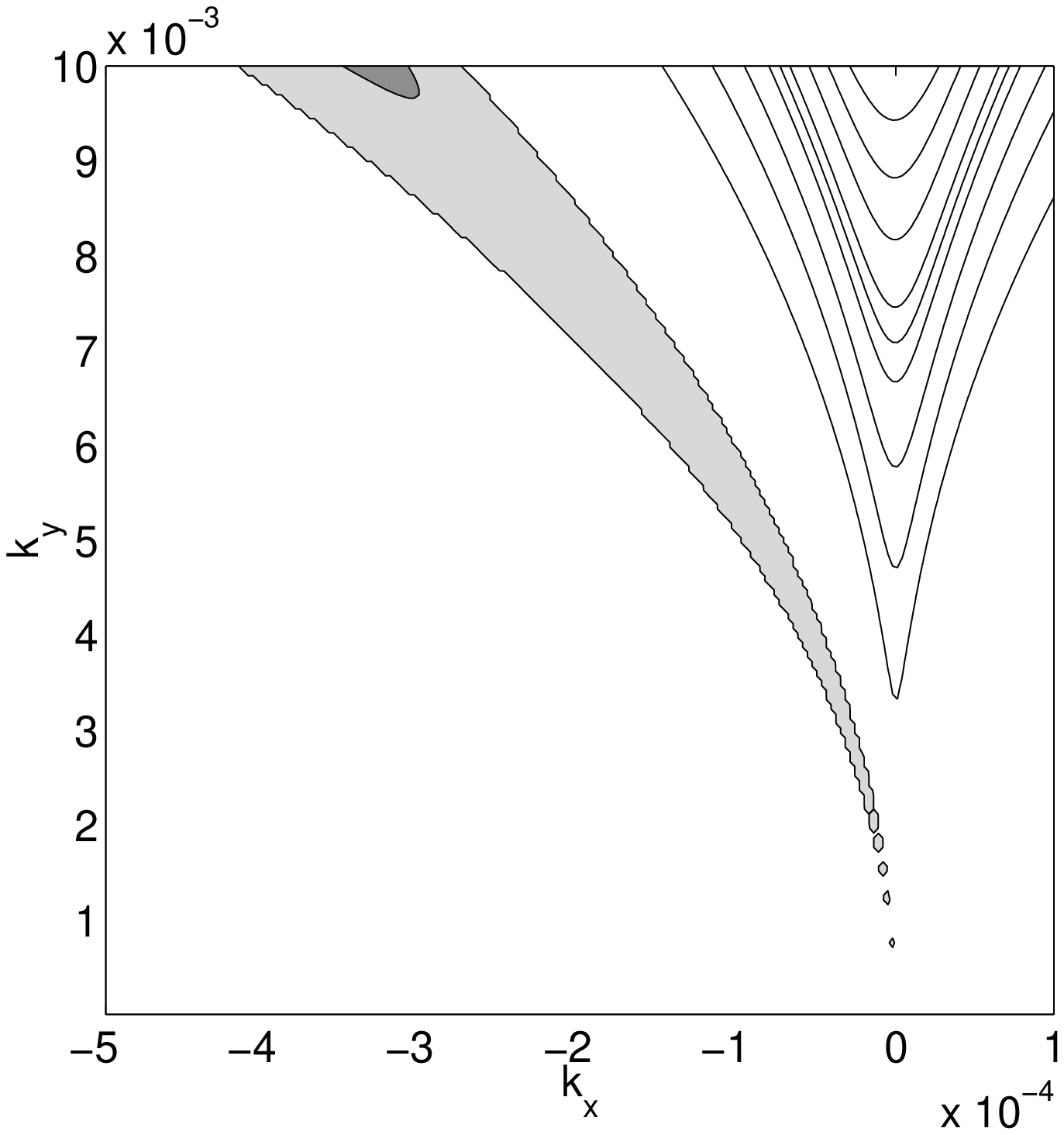}
\end{tabular}
\end{center}
\caption{
Contour plot of $\Re(\sigma(\veck;\alpha_\vecp,Re,Rh))$
for $\alpha_\vecp=\pi/4$ at $Rh=10$ and (a) $Re=1.67$ (b) $Re=8$.
In (b), note that the $x$-axis is scaled by $10^{-4}$ and
the $y$-axis is scaled by $10^{-3}$.
}\label{pic_FI}
\end{figure}

In this section,
we consider the marginal Reynolds number $Re^{0}(\veck;\alpha_\vecp,Rh)$
for $k\rightarrow 0$.
Without loss of generality, we only consider 
$ 0 \le \alpha_\vecp \le \pi/2$ and the result is the following:
\begin{eqnarray}
\lim_{k_0\rightarrow 0} \min_{|\veck|<k_0}
&&Re^0(\veck;\alpha_\vecp,Rh)
\nonumber \\
\nonumber \\
&&
= \left\{
\begin{array}{ll}
\D{
\sqrt{\frac{2}{\cos^2\alpha_\vecp(1-8\sin^2\alpha_\vecp)}}
},
&\ 0\le \alpha_\vecp \le \sin^{-1}\frac{1}{4}
\\
\\
8|\tan{\alpha_\vecp}|,
& \sin^{-1}\frac{1}{4}\le \alpha_\vecp \le \pi/2.
\end{array}
\right.
\label{cre_k0}
\end{eqnarray}
Note that (\ref{cre_k0}) depends only on $\alpha_\vecp$ but
not $Rh$ and is an increasing function in $0\le \alpha_\vecp \le \pi/2$.
For the case $\alpha_\vecp = \pi/2$, 
(\ref{cre_k0}) is infinite in the sense that for any
Reynolds number, wavevectors $\veck$ with $|\veck|<k_0$ 
are stable if $k_0$ is small enough. 
By definition, the critical Reynolds number
$Re^{c}(\alpha_\vecp,Rh)$ is less than or equal to
(\ref{cre_k0}). The expression (\ref{cre_k0}) increases to infinity
as $\alpha_\vecp$ increases to $\pi/2$. Thus, we expect that
$Re^{c}(\alpha_\vecp,Rh)$ is strictly less than (\ref{cre_k0})
if $\alpha_\vecp$ is not too close to zero. In this case, the first
unstable wavevector is observed ``away'' from the origin as the Reynolds
number is increased from zero. For example, for $\alpha_\vecp=\pi/4$ and
$Rh = 10$, instability is first observed near wavevector
$\veck = (-0.17,0.18)$ at Reynolds number $Re\approx 1.59$ (Figure~\ref{pic_FI}). 
We have
\begin{eqnarray*}
Re^{c}(\alpha=\pi/4,Rh=10) < 1.59 
< 8 = \lim_{k_0\to 0}\min_{|\veck|<k_0}Re^{0}(\veck;\alpha_\vecp=\pi/4,Rh=10).
\end{eqnarray*}
As the Reynolds number is increased, the unstable region becomes
a long strip and reaches to the origin
at the Reynolds number (\ref{cre_k0}) 
(Figure~\ref{pic_FI}(b) with $\alpha_\vecp=\pi/4$ and $Re=8$).
This is different from the 
2D Navier-Stokes equations with a Kolmogorov
base flow where the first instability is always observed at $k\to 0$
at Reynolds number $\sqrt{2}$ \cite{MS}, \cite{SY}, and \cite{WAL}.

The rest of this section is devoted to deriving (\ref{cre_k0}). 
Since $\Re(\sigma(\veck;\alpha_\vecp,Re,Rh))=
\Re(\sigma(-\veck;\alpha_\vecp,Re,Rh))$ as discussed
in Section \ref{sec_numeric}, we only consider
$0\le\alpha_\veck\le \pi$. For a fixed real number $\eta$, we 
introduce a function 
\begin{eqnarray*}
f(w;\eta) = \frac{1}{2}\frac{\sin^2{\eta}}{(1+w)^2}
\left\{ 1-8\cos^2{\eta} + w\right\} \quad \textrm{for } w \ge 0.
\end{eqnarray*}
Then, $f(Re^2\omega^2;\alpha)$ is the reciprocal of the right-hand side
of (\ref{re0_def}).
Recall that $\omega = \omega(\veck) = -Rh^{-1}k_x/k^2$
and consider all wavevectors $\veck = k(\cos\alpha_\veck,\sin\alpha_\veck)$
with $k>0$ and a fixed $\alpha_\veck$.
In the case $Rh<\infty$ and $\alpha_\vecp \ne \pi/2$,
the range of $\omega(\veck)$
consists of all positive numbers, and
in the case $\alpha_\veck = \pi/2$, the range of $\omega(\veck)$ 
is zero. In any case,
the minimum of $Re^0(\veck;\alpha_\vecp,Rh)$ 
over wavevectors $\veck$ with
a fixed $\alpha_\veck$ satisfies
\begin{eqnarray}
\min_{k>0}
\left[Re^{0}(k(\cos{\alpha_\vecp},\sin{\alpha_\vecp});\alpha_\vecp,Rh)
\right]^2
\ge
\frac{1}{\max_{w\ge 0}\ f(w;\alpha)}.
\label{f_prop1}
\end{eqnarray}
The maximum of $f(w;\eta)$ is
\begin{eqnarray}
\D{
\max_{w\ge 0} f(w;\eta) = 
\left\{
\begin{array}{ll}
\frac{1}{2}\sin^2{\eta}(1-8\cos^2{\eta}),
&\quad \textrm{for }\cos^2{\eta} \le \frac{1}{16}\\
\\
\frac{1}{64}\tan^2{\eta},
&\quad \textrm{for }\cos^2{\eta} \ge \frac{1}{16}
\end{array}
\right.
}
\label{f_prop2}
\end{eqnarray}
and achieved at
\begin{eqnarray}
\D{
w(\eta) = 
\left\{
\begin{array}{ll}
0,
&\quad \textrm{for } \cos^2{\eta} \le \frac{1}{16}\\
\\
16\cos^2{\eta}-1,
&\quad \textrm{for } \cos^2{\eta} \ge \frac{1}{16}
\end{array}
\right.
}
\label{f_prop3}
\end{eqnarray}

To show the equality in (\ref{cre_k0}), 
we demonstrate that the inequality ``$\le$'' and ``$\ge$'' both hold.
The inequality ``$\le$'' is trivial for $\alpha_\vecp=\pi/2$ since
(\ref{cre_k0}) is infinite.
If $\alpha_\vecp\le \sin^{-1}(1/4)$, then we have
\begin{eqnarray*}
\min_{|\veck|<k_0}
(Re^{0}(\veck;\alpha_\vecp,Rh))^2
\le
\min_{|k_y|<k_0}(Re^{0}((0,k_y);\alpha_\vecp,Rh))^2
=\frac{2}{\cos^{2}\alpha_\vecp(1-\sin^2\alpha_\vecp)}
\end{eqnarray*}
since $\omega(\veck) = 0$ if $k_x=0$. 
Next, we suppose that $\alpha_\vecp > \sin^{-1}(1/4)$. 
Motivated by (\ref{f_prop3}),
fix the Reynolds number
$Re = 8\tan\alpha_\vecp$ and consider the curve consisting of
$\veck$'s satisfying
\begin{eqnarray}
Re^2\omega(\veck)^2 = 16\cos^2{\alpha}-1,
\quad k_x < 0,\quad k_y > 0, \quad \alpha=\alpha_\veck-\alpha_\vecp.
\label{def_min_curve}
\end{eqnarray}
Since we have $\omega(\veck)=-Rh^{-1}k_x/k^2$, the curve
(\ref{def_min_curve})
is tangential to the line $k_x=0$ at the origin (actually, this
curve is along the unstable region in Figure~\ref{pic_FI}(b).).
At a fixed wavevector $\veck$ on curve (\ref{def_min_curve})
with $Re=8\tan\alpha_\vecp$, we have
\begin{eqnarray*}
&&-\frac{k^2}{Re}\left[
1-\frac{Re^2}{2}\frac{\sin^2{\alpha}}{(1+Re^2\omega^2)^2}
\left\{ 1-8\cos^2\alpha + Re^2\omega^2\right\}
\right]
\\
\\
&&\quad =
-\frac{k^2}{Re}\left[1-64\tan^2\alpha_\vecp\frac{\tan^2\alpha}{64}
\right]
= 
-\frac{k^2}{Re}\left[
1 - \frac{\tan^2(\alpha_\veck-\alpha_\vecp)}{\tan^2(\pi/2-\alpha_\vecp)}
\right]
> 0
\end{eqnarray*}
since we have $ \pi/2 < \alpha_\veck <\pi$ on the curve. This implies that
$Re^{0}(\veck;\alpha_\vecp,Rh)\le 8\tan\alpha_\vecp$ and establishes the
inequality ``$\le$'' in (\ref{cre_k0}).

Now, consider the opposite inequality ``$\ge$'' in (\ref{cre_k0}).
For this purpose, let
$\{\veck^n=k^n(\cos\alpha_{\veck^n},\sin\alpha_{\veck^n})\}$ 
be a sequence satisfying
\begin{eqnarray}
\left\{
\begin{array}{l}
(\textrm{a})\ \D{\lim_{n\rightarrow \infty}|\veck^n| = 0},\\
\\
(\textrm{b})\ \D{ \lim_{k_{0}\rightarrow 0 }\min_{|\veck|<k_{0}}
Re^0(\veck;\alpha_\vecp,Rh) 
= \lim_{n\rightarrow\infty} Re^0(\veck^n,\alpha_\vecp,Rh)},\\
\\
(\textrm{c})\ \D{ Re^{0}(\veck^n;\alpha_\vecp,Rh)} \textrm{ is increasing}.
\end{array}
\right.
\label{kn_prop}
\end{eqnarray}
For example, $\{\veck^{n}\}$ can be chosen by the relation
\begin{eqnarray*}
\min_{|\veck|\le n^{-1}}Re^{0}(\veck;\alpha_\vecp,Rh)
=Re^{0}(\veck^n;\alpha_\vecp,Rh)
\quad\textrm{and }|\veck^{n}| \le n^{-1}.
\end{eqnarray*}
We suppose that 
\begin{eqnarray}
\lim_{n\to \infty} \alpha_{\veck^n} = \frac{\pi}{2}
\label{derive_claim}
\end{eqnarray}
and show this at the end of this section.
From (\ref{derive_claim}) we have
\begin{eqnarray}
\lim_{n\to \infty} \alpha^n = 
\pi/2-\alpha_\vecp,
\label{derive_claim2}
\end{eqnarray}
where $\alpha^n = \alpha_{\veck^n} - \alpha_\vecp$.
Applying (\ref{kn_prop}a), (\ref{f_prop1}), (\ref{f_prop2})
and (\ref{derive_claim2}) in order,
we have
\begin{eqnarray*}
\lim_{k_0\rightarrow 0}\min_{|\veck|<k_0}
Re^0(\veck;\alpha_\vecp,Rh)^2 
&=&\lim_{n\rightarrow \infty} (Re^0(\veck^n,\alpha_{\vecp},Rh))^2\\
\\
&\ge& \lim_{n\rightarrow\infty} \frac{1}{\max_{w\ge0}f(w;\alpha^n)}
\\
\\
&=&
\lim_{n\to \infty}
\D{
\left\{
\begin{array}{ll}
\D{\frac{2}{\sin^2{\alpha^n}(1-8\cos^2{\alpha^n})}},
& \ \textrm{if } \cos{\alpha^n}\le \frac{1}{4}\\
\\
\D{\frac{64}{\tan^2{\alpha^n}}},
& \ \textrm{if } \cos{\alpha^n}\ge \frac{1}{4}\\
\end{array}
\right.
}
\\
\\
\\
&=&\D{
\left\{
\begin{array}{ll}
\D{ \frac{2}{\cos^2{\alpha_\vecp}(1-8\sin^2{\alpha_\vecp})} },
& \ \textrm{if } \sin{\alpha_{\vecp}}\le \frac{1}{4}\\
\\
\D{64\tan^2{\alpha_{\vecp}}},
& \ \textrm{if } \sin{\alpha_{\vecp}}\ge \frac{1}{4}
\end{array}
\right.
}
\end{eqnarray*}
which establishes the inequality ``$\ge$'' in (\ref{cre_k0}).

We are left to show (\ref{derive_claim}).
Suppose (\ref{derive_claim}) does not hold so that we have
\begin{eqnarray*}
\lim_{n\to \infty} \cos\alpha_{\veck^n} \ne 0. 
\end{eqnarray*}
Then, since $k^n \to 0$ by (\ref{kn_prop}a), we have
\begin{eqnarray}
\lim_{n\to \infty}\left|\omega(\veck^n)\right|
=\D{ \lim_{n\to \infty}\left|\frac{1}{Rh}\frac{\cos\alpha_{\veck^n}}{k^n}
\right|}
= \infty.
\label{derive_infty1}
\end{eqnarray}
By definition, 
$M_n=Re^{0}(\veck^n;\alpha_\vecp,Rh)^2$ satisfies
\begin{eqnarray}
1-\frac{M_n}{2}
\frac{\sin^2{\alpha^n}}{(1+M_n \omega(\veck^n)^2)^2}
\left\{
1-8\cos\alpha^n+M_n\omega(\veck^n)^2\right\}
=0.
\label{derive_infty2}
\end{eqnarray}
Using (\ref{re0gesqrt2}), (\ref{kn_prop}c),
(\ref{derive_infty1}) and (\ref{derive_infty2}),
we have
\begin{eqnarray*}
0=\lim_{n\to\infty}
\left[1-\frac{M_n}{2}
\frac{\sin^2{\alpha^n}}{(1+M_n \omega(\veck^n)^2)^2}
\left\{
1-8\cos\alpha^n+M_n\omega(\veck^n)^2\right\}
\right]
=1.
\end{eqnarray*}
This is a contradiction and thus (\ref{derive_claim}) holds.

\section{Linear Stability as $Rh$ decreases}\label{sec_mid_beta}


From the numerical
solution of (\ref{CF}), we consider the behavior of the growth rate 
of the perturbation at 
wavevectors $\veck$ over $|\veck|<1$ 
for various Rhines numbers $Rh$. 
We first give a preliminary discussion of resonant triad interactions between 
wave solutions of the $\beta$-plane equation.

\subsection{Resonant Triad Interactions}\label{sec_reson}

As stated in Section \ref{sec_intro}, the $\beta$-plane
equation has wave solutions, called Rossby waves,
\begin{eqnarray*}
\Psi(\vecx,t;\veck) 
= \hat{\Psi}(t;\veck)
\exp\left( i\theta(\vecx,t;\veck)\right) + c.c,
\end{eqnarray*}
where $c.c$ is the complex conjugate and the phase $\theta(\vecx,t;\veck)$
is given by (\ref{theta}).
We may represent the solution $\Psi(\vecx,t)$ in terms of the waves by
\begin{eqnarray*}
\Psi(\vecx,t) = \sum_{\veck}
\hat{\Psi}(t;\veck)\exp\left( i\theta(\vecx,t;\veck)\right)
\quad\textrm{with } \hat{\Psi}^{*}(t;\veck)=\hat{\Psi}(t;-\veck)
\end{eqnarray*}
where $*$ is the complex conjugate.
Then, the governing equation (\ref{beta_eqn}) with $\vecF = \veczero$
is written as
\begin{eqnarray}
&&\left(
\partial_t + \frac{1}{Re} k^2 \right)
\hat{\Psi}(t;\veck)
=
\nonumber\\
\nonumber\\
&&
\quad\sum_{\veck+\vecp+\vecq=\veczero}
C_{kpq}\hat{\Psi}^{*}(t;\vecp)\hat{\Psi}^{*}(t;\vecq)
\exp\left(
i(\omega(\veck) + \omega(\vecp) +\omega(\vecq))t
\right)\label{beta_eqn_fourier}
\end{eqnarray}
where the coefficients are given by
$\D{
C_{kpq} = (q^2-p^2)[(\vecp\times\vecq)\cdot\hat{\vecz}]/(2k^2).
}$
In this section $\vecp$ is not necessarily a unit vector.
These coefficients satisfy
\begin{eqnarray*}
\begin{array}{l}
k^2C_{kpq} + p^2C_{pqk} + q^2C_{qkp} = 0,\\
\\
k^4C_{kpq} + p^4C_{pqk} + q^4C_{qkp} = 0
\end{array}
\end{eqnarray*}
as can be deduced directly from energy and enstrophy conservation by triad
interactions. See Waleffe \cite{WAL2}, \cite{WAL3} for more discussion.
\noindent
From (\ref{beta_eqn_fourier}), in the limit $Rh\to 0$, energy transfer
between three wavevectors $\veck$, $\vecp$ and $\vecq$ is maximal when
the frequency $\omega(\veck)+\omega(\vecp)+\omega(\vecq) = 0$.
Otherwise, a nonzero frequency produces an oscillation and energy
transfer between the wavevectors will be zero on average.

\begin{figure}
\begin{center}
\begin{tabular}{c}
\includegraphics[width=6cm]{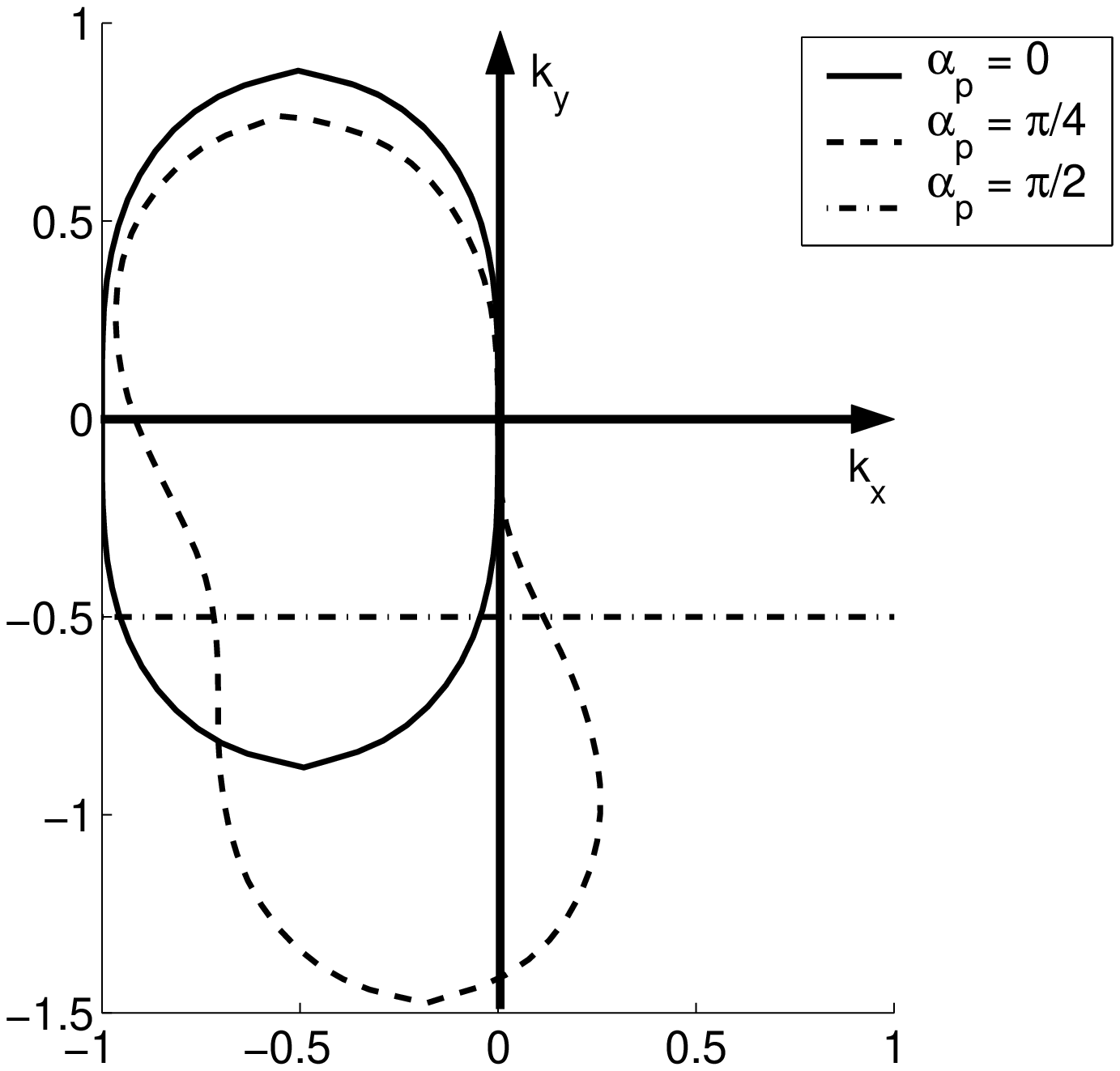}
\end{tabular}
\caption{Resonant traces with $\alpha_\vecp=0,\pi/4$
and $\pi/2$}
\label{pic_reson_trace}
\end{center}
\end{figure}

A triad $(\veck,\vecp,\vecq)$ is resonant 
if it satisfies the resonant conditions
\begin{eqnarray}
\veck+\vecp+\vecq = \veczero,\qquad 
\omega(\veck) + \omega(\vecp) + \omega(\vecq) = 0.
\label{reson_cond}
\end{eqnarray}
The dispersion relation (\ref{dispersion})
of the $\beta$-plane equation allows resonant triads
as shown in Longuet-Higgins and Gill \cite{LG}. 
They showed that for a given wavevector $\vecp$
with $p_x\ne 0$, the trace of wavevectors resonant to $\vecp$ forms
a closed curve. For $\vecp$ with $p_x=0$, the resonant trace consists
of the wavevectors $\veck$ with $k_y = -p_y/2$. 
Figure~\ref{pic_reson_trace}
shows three traces resonant to $\vecp$
for $\alpha_\vecp=0,\pi/4$ and $\pi/2$ with $p=1$.
For $0\le \alpha_\vecp < \pi/2$, as illustrated in
Figure~\ref{pic_reson_trace}, resonant traces are tangential 
to the line $\veck_x=0$ at the origin. By asymptotic expansion of
(\ref{reson_cond})
in $k_x$ and $k_y$ for $k\ll 1$, one finds that
\begin{eqnarray}
k_x = -\frac{p_xp_y}{p^2}k_y^3 + \frac{p_x(4p_y^2-1)}{p^2}k_y^4 + \cdots,
\label{reson_asymp}
\end{eqnarray}
which implies that
\begin{eqnarray}
k_x = O(k^r),\qquad \alpha_\veck = \pi/2 + O(k^{r-1})
\label{reson_asymp_tangential}
\end{eqnarray}
where $r = 3$ for $0<\alpha_\vecp <\pi/2$ and $r = 4$ for $\alpha_\vecp=0$.
The resonant trace for $\vecp$ with $\alpha_\vecp=\pi/2$ 
is different from $\alpha_\vecp\ne \pi/2$ and it consists of the modes with
$k_y=-\frac{1}{2}$. In this case,
if the triad $(\veck,\vecp,\vecq)$ is a resonant triad,
then $k=q$ which results in $C_{pqk}=0$ from (\ref{beta_eqn_fourier}).
Thus, the mode $\vecp$ with $p_x=0$ can not receive or lose energy
by direct resonant triad interactions (\cite{LG}, \cite{WAL2}).

\subsection{Growth rates for various $Rh$ and
$|\veck|<1$}\label{sec_CN}

We study the linear growth rate over $\veck$ with $|\veck|<1$ at 
various values of Rhines numbers by solving the eigenvalue problem
(\ref{ILS}) numerically as discussed in Section \ref{sec_numeric}.
Equation (\ref{ILS}) is truncated at $N=15$, resulting
in a $31\times 31$ matrix.
The Reynolds number is finite and large enough
to exhibit instabilities. In most figures, the Reynolds number is
fixed at $10$ unless otherwise specified.

\begin{figure}
\begin{center}
\begin{tabular}{ccc}
{\small (a) $Re = 10$, $Rh = \infty$}
&
{\small (b) $Re= 10$, $Rh = 2$}
&
{\small (c) $Re = 10$, $Rh = 1/2$}
\\
\includegraphics[height= 3.8cm]{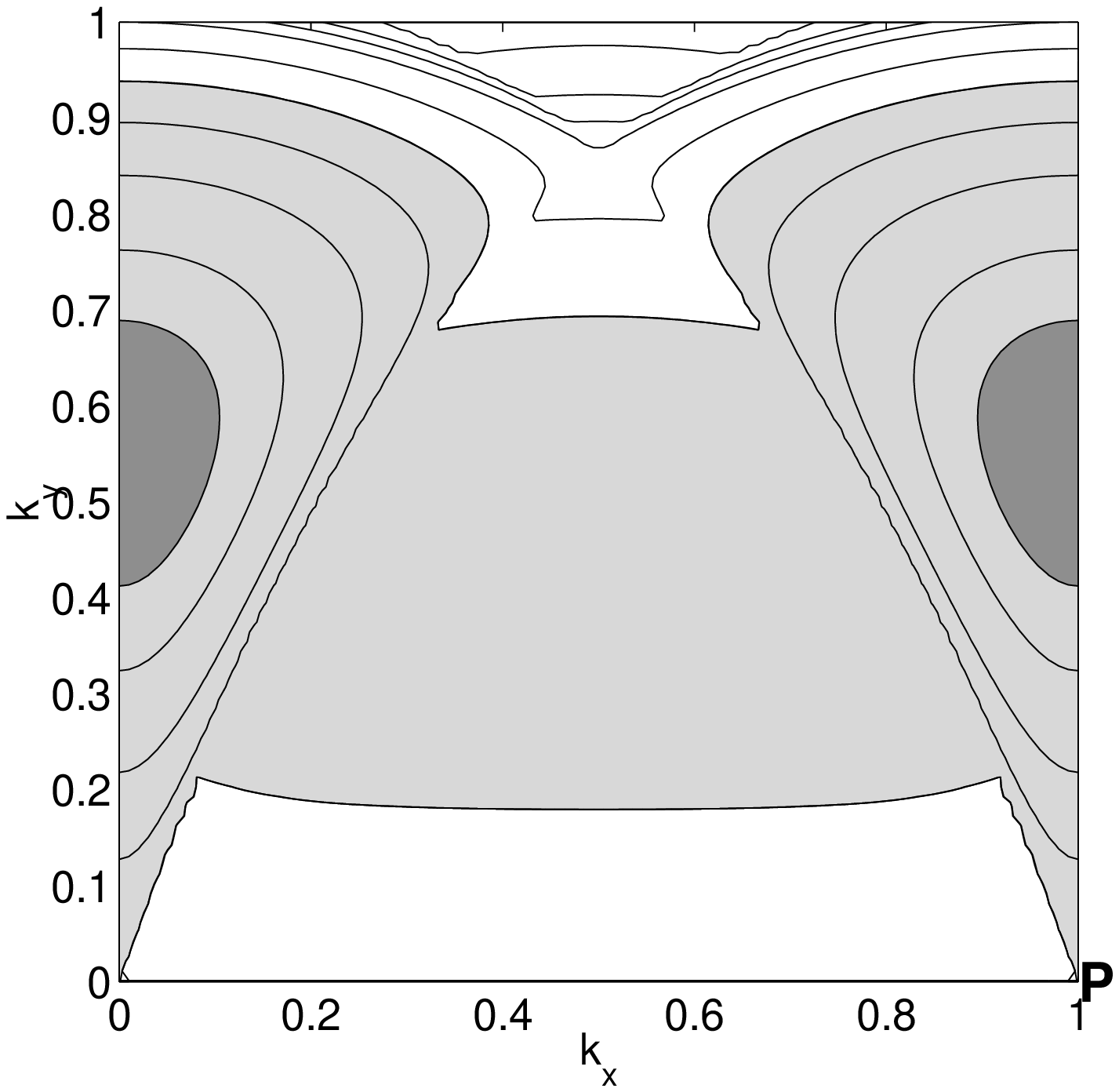}
&
\includegraphics[height=3.8cm ]{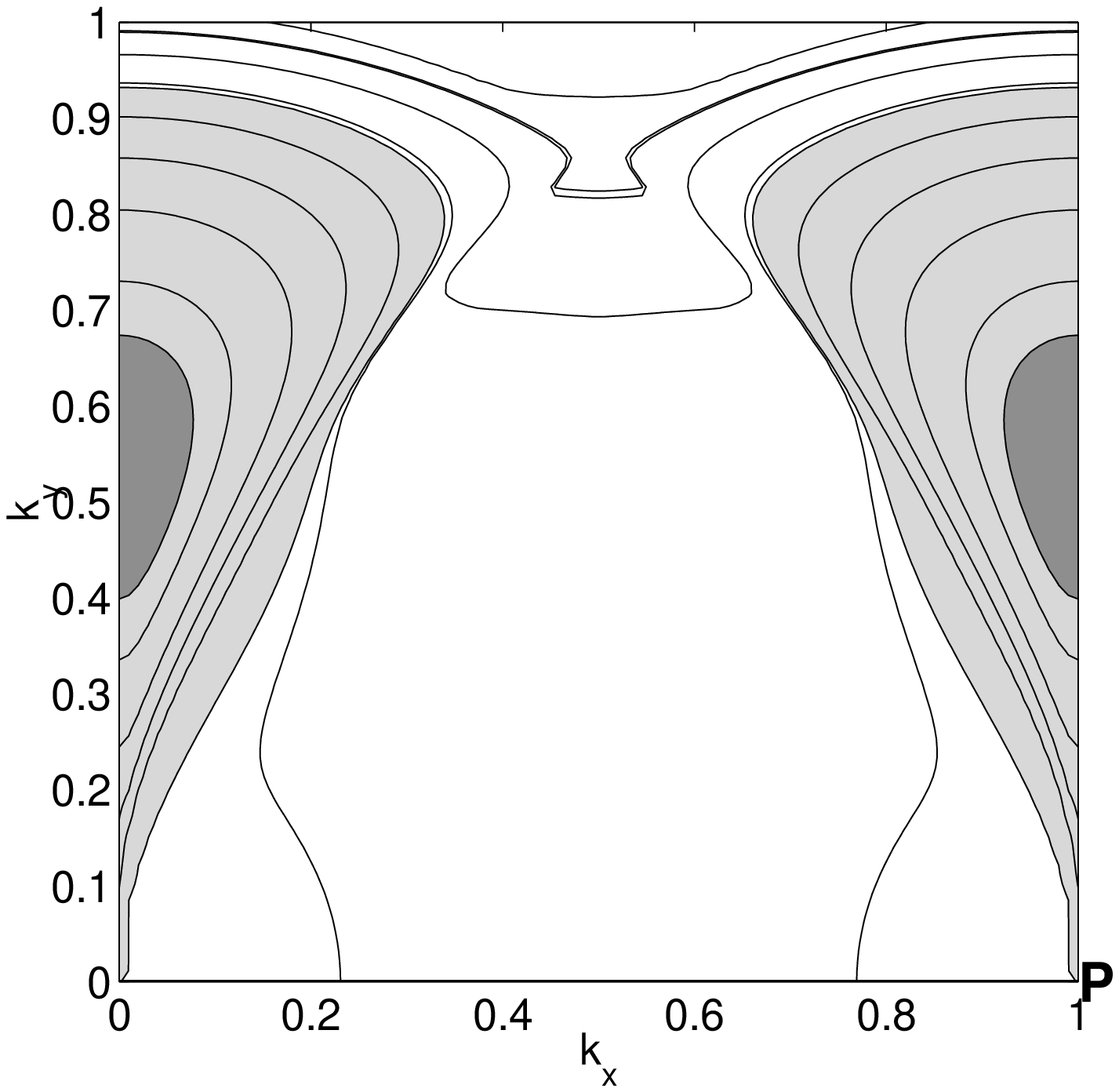}
&
\includegraphics[height=3.8cm ]{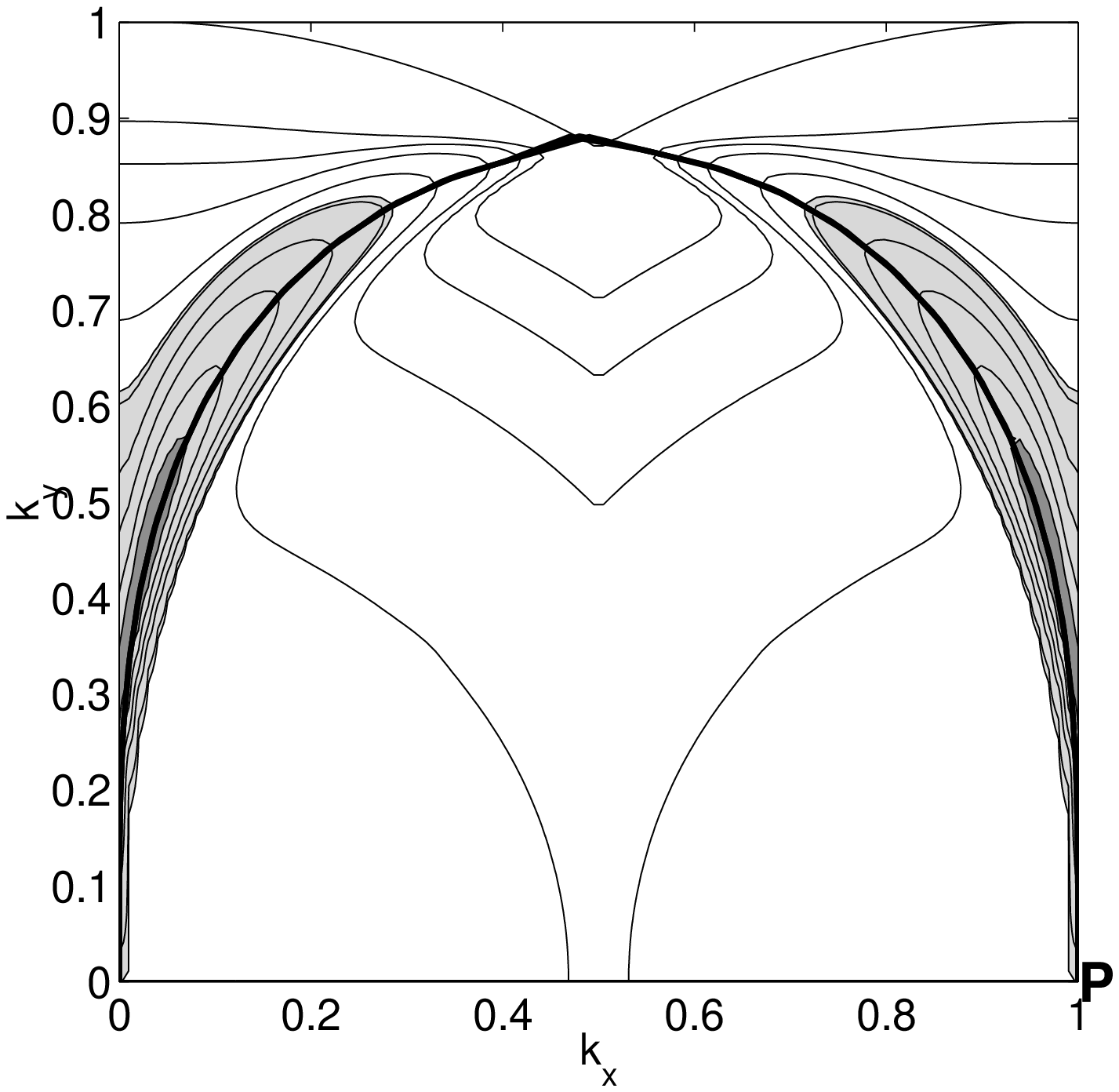}
\\
{\small (d) $Re = 10$, $Rh = 1/10$}
&
{\small (e) $Re = 1000$, $Rh = 2$}
&
{\small (f) $Re = 1000$, $Rh = 1/2$}
\\
\includegraphics[height= 3.8cm]{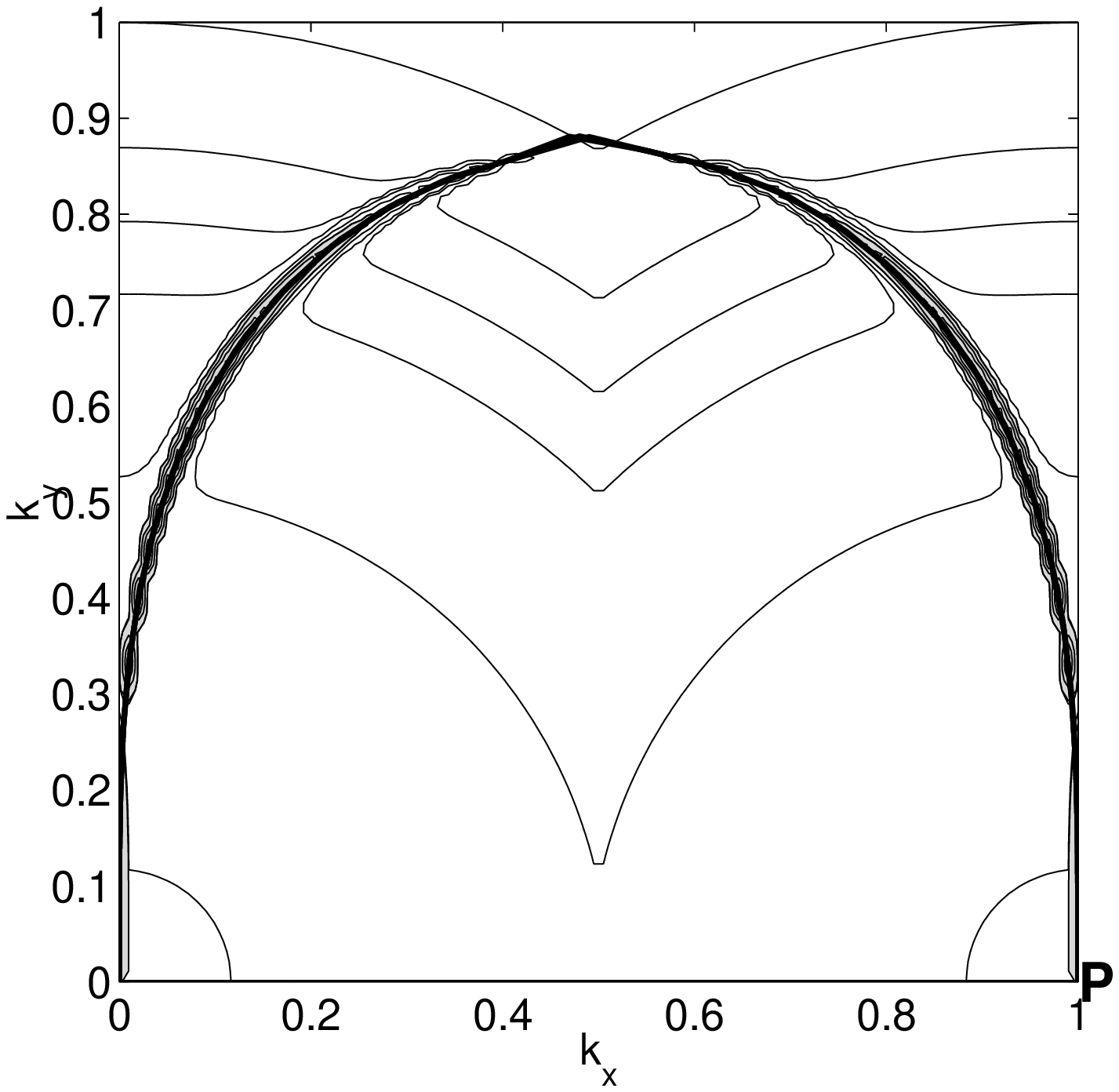}
&
\includegraphics[height=3.8cm ]{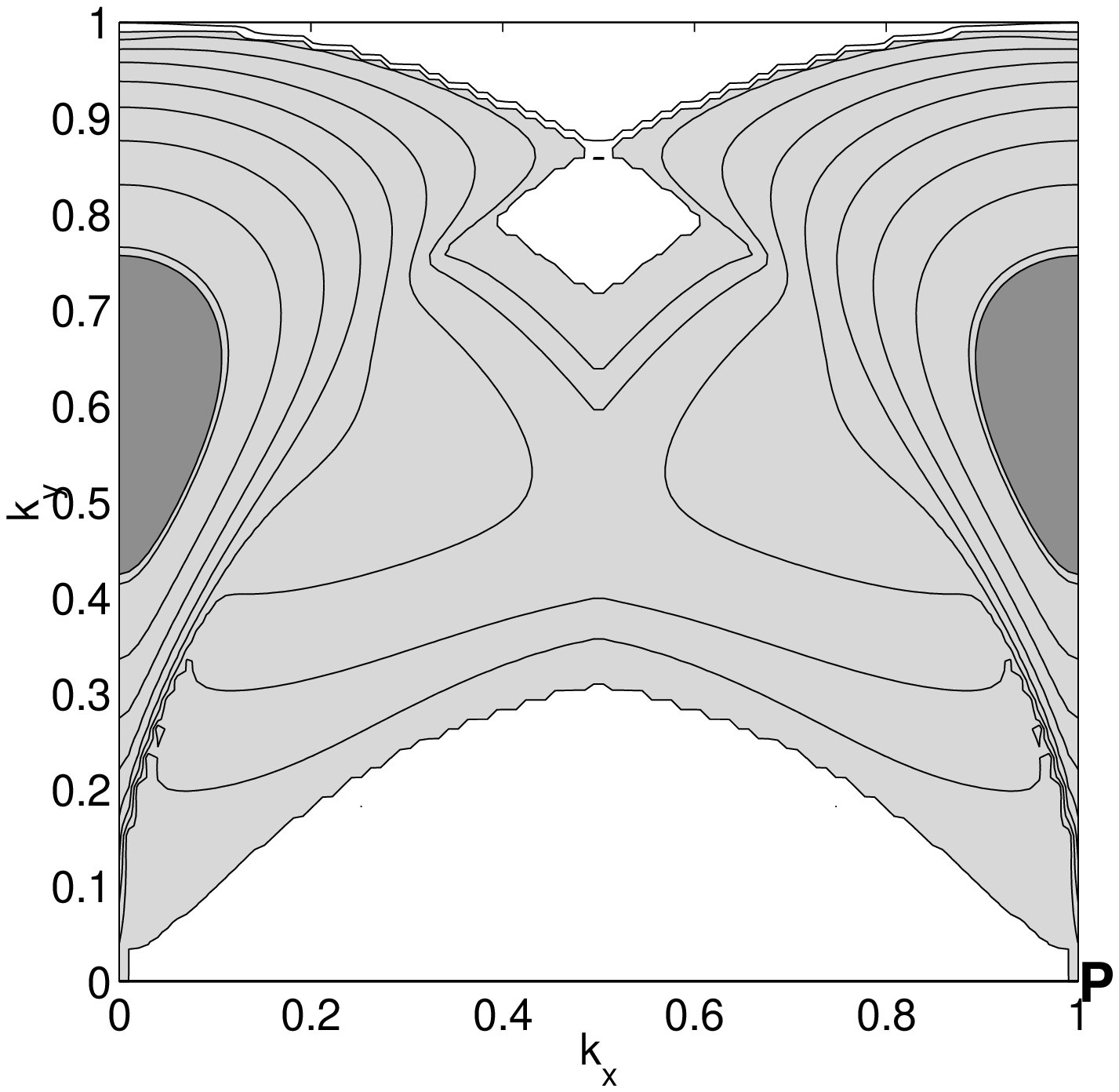}
&
\includegraphics[height=3.8cm ]{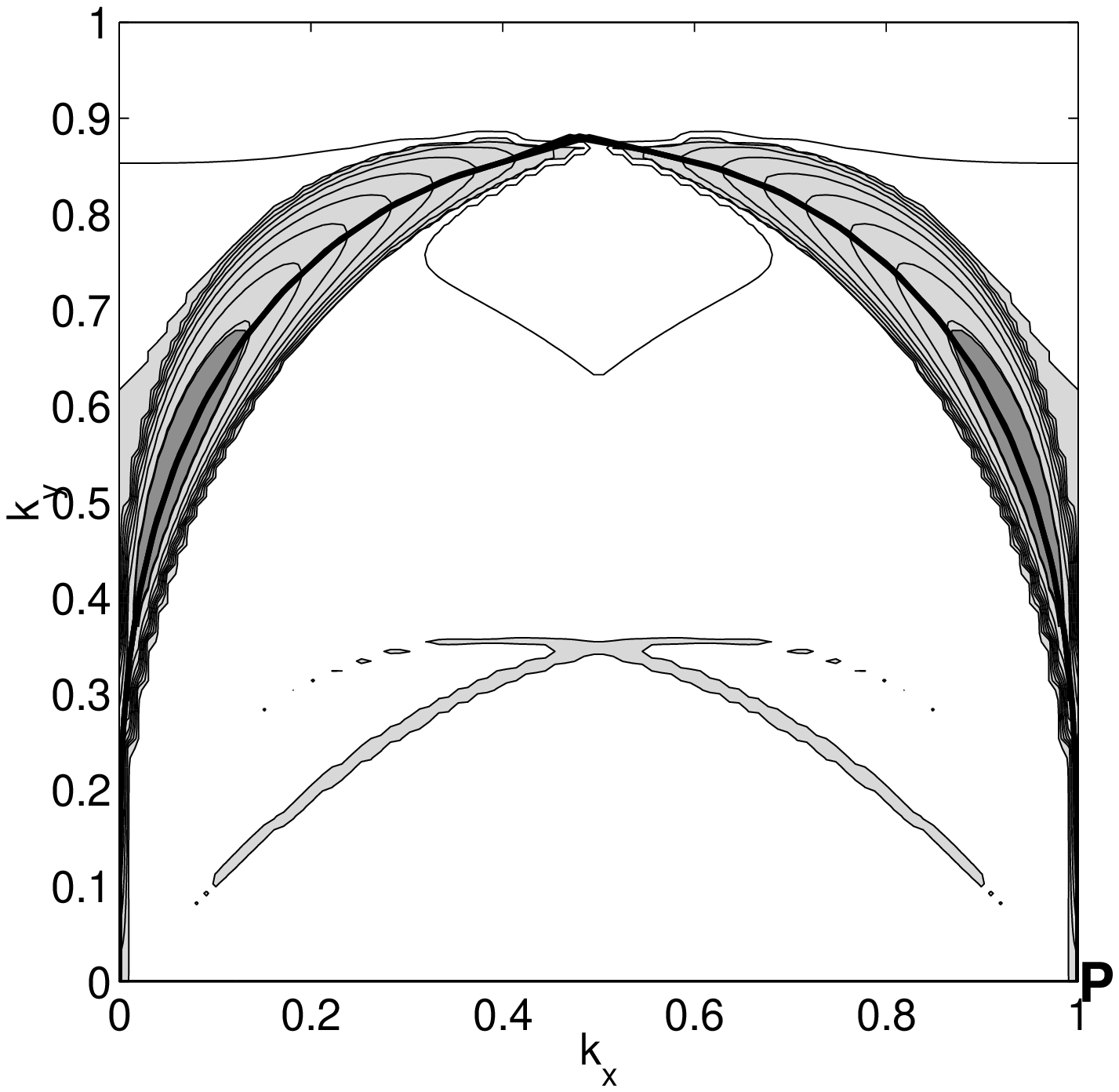}
\end{tabular}
\end{center}
\caption{
Contour plot of $\Re(\sigma(\veck;\alpha_\vecp,Re,Rh))$
for $\alpha_\vecp=0$. Resonant traces including $\pm \vecp$ are solid
lines for (c), (d) and (f).
}\label{pic_CN_0}
\end{figure}

First, we consider the case $\alpha_\vecp=0$ 
in Figure~\ref{pic_CN_0}. 
Recall that growth rate $\Re(\sigma(\veck;\alpha_\vecp=0,Re,Rh))$ has 
symmetries about both the $x$-axis and the $y$-axis 
(Section \ref{sec_numeric}). Thus, it is enough to consider
wavevectors $\veck$ with $0<k_x<1$ and $0<k_y<1$ in each
plot of Figure~\ref{pic_CN_0}. We use a grid size of
$100\times 100$.
At $Rh=\infty$ (Figure~\ref{pic_CN_0}(a)), 
the most unstable wavevectors are along the line of $k_x=0$
which is perpendicular to the base flow wavevector $\vecp$,
consistent with
the results by Sivashinsky and Yakhot \cite{SY} and 
by Waleffe \cite{WAL}. 
We refer to the instability at wavevectors near
perpendicular to $\vecp$ at large Rhines numbers ($Rh \ge 2$)
as the inflectional instability. This terminology is motivated
by the existence of inflection points in the profile of the sinusoidal
base flow, which can be considered as the source of
instability.\footnote{By Rayleigh's Inflection Point Criterion.
See \cite{KUN} for example.}

As $Rh$ is decreased, the maximum growth rate decreases
(Figure~\ref{pic_inf_line}) 
and the most unstable wavevector is
not exactly at $k_x=0$ but near $k_x=0$ (Figure~\ref{pic_CN_0}(c)). As $Rh$ is 
decreased more, the unstable region becomes
much narrower (Figure~\ref{pic_CN_0}(d)). In fact, this narrow 
band lies along the traces resonant to the base flow
wavevector $\vecp$ and its conjugate $-\vecp$. 
The resonant traces correspond to black lines 
in Figure~\ref{pic_CN_0}(c), (d) and (f).
Thus, the nature of the instability changes 
from the inflectional instability
to the resonant triad interaction as $Rh$ is decreased. 
This transition of instability
is observed at Rhines number between $2$ and $1/2$ for Reynolds number
$Re = 10$.
Actually, the order of the Rhines number of the transition
appears to be 
independent of the Reynolds number. Figure~\ref{pic_CN_0}(e) and (f)
exhibit the transition at Rhines number between $2$ and $1/2$
with Reynolds number $Re=1000$. 
See also Gill \cite{G}.

\begin{figure}
\begin{center}
\begin{tabular}{cc}
(a) $Re = 10$, $Rh = 2$
&
(b) $Re= 10$, $Rh = 1/2$
\\
\includegraphics[height= 3.5cm]{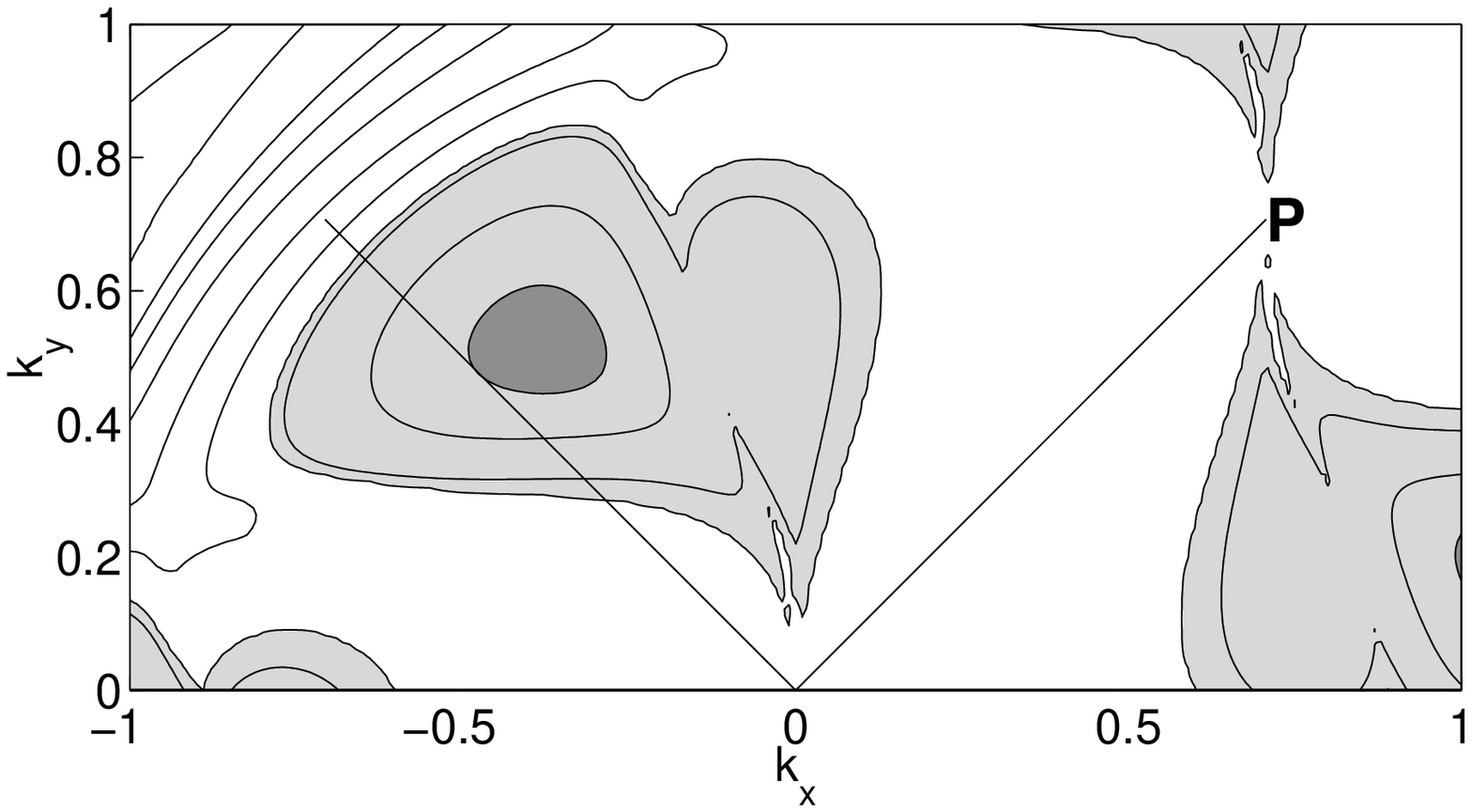}
&
\includegraphics[height=3.5cm ]{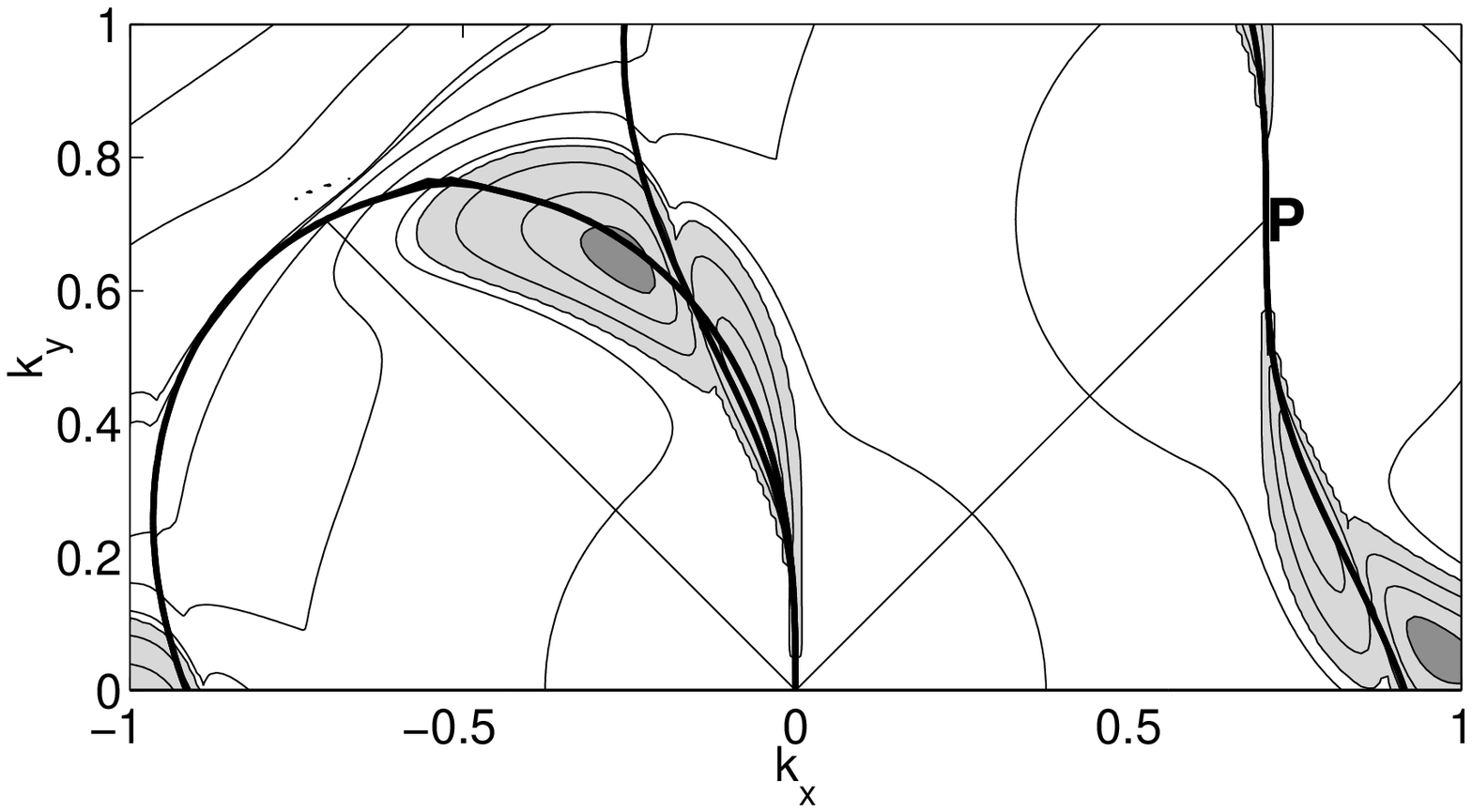}
\end{tabular}
\end{center}
\caption{
Contour plot of $\Re(\sigma(\veck;\alpha_\vecp,Re,Rh))$
for $\alpha_\vecp=\pi/4$ at $Re=10$. Resonant traces including $\pm \vecp$
are solid lines in (b).
}\label{pic_CN_45}
\end{figure}

A similar transition is observed for any $0<\alpha_\vecp<\pi/2$ as
demonstrated in Figure~\ref{pic_CN_45} for $\alpha_\vecp=\pi/4$
at $Re = 10$ over
$-1<k_x<1$ and $0<k_y<1$ with a grid size of $200\times 100$.
As $Rh$ is 
decreased from $2$ to $1/2$, the most unstable wavevector occurs near
the traces resonant to $\pm \vecp$. 
Thus, the most unstable wavevector
moves from a wavevector perpendicular to $\vecp$
(an inflectional instability) to a wavevector resonant to $\pm \vecp$.

The transition of instability to the resonant triad interaction can be
shown from (\ref{CF}) for $Rh\to 0$ as follows. Suppose that the 
wavevector $\veck$ is not resonant to either $\pm \vecp$ so that
$\omega_{\pm 1}\ne 0$ in (\ref{an_dn}). 
Then, since $|\omega_{\pm 1}| \to \infty$ in the $Rh\to 0$ limit, 
we have $|a_{\pm 1}|\to \infty$ in (\ref{an_dn}). Thus, 
(\ref{CF}) is reduced to 
\begin{eqnarray*}
a_0=0,
\end{eqnarray*}
which is equivalent to
\begin{eqnarray*}
\sigma(\veck;\alpha_\vecp,Re,Rh)=-k^2/Re < 0.
\end{eqnarray*}
Thus only wavevectors resonant
to the base flow wavevector $\vecp$ (or its conjugate) can be unstable. 

\begin{figure}
\begin{center}
\begin{tabular}{cc}
(a) $Re = 10$, $Rh = 2$
&
(b) $Re= 10$, $Rh = 1/2$
\\
\includegraphics[height= 5cm]{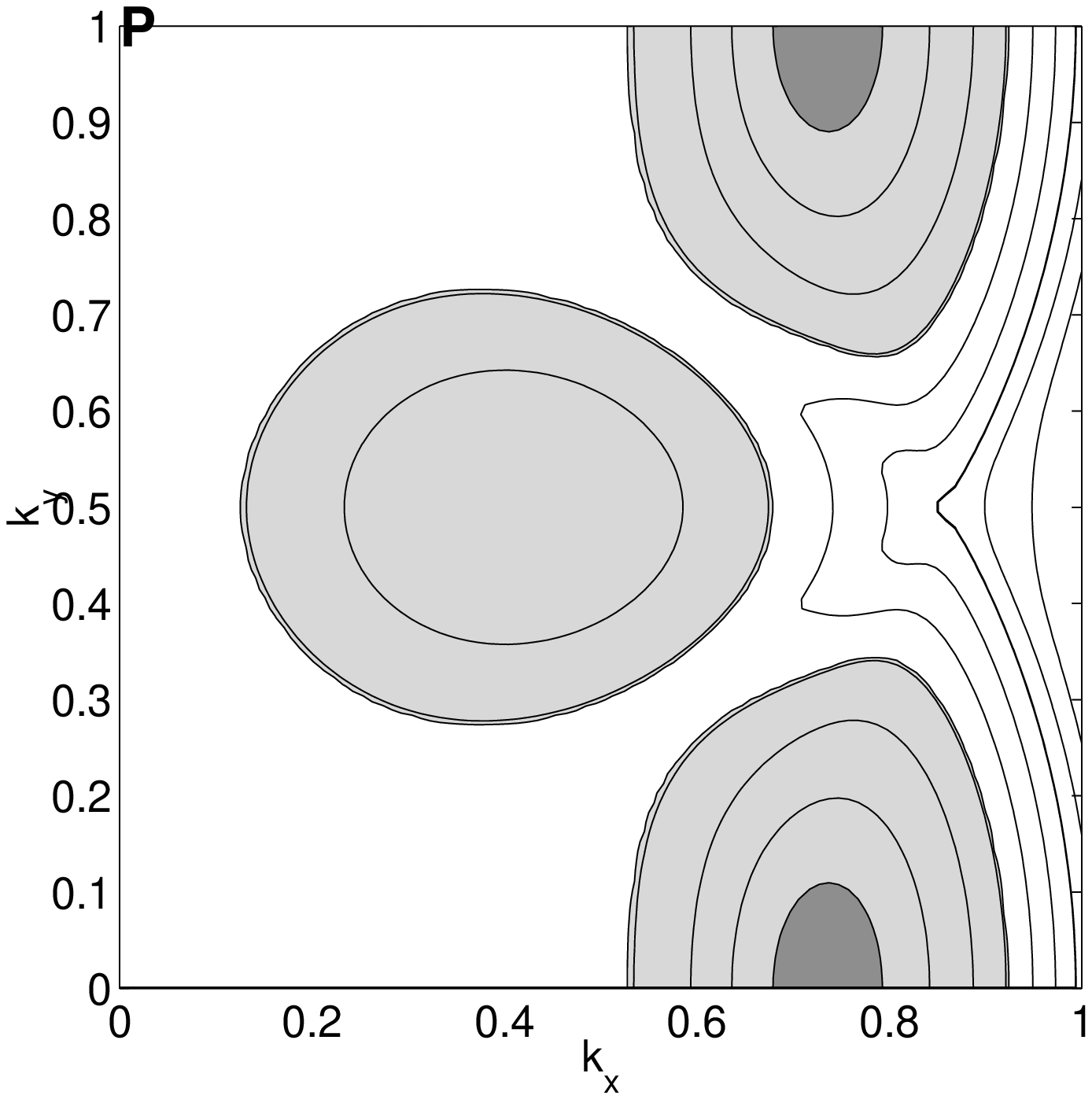}
&
\includegraphics[height=5cm ]{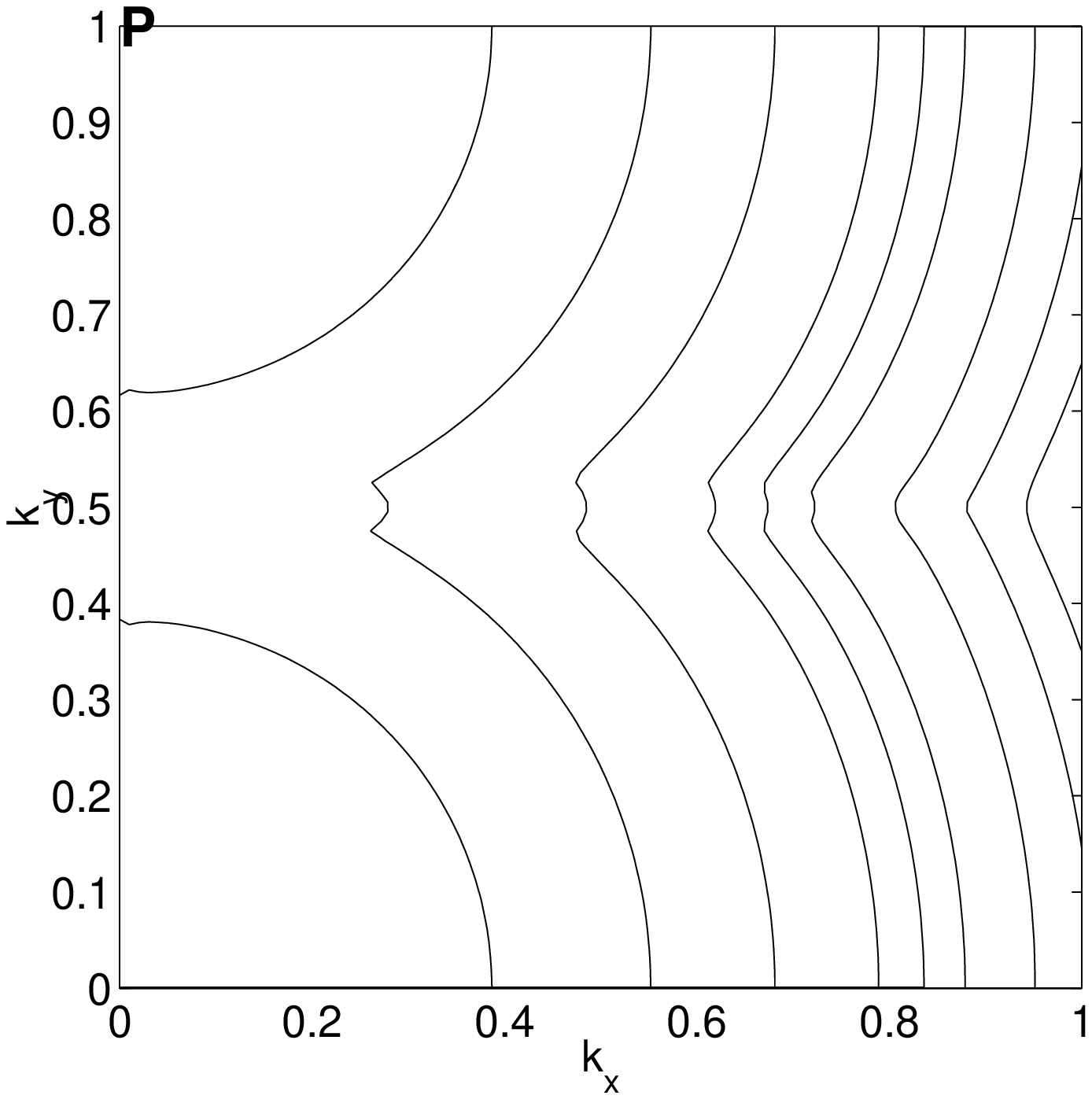}
\end{tabular}
\end{center}
\caption{
Contour plot of $\Re(\sigma(\veck;\alpha_\vecp,Re,R))$
for $\alpha_\vecp=\pi/2$ at $Re=10$.
}\label{pic_CN_90}
\end{figure}

The case of $\alpha_\vecp=\pi/2$ must be considered separately.
As discussed in Section \ref{sec_reson}, the base flow wavevector
$\vecp = (0,1)$ cannot lose energy directly by resonant triad interactions,
and as the Rhines number is decreased, the flow becomes stable as show in
Figure~\ref{pic_CN_90} at $Rh=1/2$. Indeed, at any finite Reynolds number,
the zonal base flow ($\alpha_\vecp=\pi/2$) 
becomes stable at sufficiently small Rhines number.
Stability of zonal flows for small Rhines number was shown
by Kuo \cite{KUO} and is discussed further in Section \ref{sec_large_beta}.

\begin{figure}
\begin{center}
\begin{tabular}{cc}
(a) $Re = 10$, $Rh = 2$
&
(b) $Re= 10$, $Rh = 1/2$
\\
\includegraphics[height= 5cm]{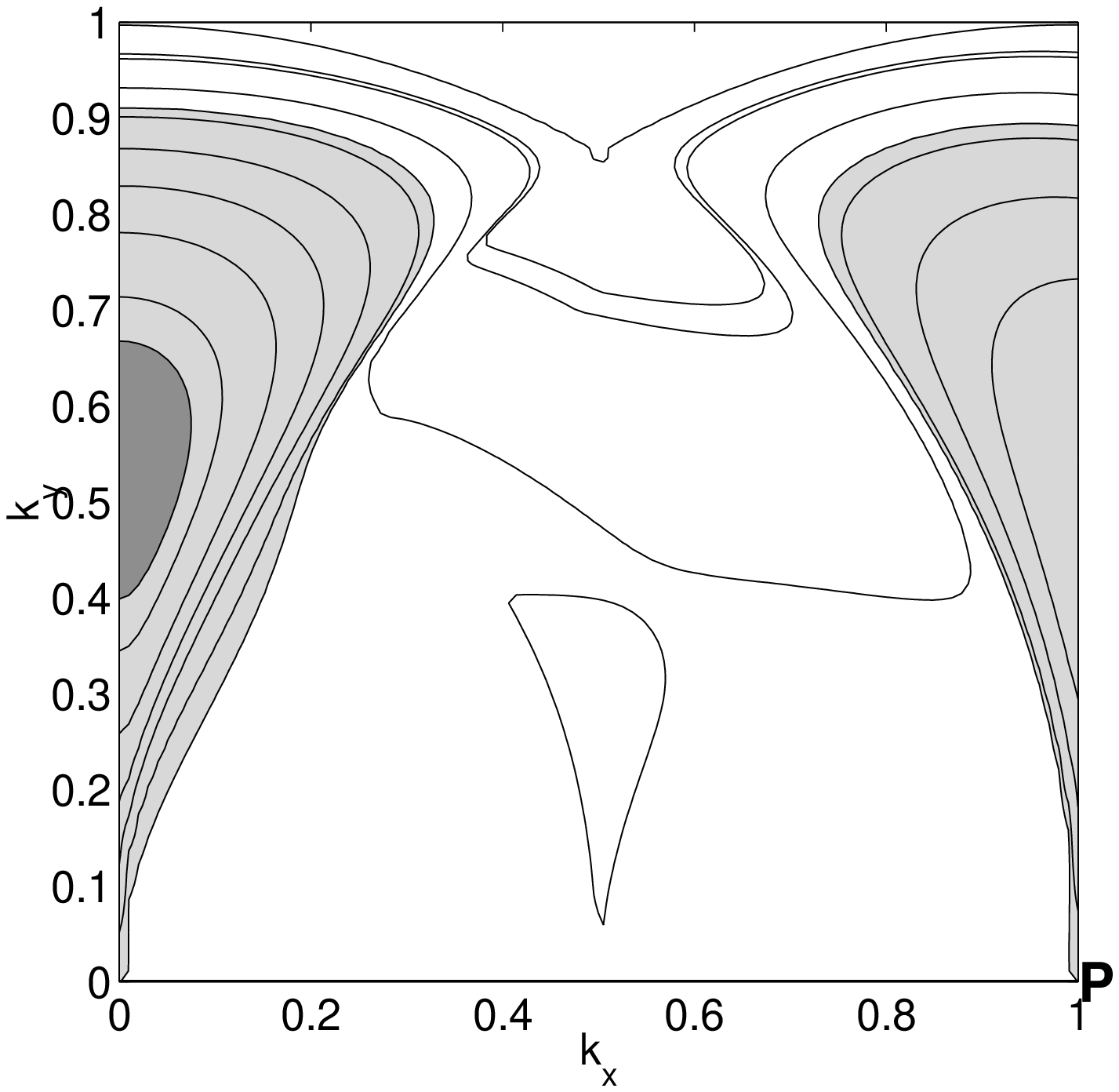}
&
\includegraphics[height=5cm ]{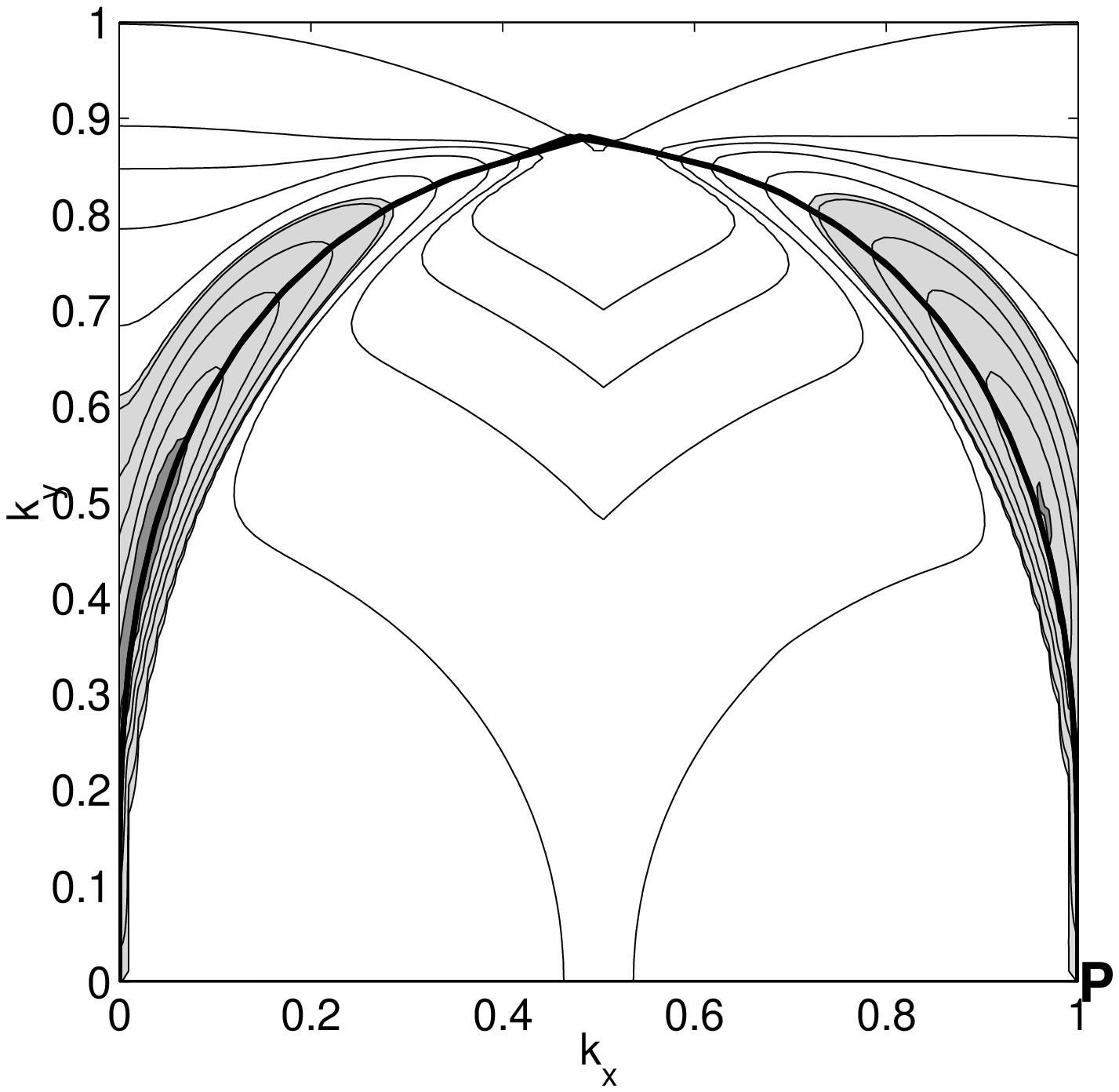}
\end{tabular}
\end{center}
\caption{
Contour plot of $\Re(\sigma(\veck;\alpha_\vecp,Re,Rh))$ obtained from
(\ref{TCF_one})
for $\alpha_\vecp=0$ at $Re=10$. Resonant traces including $\pm \vecp$
are solid lines in (b).
}\label{pic_CN_one}
\end{figure}

The transition from inflectional to resonant triad instability
can be observed with only two triads:$\veck,\vecp,-(\veck+\vecp)$
and $\veck,-\vecp,-(\veck-\vecp)$. This corresponds to the severe 
truncation of (\ref{CF}) to 
\begin{eqnarray}
a_0+\frac{1}{a_1} + \frac{1}{a_{-1}} = 0
\label{TCF_one}
\end{eqnarray}
which is same as the eigenvalue problem obtained from (\ref{ILS})
after $3\times 3$ truncation with $n = -1,0,1$.
Figure~\ref{pic_CN_one} shows the linear growth rate 
$\Re(\sigma(\veck;\alpha_\vecp,Re,Rh))$ by solving (\ref{TCF_one})
and the change in the nature of the instability 
is observed as $Rh$ is decreases.
In particular, the Rhines number for the transition for $\alpha_\vecp=0$ is
also between $2$ and $1/2$, as in Figure~\ref{pic_CN_0}. 
Because of the severe truncation,
$\Re(\sigma(\veck;\alpha_\vecp,Re,R))$ is not invariant 
under the shift $j\vecp$. 
Comparing Figure~\ref{pic_CN_0}(b),(c) and Figure~\ref{pic_CN_one},
at least qualitatively, (\ref{TCF_one}) is a good approximation of (\ref{CF})
for wavevectors $\veck$ with $k<1$. 
This observation is the
background of the numerical computation in Section \ref{sec_manyf}
to solve (\ref{ILS_m}) approximately with $m\ge 2$. 
Note that such truncation ($3\times 3$) 
was also used in Lorenz \cite{L} and
Gill \cite{G}.

\subsection{Inflectional instability with $\veck\perp \vecp$}
\label{sec_inf}

As defined in section \ref{sec_CN}, we refer
to an inflectional instability 
as the instability at wavevectors near perpendicular to 
$\vecp$ at large Rhines number 
($Rh \ge 2$).
Inflectional instability with $\veck \perp \vecp$ is observed
to be strong for large $Rh$. 
For example,
for $\alpha_\vecp=0$ and $Rh \ge 2$, 
inflectional instabilities are dominant (Figure~\ref{pic_CN_0}). 
For $\alpha_\vecp=\pi/4$ and $Rh \ge 2$,
the wavevector with the strongest growth rate is almost perpendicular
to $\vecp$, although they are not exactly perpendicular to $\vecp$
(Figure~\ref{pic_CN_45}).
Thus, studying the inflectional instability with $\veck\perp\vecp$ can give
a qualitative picture of
the behavior of instability for large Rhines number.

\begin{figure}
\begin{center}
\begin{tabular}{c}
\includegraphics[width = 7cm]{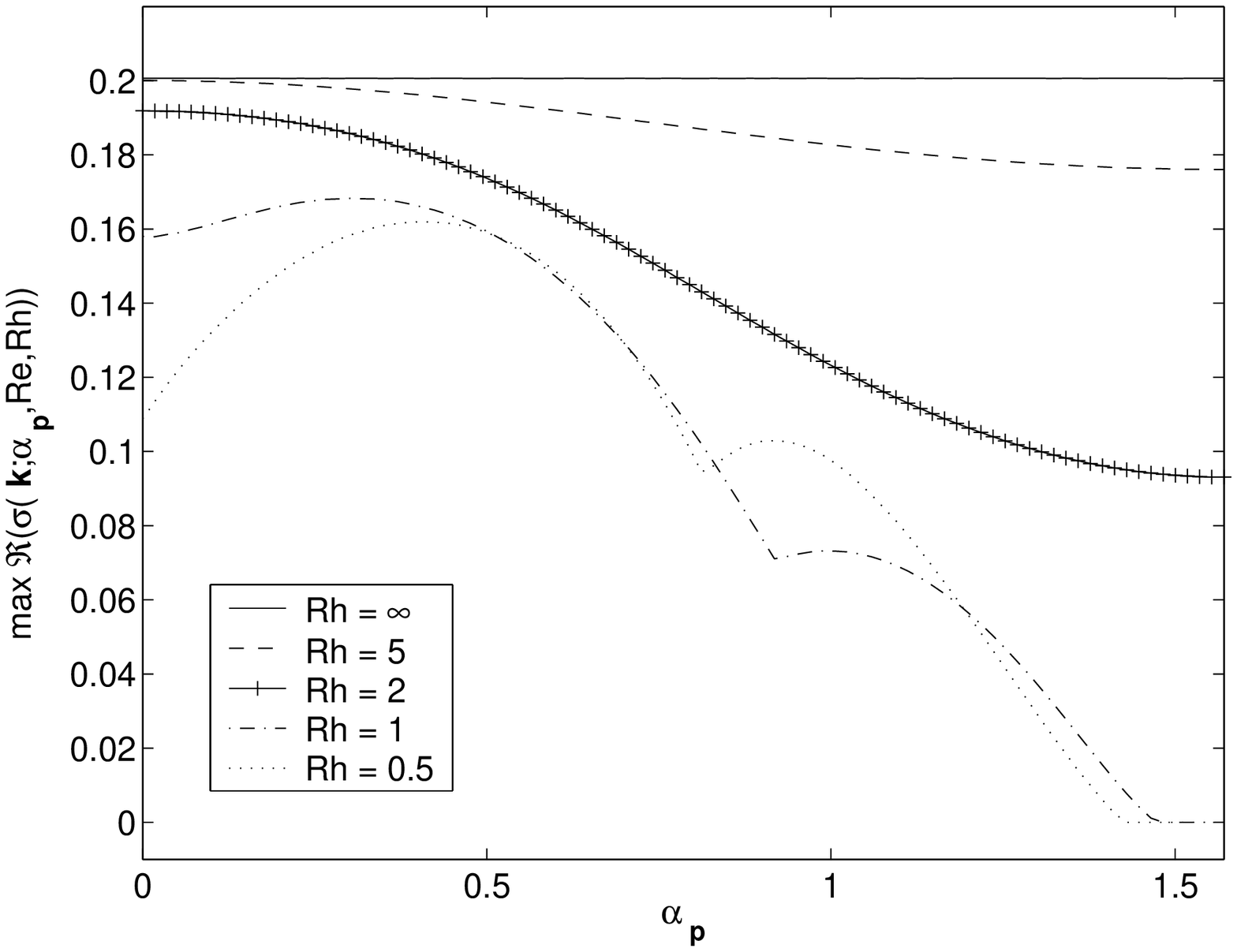}
\end{tabular}
\caption{Maximum growth rate 
$\D{\max_{|\veck|<1}}\Re(\sigma(\veck;\alpha_\vecp,Re,Rh))$ vs $\alpha_\vecp$
at $Re = 10$ and $Rh = \infty$, $5$, $2$, $1$, and $0.5$.
}
\label{pic_inf_line}
\end{center}
\end{figure}

\begin{figure}
\begin{center}
\begin{tabular}{ccc}
(a) $Rh=\infty$
&
(b) $Rh=2$
&
(c) $Rh=0.5$
\\
\includegraphics[width=4cm]{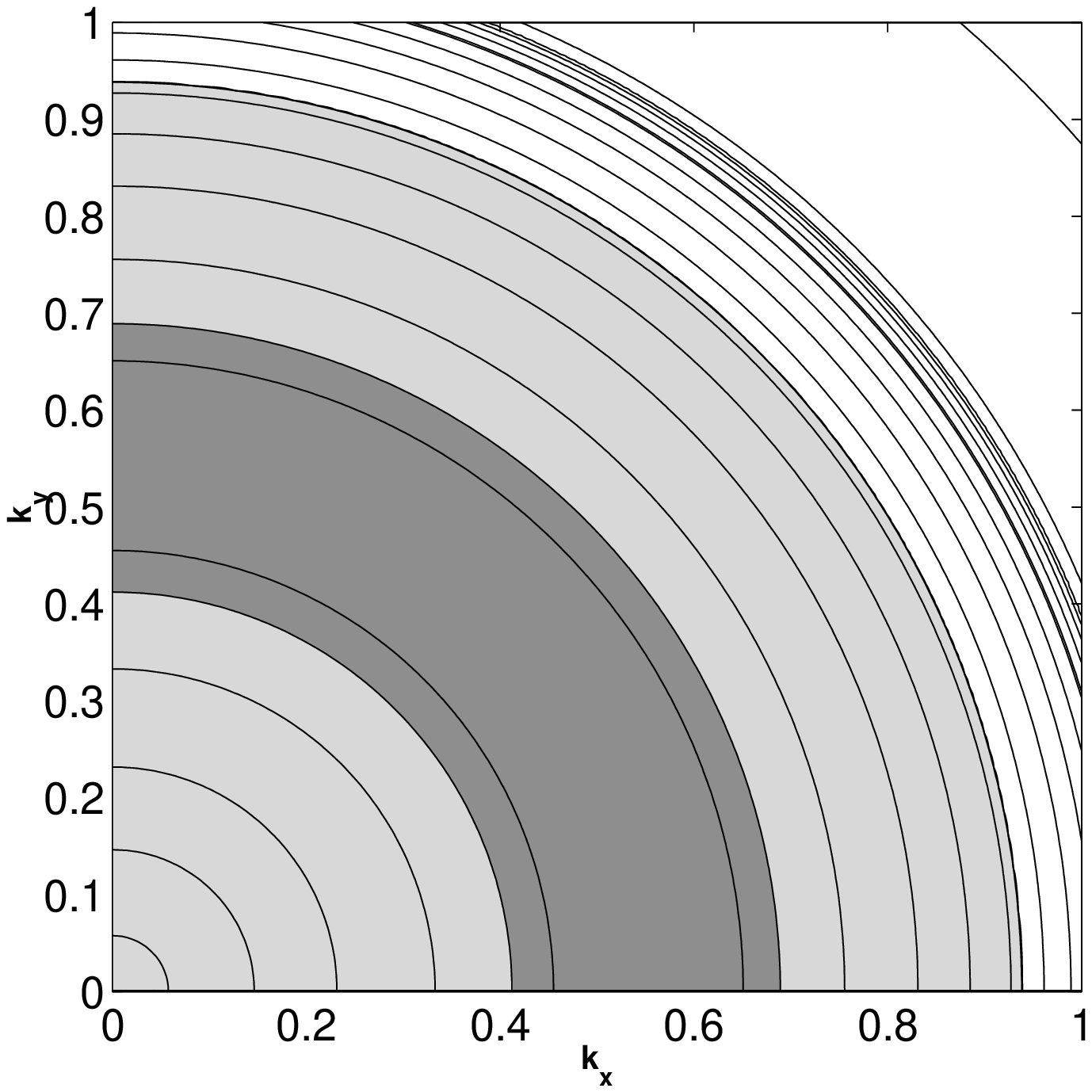}
&
\includegraphics[width=4cm]{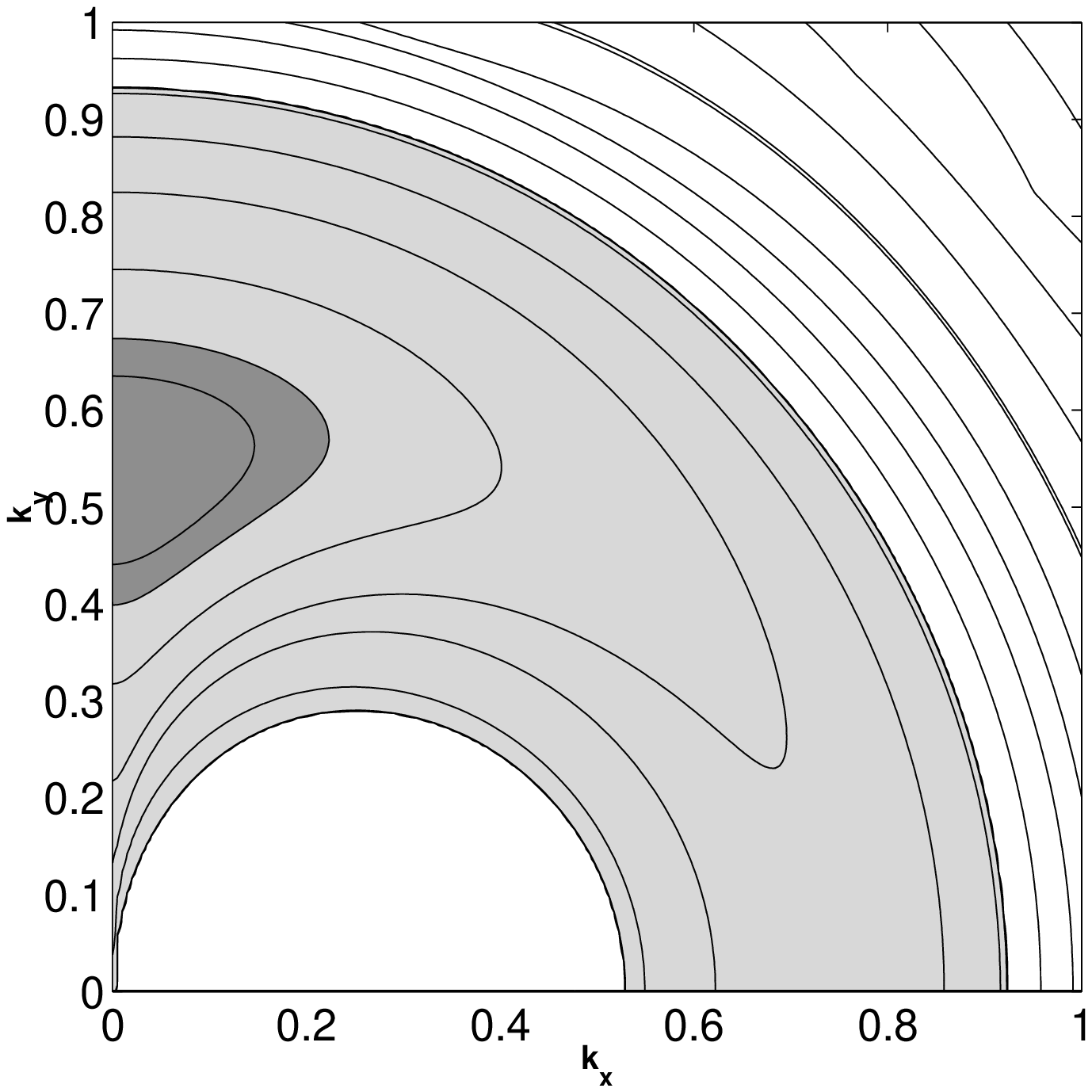}
&
\includegraphics[width=4cm]{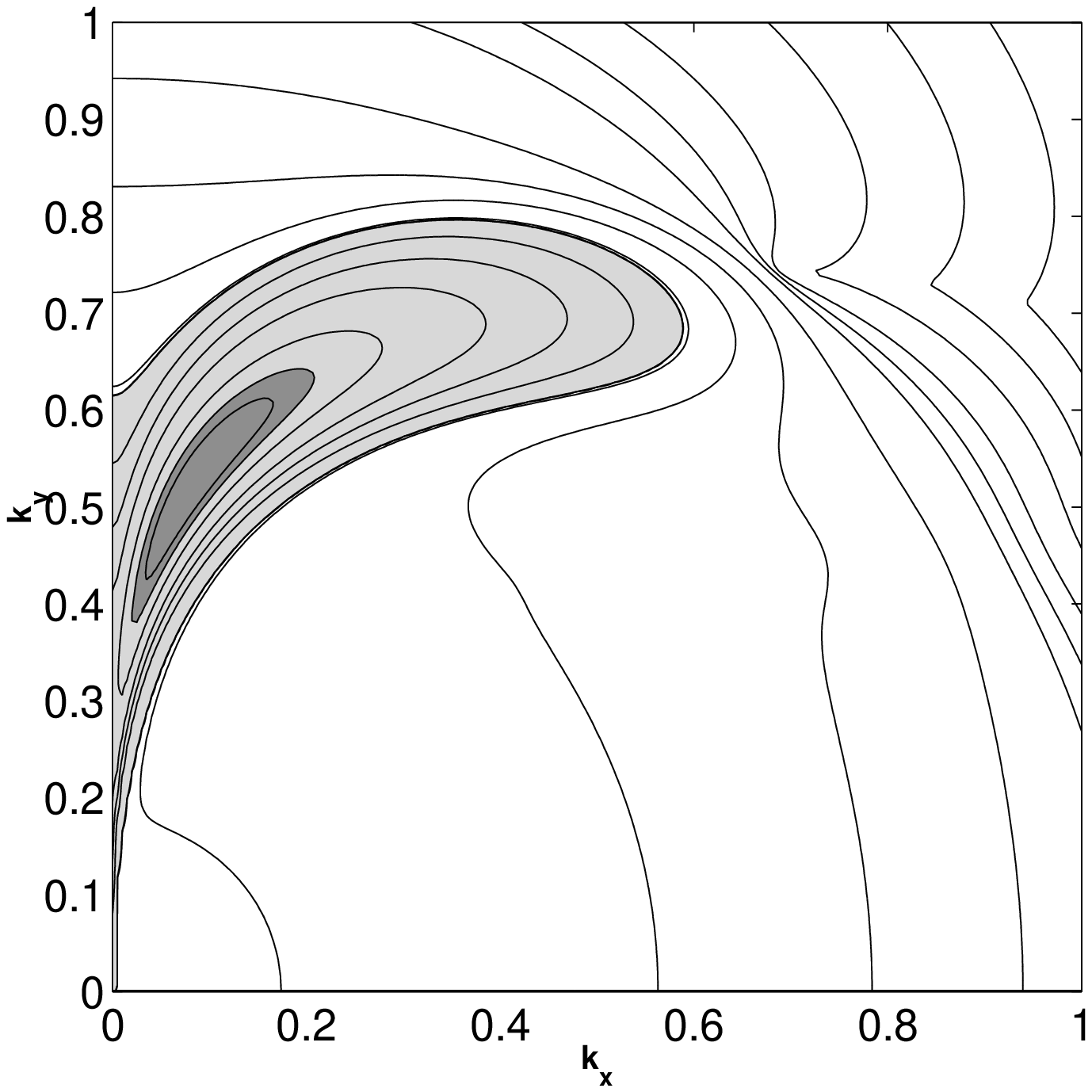}
\end{tabular}
\caption{Contour plot of $\Re(\sigma(\veck;\alpha_\vecp,Re,Rh))$. 
For each $\veck$, 
$\vecp$ is chosen to be $\alpha_\vecp = \alpha_\veck - \pi/2$.
}\label{pic_inf_plane}
\end{center}
\end{figure}

Figure~\ref{pic_inf_plane} shows contours
of growth rates $\Re(\sigma(\veck;\alpha_\vecp,Re,Rh))$
due to the inflectional instability with $\veck\perp\vecp$
at Reynolds number $Re=10$. For each
$\veck$ with $0<k_x<1$ and $0<k_y<1$, 
$\vecp$ is chosen so that $\alpha_\vecp = \alpha_\veck-\pi/2$ 
($\veck\perp\vecp$). For $Rh=\infty$, the growth rate
is invariant under rotations since the Navier Stokes
equations are isotropic (Figure~\ref{pic_inf_plane}(a)). 
For $Rh = 2$, note that the maximum growth rate at $Re = 10$ is strongest
for $\alpha_\vecp = 0$ and monotonically decreasing
for $0<\alpha_\vecp<\pi/2$ (Figure~\ref{pic_inf_line}).
Thus, 
the inflectional instabilities with $\veck \perp \vecp$ for $Rh=2$
are strong for $\veck$ with $k_x=0$ ($\alpha_\vecp = 0$) 
in Figure~\ref{pic_inf_plane}(b)
since
the most unstable wavevectors are near perpendicular to $\vecp$
for large $Rh$.
For $Rh = 1/2$, wavevectors $\veck$ resonant to
$\pm \vecp$  have strong growth rates and such wavevectors can not
have $k_x = 0$ (Section \ref{sec_reson} and \ref{sec_CN}). 
In Figure~\ref{pic_inf_plane}(c) the strongest inflectional instability
with $\veck \perp \vecp$ is observed at wavevectors 
$\veck$ with $k_x \ne 0$.

We conclude that for large Rhines numbers ($Rh\ge 2$ for example),
the inflectional instability is the dominant mechanism for instability 
(Section \ref{sec_CN}) and the strongest inflectional
instability is observed when $\alpha_\vecp=0$ and $\alpha_\veck=\pi/2$.
This observation is consistent with the picture of anisotropic
transfer of energy to zonal flows observed in Chekhlov {\it et al.} \cite{CH}
(See Section \ref{sec_disc} for more discussion).
The inflectional instability is no longer dominant for 
smaller Rhines number
($Rh\le 1/2$ for example) when the resonant triad interaction becomes
dominant.

Figure~\ref{pic_inf_line} and \ref{pic_inf_plane} suggest that 
for an isotropic base flow (\ref{psi_general}) with large $m$,
the strongest linear growth rate could be observed
along $k_x=0$ for large Rhines number ($Rh\ge 2$ for example).
Thus, the modes with $\alpha_{\vecp_j}$ close to $0$ may play
more important roles than the modes with $\alpha_{\vecp_j}$ close to
$\pi/2$.
In Section \ref{sec_manyf}, we consider the linear growth rate
for more isotropic base flows (\ref{base_flow}) with $m\ge 2$
by solving 
a severe truncation of (\ref{ILS_m})
to verify this suggestion.

\section{Critical Reynolds number in $Rh\rightarrow 0$
limit}\label{sec_large_beta}

Here we consider the critical Reynolds number $Re^{c}(\alpha_\vecp,Rh)$
in the limit $Rh\to 0$. The main result is
\begin{eqnarray}
\lim_{Rh \to 0 } 
Re^{c}(\alpha_\vecp,Rh) = 
\left\{
\begin{array}{ll}
\sqrt{2}, &\quad \alpha_\vecp = 0 \\
0,       &\quad 0 < \alpha_\vecp  < \pi/2 \\
\infty,   &\quad \alpha_\vecp = \pi/2.
\end{array}
\right.
\label{cre_Rh0}
\end{eqnarray}
For $\alpha_\vecp = \pi/2$, the critical Reynolds number is
$\infty$ in the sense that for any Reynolds number, the base flow becomes
stable if the Rhines number is small enough.
Similarly, the critical Reynolds number is $0$ (or $\sqrt{2}$) for
$0<\alpha_\vecp<\pi/2$ (or $\alpha_\vecp=0$) in the sense that for
any Reynolds number larger than $0$ (or $\sqrt{2}$) the base flow
is unstable if $Rh$ is small enough. As discussed in Section
\ref{sec_mid_beta},
resonant triad interactions are the dominant mechanisms for 
instability in the limit $Rh \to 0$. 
In fact, the result for $0 < \alpha_\vecp <\pi/2$ in (\ref{cre_Rh0})
is obtained by considering wavevectors $\veck$ resonant to $\vecp$
as shown later in this section. 
We will show that the expression for $Re^c(\alpha_\vecp,Rh)$
in the limit $Rh\to 0$ is asymptotic to the critical Reynolds number
obtained from a single resonant triad for $0 < \alpha_\vecp <\pi/2$.
Thus, we first consider linear stability
of a single triad interaction. 

\subsection{A single resonant triad interaction and the critical Reynolds
number}\label{sec_triad}

Linear stability of a single triad $\veck$, $\vecp$ and
$-(\veck+\vecp)$ with a forced wavevector $\vecp$ and
a response wavevector $\veck$ can be treated by
a truncation of (\ref{ILS}) with $n=0,1$ (See \cite{SW1}). 
This corresponds to
\begin{eqnarray}
a_0+\frac{1}{a_1} = 0
\label{TCF_large_beta}
\end{eqnarray}
obtained by a truncation of (\ref{CF}).
Using (\ref{an_expression}), 
(\ref{TCF_large_beta}) can be written as
\begin{eqnarray}
\frac{2k(\sigma + k^2/Re)}{\sin{\alpha}(1-k^2)}
-\frac{k^2\sin\alpha(2\cos{\alpha} + k)}
{2q^2(\sigma+q^2/Re)+i\omega_1} = 0,
\label{triad_sigma_complex}
\end{eqnarray}
where $q = \sqrt{1+2k\cos\alpha+k^2} = | \veck+\vecp |$.
Taking the real part of (\ref{triad_sigma_complex}),
we find that the growth rate 
$\sigma_r= \Re(\sigma)$ satisfies
\begin{eqnarray}
&&(\sigma_r + k^2/Re)(\sigma_r+q^2/Re)
+ \omega_1^2
\frac{(\sigma_r+k^2/Re)(\sigma_r+q^2/Re)}{2\sigma_r+(k^2+q^2)/Re}
\nonumber\\
\nonumber\\
&&\quad =\frac{k\sin^{2}\alpha(2\cos\alpha+k)(1-k^2)}{4 q^2}.
\label{triad_sigma}
\end{eqnarray}
The critical Reynolds number for instability is obtained by setting
$\sigma_r = 0$ in (\ref{triad_sigma}) and is denoted by
$Re^c_T$;
$Re^c_T$ satisfies
\begin{eqnarray}
\left[ Re^c_T\right]^{-2}
=
\frac{\sin^2\alpha(2\cos\alpha+k)(1-k^2)}{4kq^4}
-\frac{\omega_1^2}{Re^c_T(k^2+q^2)}.
\label{cre_triad}
\end{eqnarray}
If $\veck$ is not resonant to $\vecp$, nonzero $\omega_1$
increases the critical Reynolds number. Thus, the smallest critical
Reynolds number is observed at $Rh=\infty$ ($\omega_1=0$)
for $\veck$ not resonant to $\vecp$.
In the case that $\veck$ is resonant to $\vecp$, we have
$\omega_1=0$, and (\ref{cre_triad}) is reduced to
\begin{eqnarray}
\left[ Re^c_T(\veck;\alpha_\vecp,Rh=\infty) \right]^{2}
=
\frac{4k(1+2k\cos\alpha+k^2)^2}{\sin^2\alpha(2\cos\alpha+k)(1-k^2)}
\label{cre_reson_triad}
\end{eqnarray}
yielding the expression obtained in 
Smith and Waleffe \cite{SW1}.

\subsection{$0<\alpha_\vecp<\pi/2$}\label{sec_cre_interior}

\begin{figure}
\begin{center}
\begin{tabular}{cc}
(a) $Re^c(\veck;\alpha_\vecp,Rh=k^s)$ 
& (b) $Re^c(\veck;\alpha_\vecp,Rh=k^s)$
\\
\includegraphics[height=4.5cm]{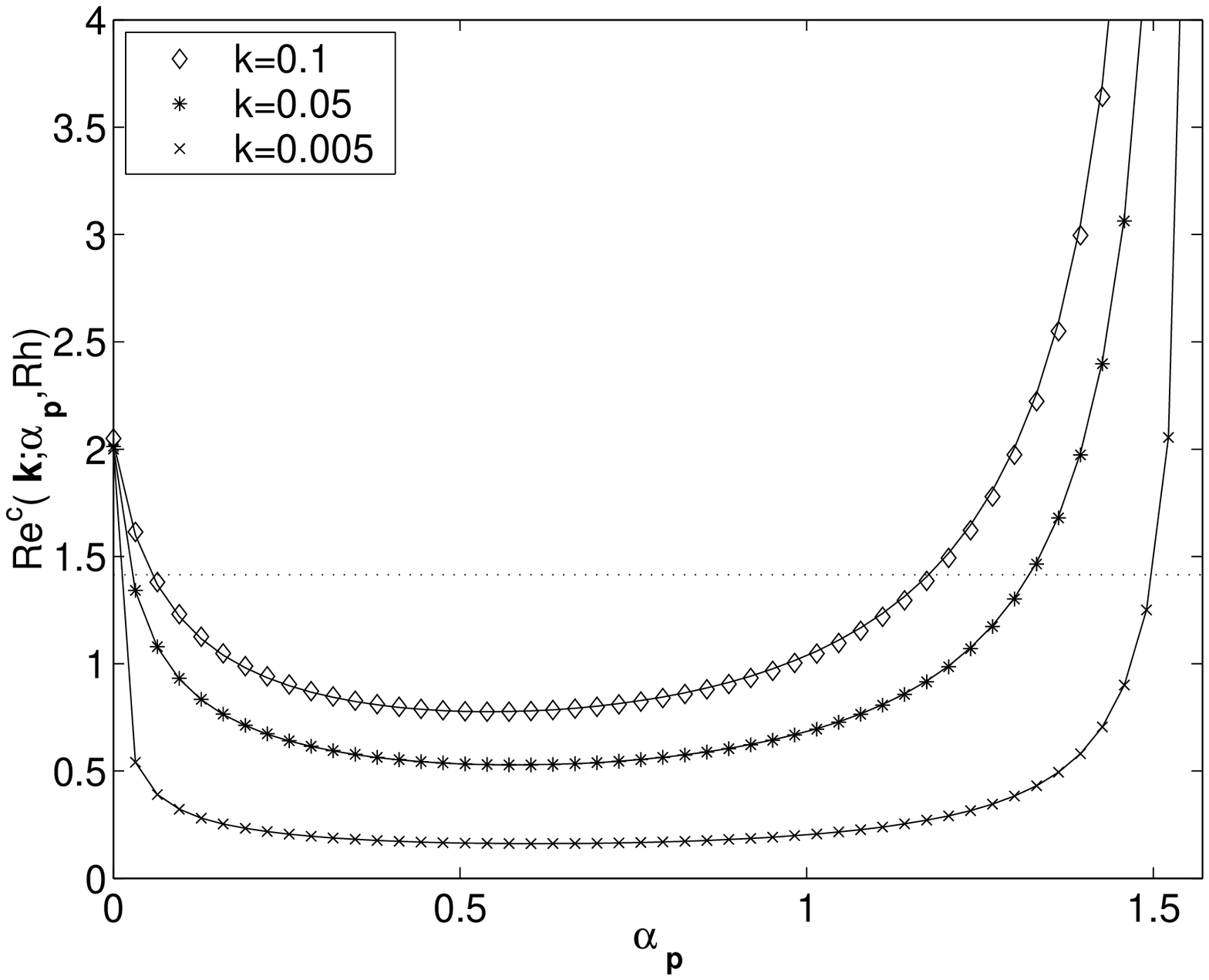}
&
\includegraphics[height=4.5cm]{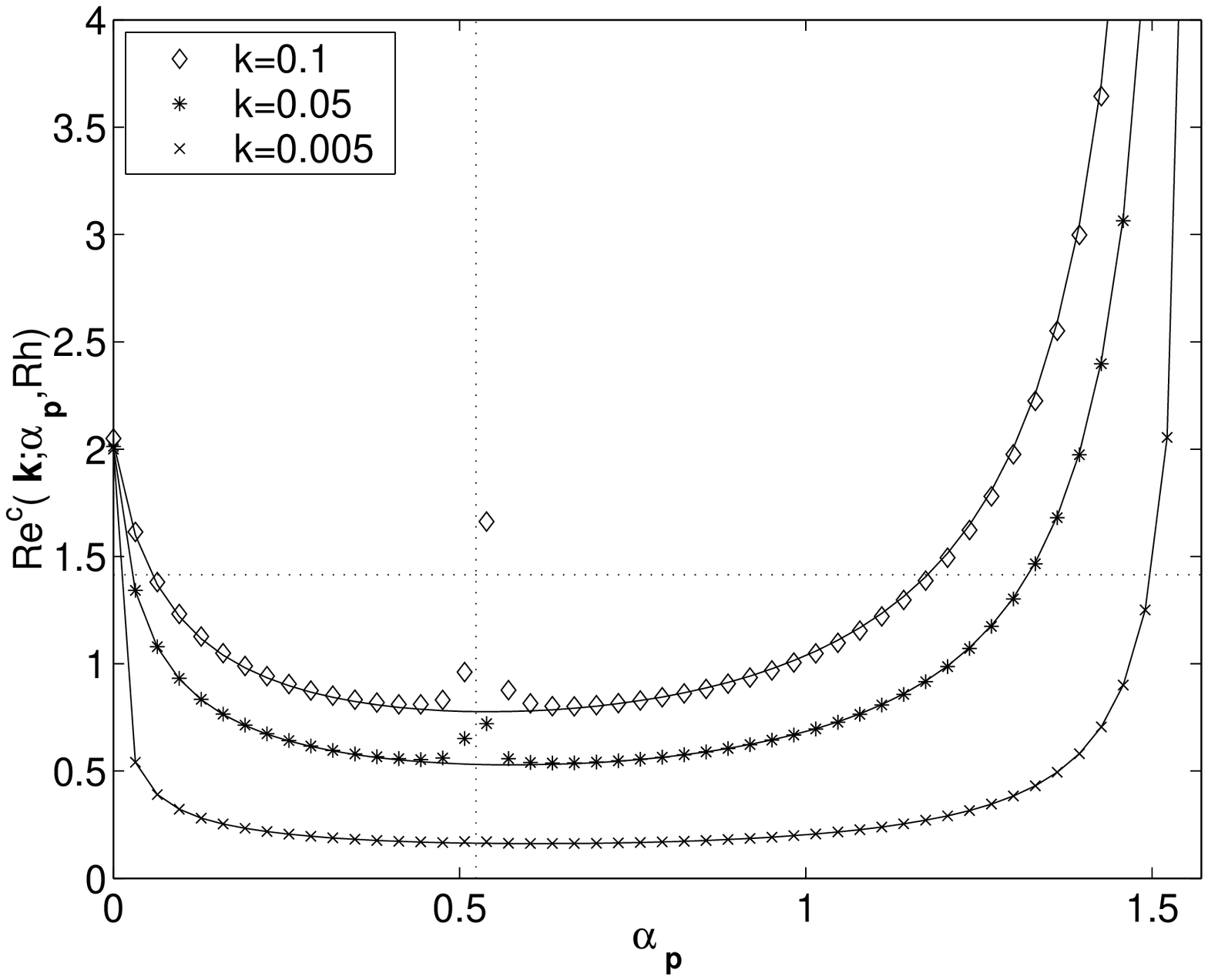}
\\
(c) $Re^c(\veck;\alpha_\vecp,Rh=k^s)$
& (d) $\Re(\sigma(\veck;\alpha_\vecp,Re,Rh))$ 
\\
\includegraphics[height=4.5cm]{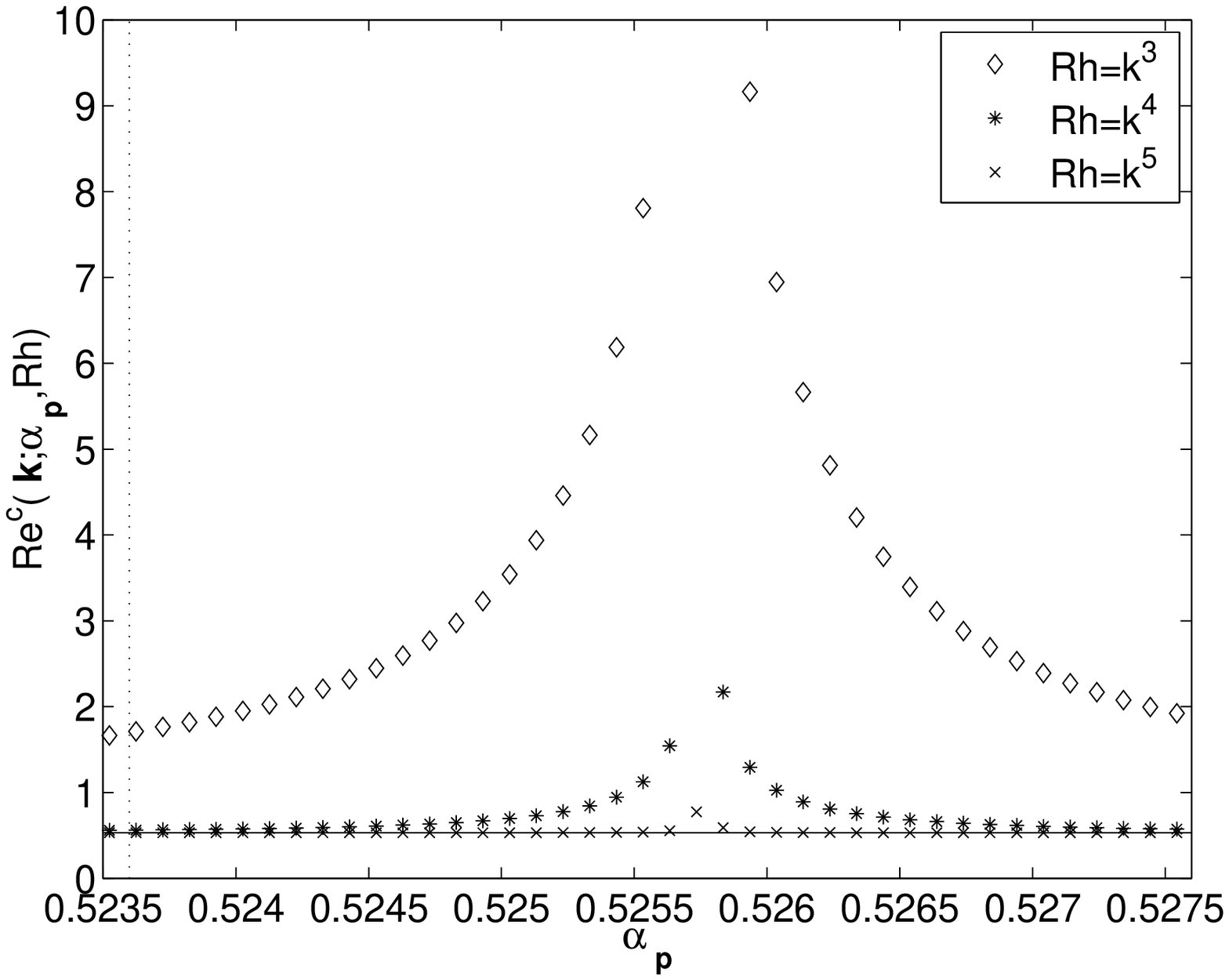}
&
\includegraphics[height=4.5cm]{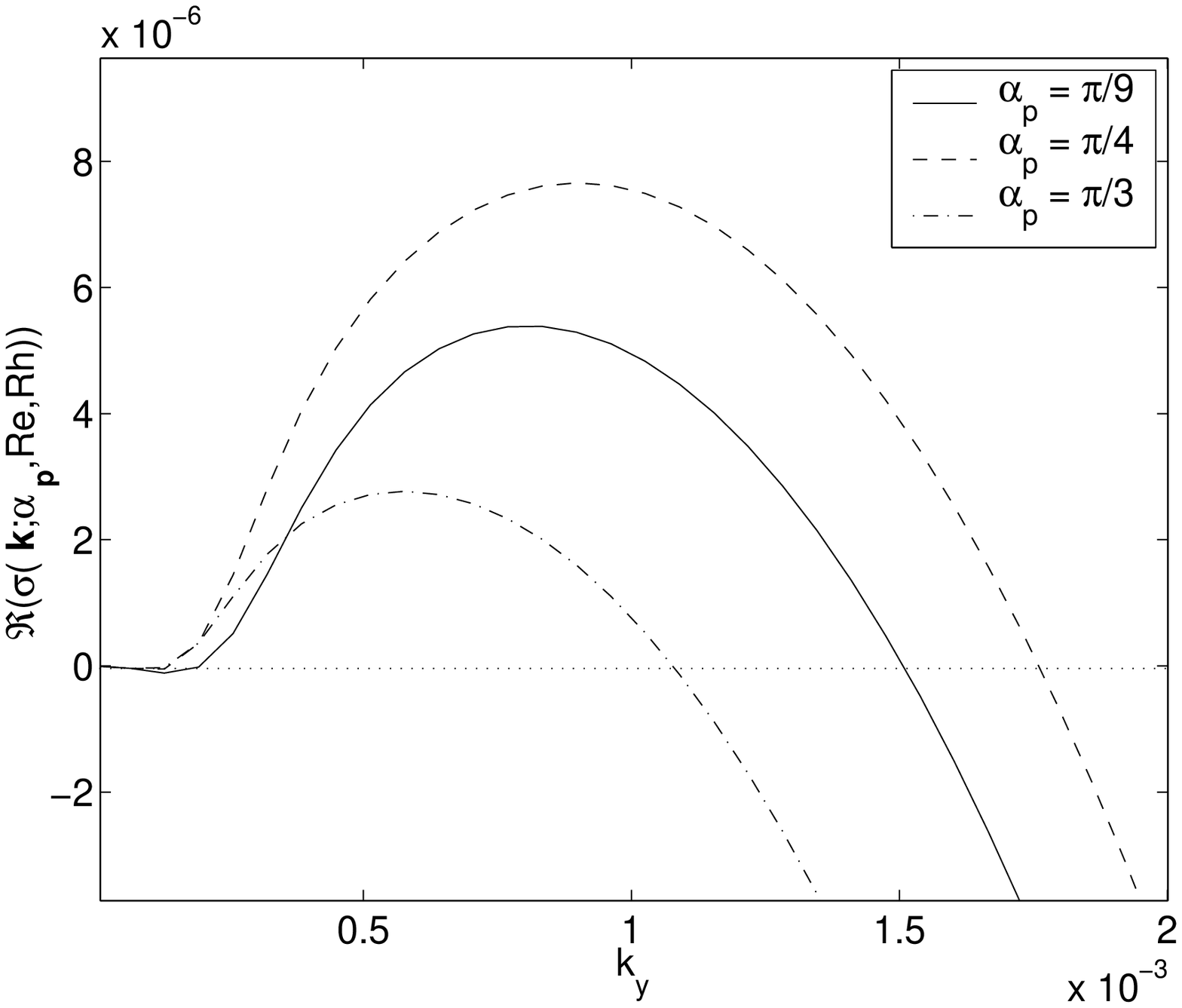}
\end{tabular}
\caption{
In all figures, $\veck$ is resonant to $\vecp$.
In (a-c), the solid lines
are the leading order critical Reynolds numbers
in (\ref{smallRh_claim_2}), and the symbols are numerically computed
critical Reynolds numbers
In (a) and (b), the horizontal dotted lines
are $\sqrt{2}$. In (b) and (c), the vertical dotted lines are
$\alpha_\vecp = \pi/6$. (a) $s = 5$. (b) $s=3$. (c) $k = 0.05$.
In (d), the horizontal axis $k_y$ is the $y$-coordinate
of $\veck$ along resonant traces.
}\label{pic_cre_largeB}
\end{center}
\end{figure}

For $0 < \alpha_\vecp<\pi/2$, consider wavevectors $\veck$
resonant to $\vecp$ and with $k_y > 0$.
We show (\ref{cre_Rh0}) by deriving that,
with a fixed $\alpha_\vecp$,
\begin{eqnarray}
\left[Re^c(\veck;\alpha_\vecp,Rh=k^s)\right]^2
&=& \frac{4k(1+2k\cos\alpha+k^2)^2}
{\sin^2\alpha(2\cos\alpha+k)(1-k^2)} + o(k)
\label{smallRh_claim_1}\\
\nonumber\\
&=& \frac{4k(1+2k\sin\alpha_\vecp+k^2)^2}
{\cos^2\alpha(2\sin\alpha_\vecp+k)(1-k^2)} + o(k),
\label{smallRh_claim_2}
\end{eqnarray}
where $s$ is a positive number depending on $\alpha_\vecp$
to be determined later. The remainder terms $o(k)$ depend on 
$s$ and $\alpha_\vecp$.
Relation (\ref{smallRh_claim_2}) implies (\ref{cre_Rh0}) 
since (\ref{smallRh_claim_2}) approaches zero as $k\to 0$.
The leading order term in (\ref{smallRh_claim_1})
is the same expression for the critical Reynolds number
(\ref{cre_reson_triad}) of a resonant triad,
and (\ref{smallRh_claim_2}) is immediate from 
(\ref{smallRh_claim_1}) by (\ref{reson_asymp_tangential}). 
In Figure~\ref{pic_cre_largeB}(a), the leading order term 
in (\ref{smallRh_claim_2}) (solid lines)
and numerically computed critical Reynolds numbers (symbols)
are plotted for $k=0.1$, $0.05$, and $0.005$ with $s=5$ and
$0\le \alpha_\vecp < \pi/2$. 
\footnote{The Matlab function `fzero' is applied to 
(\ref{ILS}) to find critical
Reynolds numbers numerically.}
For each $k$, the lines and symbols 
are almost identical
verifying (\ref{smallRh_claim_2}). 
Figure~\ref{pic_cre_largeB}(a) also verifies that the critical Reynolds
numbers are decreased to $0$ for $0<\alpha_\vecp<\pi/2$ as $k$ is
decreased. 
In particular, the critical Reynolds numbers become less than $\sqrt{2}$,
the critical Reynolds number for the 2D Navier Stokes equations.
An interesting observation is that the critical Reynolds
number for $\alpha_\vecp = 0$ is $2$ in the limit $k\to 0$ yielding
a boundary layer at $\alpha_\vecp = 0$. Also, for a fixed $k$, the
critical Reynolds number in the limit $\alpha_\vecp \to \pi/2$ is
infinity resulting another boundary layer at $\alpha_\vecp = \pi/2$. 

Figure~\ref{pic_cre_largeB}(a) suggest that $s=5$ is large enough 
to obtain (\ref{smallRh_claim_2}) for $0<\alpha_\vecp<\pi/2$.
In Figure~\ref{pic_cre_largeB}(b), the critical Reynolds
numbers are computed with $s=3$, and agree with the leading order term
in (\ref{smallRh_claim_2}) except near $\alpha_\vecp = \pi/6$.
This observation can be explained as follows.
From (\ref{omega_taylor}) and the fact that $\omega_{1} = 0$, we have
\begin{eqnarray}
\omega_{-1} &=& \omega_{-1} - \omega_{1} \nonumber\\
&=& -\veck^{T}D^{2}\omega(\vecp)\cdot\veck + O(k^4/Rh)\nonumber\\
&=& -(0,k_y)^{T}D^{2}\omega(\vecp)\cdot (0,k_y) + O(k^4/Rh)\label{p30}.
\end{eqnarray}
The last equality is a consequence of $k_x = O(k_y^3)$ along the
resonant trace. 
One can show that for $\alpha_\vecp = \pi/6$,
the first term in (\ref{p30}) is zero.
Thus we have
\begin{eqnarray}
\omega_{-1} = 
\left\{
\begin{array}{ll} 
O(k^4/Rh), & \quad \textrm{if } \alpha_\vecp = \pi/6,
\\
O(k^2/Rh), & \quad \textrm{otherwise}.
\end{array}
\right.
\label{reson_leading_order}
\end{eqnarray}
To have $\omega_{-1}\to \infty$
in the limit $k\to 0$, we need $s > 4$ for $\alpha_\vecp =
\pi/6$. For $\alpha_\vecp\ne \pi/6$, $s > 2$ is enough to have
$\omega_{-1}\to \infty$. However, to derive (\ref{smallRh_claim_2}),
we need to assume that $ s > 4.5$ for $\alpha_\vecp = \pi/6$
and $s > 2.5$ otherwise, as we will see at the end of this section.
For $\pi/6 - 0.0001 < \alpha_\vecp < \pi/6 + 0.004$, 
Figure~\ref{pic_cre_largeB}(c) shows the critical Reynolds numbers
for $k = 0.05$ with $s = 3,4$ and $5$. 
The divergence near $\alpha_\vecp = \pi/6$ is strong
for $s = 3$, and almost disappears for $s=5$.
Relation (\ref{cre_Rh0}) for $0<\alpha_\vecp<\pi/2$
is also verified in Figure~\ref{pic_cre_largeB}(d)
where the growth rates $\Re(\sigma(\veck;\alpha_\vecp,Re,Rh))$ are
obtained numerically from (\ref{ILS}) 
along wavevectors $\veck$ resonant to $\vecp$
with $\alpha_\vecp = \pi/9, \pi/4$ and $\pi/3$.  
The horizontal line is the $k_y$-coordinate of $\veck$.
We choose a small Reynolds number $Re =0.1$ and 
a small Rhines number $Rh=10^{-8}$, and a positive growth rate 
is observed along wavevectors $\veck$ resonant to $\vecp$ for
each $\alpha_\vecp$.

In the rest of this section,  we sketch the derivation
of (\ref{smallRh_claim_1}). Since we are interested in $k\ll 1$,
we use (\ref{TCF}) and set $\Re(\sigma(\veck;\alpha_\vecp,Re,Rh)) = 0$
to find the critical Reynolds number. Omitting details, we find that
the critical Reynolds number $Re^{c}(\veck;\alpha_\vecp,Rh)$ satisfies
\begin{eqnarray}
&&iRe^c\rho + k^2 = \frac{[Re^c]^{2}k\sin^2{\alpha}}{4}
\left[ \frac{k+2\cos\alpha}{q^2(1)(i\rho Re^c+q^2(1))}
\right.
\nonumber\\
\nonumber\\
&&
\quad\quad 
\left.+\frac{k-2\cos\alpha}{q^2(-1)(i\rho Re^c+q^2(-1) + i\omega_{-1}Re^c)}
\right]
+O([Re^c]^2 k^4)\label{smallRh_cre_eqn},
\end{eqnarray}
where $\rho = \Im(\sigma(\veck;\alpha_\vecp,Re^c,Rh))$ and
the fact that $\veck$ is resonant to $\vecp$ is used to set $\omega_{1}=0$.
Recall that 
\begin{eqnarray*}
q^2(\pm 1) = |\veck \pm \vecp|^2 = 1\pm k\cos\alpha + k^2.
\end{eqnarray*}
By choosing $s$ large enough in $Rh = k^s$, suppose that 
\begin{eqnarray}
\frac{1}{\omega_{-1} Re^c(\veck;\alpha_\vecp,Rh=k^s)} = O(k^{1/2}).
\label{smallRh_assumption}
\end{eqnarray}
Then, the the second term in the bracket of the right-hand side of
(\ref{smallRh_cre_eqn})
becomes higher order than the first term. Thus, balancing the lowest order
terms, we obtain
\begin{eqnarray}
iRe^c\rho + k^2 = 
\frac{[Re^c]^2k\sin^2\alpha(k+2\cos\alpha)}
{4q^2(1)(i\rho Re^c + q^2(1))}
+ o(k^2).
\label{smallRh_CF_asymp}
\end{eqnarray}
From the imaginary part of (\ref{smallRh_CF_asymp}), one can see
that $Re^c\rho = o(k^2)$. Thus, the real part balance in
(\ref{smallRh_CF_asymp})
gives (\ref{smallRh_claim_1}). 
Finally, we have $Re^{c}(\veck;\alpha_\vecp,Rh) = O(k^{1/2})$
from (\ref{smallRh_claim_1}). Thus, by (\ref{reson_leading_order}),
(\ref{smallRh_assumption}) is true for $s>4.5$ for $\alpha_\vecp = \pi/6$
and $s>2.5$ otherwise.

\subsection{$\alpha_\vecp = 0$ and $\alpha_\vecp = \pi/2$}\label{sec_0_90}

For $\alpha_\vecp= 0 $, in Figure~\ref{pic_cre_largeB}(a) and (b), 
the critical Reynolds number along wavevectors resonant to $\vecp$
in the limit $Rh\to 0$ is two, which larger than (\ref{cre_Rh0}).
In fact, we need to consider wavevectors of the form $\veck = (0,k_y)$.
One can show that for sufficiently small $|k_y|\ll 1$, 
\begin{eqnarray}
Re^{c}((0,k_y);\alpha_\vecp=0,Rh)=\sqrt{2}.
\label{claim_alp_0}
\end{eqnarray}
for any fixed $Rh$.
Relation (\ref{claim_alp_0}) implies by definition that for any $Rh$,
\begin{eqnarray*}
Re^{c}(\alpha_\vecp=0,Rh) \le \sqrt{2},
\end{eqnarray*}
which in turn implies inequality ``$\le$'' in (\ref{cre_Rh0}).
To show (\ref{claim_alp_0}), note that (\ref{growth_rate})
at $\veck=(0,k_y)$
is reduced to the growth rate for the 2D Navier Stokes equations with
$\veck\perp\vecp$ since $\omega(\veck) = 0$. 
Thus, the critical Reynolds number at $\veck=(0,k_y)$
is $\sqrt{2}$ (see \cite{MS}, \cite{SY} and \cite{WAL}).
It remains to show the other inequality ``$\ge$'' in (\ref{cre_Rh0})
for $\alpha_\vecp = 0$. This requires rather complicated analysis, and
is shown in Appendix \ref{sec_app_Rh0}.

For $\alpha_\vecp=\pi/2$, we know that the resonant triad interaction can not
transfer energy from the wavevector $\vecp=(1,0)$ to other wavevectors
(\cite{LG}, \cite{WAL2}, \cite{WAL3}). This fact can be also verified from 
(\ref{cre_reson_triad}). 
As shown in Figure~\ref{pic_reson_trace}, wavevectors $\veck$
resonant to $\vecp$ have $k_y=-1/2$. Thus, we have
$\cos\alpha = -1/2$. 
From (\ref{cre_reson_triad}), we obtain
\begin{eqnarray*}
\left[Re^{c}(\veck;\alpha_\vecp=\pi/2,Rh\to 0)\right]^2
= -\frac{4k^5}{\sin^2\alpha(1-k)^2(k+1)} < 0
\end{eqnarray*}
which is impossible. Thus, it is expected that the base flow becomes
stable in the limit $Rh\to 0$ when $\alpha_\vecp = \pi/2$
since the dominant instability mechanisms
- resonant triad interactions -
can not transfer energy to other wavevectors. 
We omit the derivation of (\ref{cre_Rh0}) for $\alpha_\vecp=\pi/2$,
which is similar to that for $\alpha_\vecp = 0$ in Appendix
\ref{sec_app_Rh0}.

\section{Comparison with Manfroi and Young's Results}\label{sec_comp_my}

Manfroi and Young \cite{MY} studied the large-scale instability
of a sinusoidal Kolmogorov base flow $\Psi_K^{0}$ given by
\begin{eqnarray}
\Psi_K^{0} = -\cos(\vecp\cdot\vecx).
\label{base_flow_kol}
\end{eqnarray}
With this base flow, the main result
is that the critical Reynolds number in the limit $Rh \to \infty$ is given
by 
\begin{eqnarray}
\lim_{
Rh\to \infty
} 
\min_{\veck}
Re_{MY}^0(\veck;\alpha_\vecp,Rh) =
\left\{
\begin{array}{ll}
\frac{4}{5}\sqrt{2}, &\textrm{ for } \alpha_\vecp = 0 
\\
0, &\textrm{ for } 0<\alpha_\vecp< \pi/2
\\
\sqrt{2}, &\textrm{ for } \alpha_\vecp =  \pi/2\\
\end{array}
\right.
\label{MY_cre}
\end{eqnarray}
Here, $Re_{MY}^{0}(\veck;\alpha_\vecp,Rh)$ is 
the Reynolds number which makes the leading order term
for the growth rate equal to zero. (The formula for the growth rate 
by Manfroi and Young \cite{MY} is
given by (\ref{MY_growth_rate}), derived at the end of this section.) 
$Re^{0}_{MY}(\veck;\alpha_\vecp,Rh)$ 
is analogous to $Re^{0}(\veck;\alpha_\vecp,Rh)$ in
Section \ref{sec_small_beta} and is understood as the critical Reynolds
number in the limit $k\to 0$.  Note that
Manfroi and Young use the nondimensional number
`$\beta$' which is same as $Re/Rh$ in this paper. 

For $0\le\alpha_\vecp <\pi/2$ in (\ref{MY_cre}),
an interesting observation is that the critical Reynolds in the
limit $Rh\to \infty$ is less than $\sqrt{2}$, 
the critical Reynolds number for the case $Rh = \infty$. In other words,
there is a discontinuity in the critical Reynolds number at $Rh = \infty$.
The result (\ref{MY_cre}) is different from our result for
a Rossby wave base flow $\Psi_R^{0}(\vecx,t;\vecp)$ given by
\begin{eqnarray}
\Psi_R^{0}(\vecx,t;\vecp) = -\cos(\vecp\cdot\vecx - \omega(\vecp) t).
\label{base_flow_rossby}
\end{eqnarray}
For (\ref{base_flow_rossby}), it is observed 
that the marginal Reynolds number $Re^{0}(\veck;\alpha_\vecp,Rh)$
satisfies (\ref{contsqrt2}) for any $\alpha_\vecp$.
Thus, there is no
discontinuity in the critical Reynolds number in the limit $Rh\to \infty$
for (\ref{base_flow_rossby}).

To see the difference between two base flows $\Psi_K^{0}$ and $\Psi_R^{0}$,
consider the forces $\vecF_K$ and $\vecF_R$ that respectively maintain 
the base flows. For the Kolmogorov base flow $\Psi_K^{0}$, the force $\vecF_K$
satisfies
\begin{eqnarray}
&&\frac{1}{Re}(\nabla\times\vecF_K)\cdot\vecz
=\sqrt{\left(\frac{p_x}{Rh}\right)^2 + \left(\frac{1}{Re}\right)^2}
\cos(\vecp\cdot\vecx - \chi)
\nonumber\\
\nonumber\\
&&
\textrm{and } \quad \chi = \tan^{-1}(p_xRe/Rh)\label{MY_FK}
\end{eqnarray}
For the Rossby wave base flow
$\Psi_R^{0}$, the force $\vecF_R$ satisfies
\begin{eqnarray}
\frac{1}{Re}(\nabla\times\vecF_R)\cdot\vecz
=\frac{1}{Re}\cos(\vecp\cdot\vecx - \omega(\vecp)t).
\label{MINE_cre}
\end{eqnarray}
From (\ref{MY_FK}) the phases between $\vecF_K$ and $\Psi_K^{0}$ are different
by $\chi$ if $p_x \ne 0$ ($\alpha_\vecp\ne\pi/2$)
and $Rh < \infty $. 
In particular, since (\ref{MY_FK}) does not
hold for $Re = \infty$ ( $\chi$ becomes infinite at $Re = \infty$),
$\Psi_K^{0}$ is not
a solution of the unforced, inviscid $\beta$-plane equation.
In contrast, 
the force $\vecF_R$ maintaining the Rossby wave base flow
$\Psi_R^{0}$ has the same phase as $\Psi_R^{0}$,
and is a solution
of the unforced, inviscid $\beta$-plane equation.
In the case $\alpha_\vecp=\pi/2$, two base flows (\ref{base_flow_kol})
and (\ref{base_flow_rossby}) are identical and 
the critical Reynolds number is $\sqrt{2}$ 
for both cases in the limit $Rh\to \infty$.

The discontinuity in the critical Reynolds number at $Rh=\infty$
for base flow $\Psi_K^{0}$ is probably linked to the fact that
$\Psi_K^{0}$ is not a solution to the inviscid, unforced $\beta$-plane
equation. 
Manfroi and Young pointed out that 
a discontinuity in the
critical Reynolds number is observed in the 2D
Navier-Stokes equations with base flow 
\begin{eqnarray*}
\Psi_N^{0}(\vecx;\delta) = \sin{x} + \cos(\delta y).
\end{eqnarray*}
In \cite{JAP}, Gotoh and Yamada 
showed that the critical Reynolds number for $\delta = 1$
is different from the critical Reynolds number in the limit $\delta \to 1 $.
Similar to $\Psi_{K}^{0}$, for $\delta\ne 1$ 
the base flow $\Psi_N^{0}$
is not a solution of the 2D unforced 
Euler equation. This is immediate since
$ J(\nabla^{2}\Psi_N^{0},\Psi_N^{0}) \ne 0$ for $\delta \ne 1$.

In the rest of this section, we sketch the derivation of
large-scale growth rates for both 
$\Psi_{K}^{0}$ and $\Psi_{R}^{0}$ by Multiple Scales analysis
following Manfroi and Young \cite{MY}. 
Further details are given in Appendix \ref{sec_app_MS}.
To this end, we consider a base
flow 
\begin{eqnarray}
\Psi_M^{0}(\vecx,t;\vecp,c) = -\cos(\vecp\cdot\vecx - c\omega(\vecp) t).
\label{base_flow_MY}
\end{eqnarray}
Note that $\Psi_M^{0}(\vecx,t;\vecp,0) = \Psi_K^{0}(\vecx;\vecp)$
and $\Psi_M^{0}(\vecx,t;\vecp,1) = \Psi_R^{0}(\vecx,t;\vecp)$.
Following Manfroi and Young, we consider a rotation of the coordinates
by $\alpha_\vecp$. As a result, the base flow becomes periodic in
$x$, and the Multiple Scales analysis becomes simpler.

With rotation by $\alpha_\vecp$, the $\beta$-plane equation is given by
\begin{eqnarray}
&&(\nabla^2\psi)_t + J(\nabla^2\psi,\psi)
+ (1/Rh_x)\psi_x - (1/Rh_y)\psi_y
\nonumber\\
\nonumber\\
&=& 
(1/Re)\nabla^4\psi + 
(1/Re)(\nabla\times\vecF)\cdot\vecz
\label{beta_eqn_rot}
\end{eqnarray}
and the base flow $\Psi_M^{0}$ becomes
\begin{equation}\label{base_flow_c}
\tilde{\Psi}_{K}^{0}(\vecx,t;\vecp,c) = 
-\cos( x - c\omega((1,0);\alpha_\vecp) t)
\end{equation}
where $(1/Rh_x,1/Rh_y) = (1/Rh)(\cos\alpha_\vecp,\sin\alpha_\vecp)$.
The rotated dispersion relation $\omega(\veck;\alpha_\vecp)$ 
is
\begin{eqnarray*}
\omega(\veck;\alpha_\vecp)
= \frac{ -k_x/Rh_x + k_y/Rh_y }{ k^2 }.
\end{eqnarray*}

We consider scalings similar to \cite{MY}:
\begin{eqnarray*}
\begin{array}{l}
\partial_t \rightarrow \partial_t + \eps\partial_T + \eps^2\partial_\tau,
\quad \partial_x \rightarrow \partial_x + \eps\partial_X, 
\quad \partial_y \rightarrow \eps\partial_Y \\
\\
Rh =\eps^{-1}Rh^1,
\quad Rh_x = \eps^{-1}Rh_x^1,\quad Rh_y = \eps^{-1}Rh_y^1\\
\\
\psi = \psi^{0} + \eps\psi^{1} + \eps^2\psi^{2}+\cdots.\\
\end{array}
\end{eqnarray*}
Our interest is in large Rhines number $Rh = \eps^{-1}Rh^1$, and large
scales $\vecX = (X,Y)$. At such spatial scales, the growth rate
is expected to be $O(\eps^2)$, thus the $O(\eps^{2})$ time scale
$\tau$ is introduced (see \cite{S}, 
\cite{SY}, \cite{FLV}, \cite{MY}, and \cite{MY2}).
Large scales correspond to wavevectors $\veck$ with $k=O(\eps)$.
Thus, at large scales
the dispersion relation, which is an inverse time scale, satisfies
$\omega(\veck;\alpha_\vecp) = (-Rh_xk_x + Rh_yk_y)/k^2 = O(1)$
since $Rh=\eps^{-1}Rh^1$.
This motivates the $O(1)$ time scale $t$.
Finally, since the base flow $\tilde{\Psi}_M^{0}$ depends on
the time scale $Rh^{-1}t = O(\eps)$, the $O(\eps)$ time scale
$T$ is introduced. Since the Kolmogorov flow $\Psi_K^{0}$
(\ref{base_flow_kol}) does not depend on $T$, it is sufficient to
consider $\partial_t \to \partial_t + \eps^{2}\partial_\tau$
in Manfroi and Young \cite{MY}.

To find the linear growth rate, we set the leading order term 
$\psi^{0}$ to
\begin{eqnarray*}
\psi^{0} = A(T,\tau)e^{i(\vecK\cdot \vecX - \Omega(\vecK;\alpha_\vecp) t)},
\end{eqnarray*}
where  the large-scale dispersion relation is 
\begin{eqnarray*}
\Omega(\vecK;\alpha_\vecp) =  \D{ \frac{- K_x/Rh_x^1 + K_y/Rh_y^1}{K^2}}.
\end{eqnarray*}
Applying Multiple Scales analysis, we find that the coefficient
$A(\tau)$ satisfies
\begin{eqnarray*}
A_T = 0 \ \quad\textrm{and}\quad A_\tau = \sigma A,
\end{eqnarray*}
for a complex number $\sigma$.
The real part of $\sigma$ is the growth rate and is given by
\begin{eqnarray}
&&\Re(\sigma(\veck;\alpha_\vecp,Re,Rh^1,c)) =
- \frac{K^2}{Re} \left( 1 \right.
\nonumber\\
\nonumber\\
&&
\left.
\quad -
Re^2 \frac{K_y^2}{K^2}
\frac{K_y^2-7K_x^2+ (K \Omega Re)^2
-4K_x \Omega Re^2 (-c+1)/Rh_x^1}
{2K^2(1+(\Omega Re)^2)^2}
\right)\label{MY_growth_rate}
\end{eqnarray}
where $\Omega = \Omega(\vecK;\alpha_\vecp)$.
The derivation of this growth rate is very similar to the one
found in Manfroi and Young \cite{MY}, and the main steps are
given in Appendix \ref{sec_app_MS}.

The expression (\ref{MY_growth_rate}) with $c=0$
is Manfroi and Young's result for the large-scale
growth rate. Then, 
$Re^0_{MY}(\veck;\alpha_\vecp,Rh)$ in (\ref{MY_cre}) is found
by setting the growth rate (\ref{MY_growth_rate}) 
with $c=0$ equal to zero.
By taking a special path for $\veck\to \veczero$ and
$Rh\to 0$ for each $\alpha_\vecp$, Manfroi and Young \cite{MY} obtained
the critical Reynolds number (\ref{MY_cre}).
For $c=1$, 
after adjusting the scales and rotating the expression by $\alpha_\vecp$,
(\ref{MY_growth_rate}) is identical
with the leading order term in the growth rate (\ref{growth_rate})
obtained from the continued
fraction. Mathematically, it is the term $4K_x\Omega Re^2(-c+1)/Rh_x^1$
that leads to the difference between growth rates for
base flows $\Psi_K^{0}$ and $\Psi_R^{0}$.

In earlier work of Manfroi and Young \cite{MY2}, 
they considered a base flow of the form 
\begin{eqnarray}
\Psi^{0}_G = \cos{x} + Uy.
\label{base_flow_Gal}
\end{eqnarray}
This base flow is obtained by adding a term 
$Uy$ to the Kolmogorov base flow
(\ref{base_flow_kol})
with $\alpha_\vecp = 0$. As Manfroi and Young \cite{MY2} pointed out,
this term accounts for a Galilean translation in the $x$-direction
with the speed U.
Through this Galilean translation the Rossby wave base flow
(\ref{base_flow_rossby}) becomes (\ref{base_flow_Gal}) if $1/Rh = U$.
Under this translation, one can show that 
the $\beta$-plane equation remains same. 
Thus, the base flows (\ref{base_flow_Gal}) and (\ref{base_flow_rossby})
are equivalent under the translation if $1/Rh = U$.
With base flow (\ref{base_flow_Gal}),
Manfroi and Young obtained the critical Reynolds number $\sqrt{2}$
if $1/Rh = U$. This is consistent with our result that the critical
Reynolds number is $\sqrt{2}$ with the Rossby wave base flow 
(\ref{base_flow_rossby}).

\section{Base flows with more than one Rossby wave}\label{sec_manyf}

\subsection{Growth rates and critical Reynolds numbers 
for the special flows
with $m=2$ or $m=3$}\label{sec_many_f_cre}

In this section, we consider large-scale ($k \ll 1$) growth rates
for two special base flows whose stream functions
are given by
\begin{eqnarray}
\Psi^{0}_2(\vecx,t;\vecp_1,\vecp_2)
= -\sum_{j=1}^{2}\cos\theta(\vecx,t;\vecp_j)
&&\ \textrm{where }\alpha_{\vecp_2} - \alpha_{\vecp_1} = \pi/2, 
\label{base_flow_two} \\
\nonumber \\
\Psi^{0}_3(\vecx,t;\vecp_1,\vecp_2,\vecp_3)
= -\sum_{j=1}^{3}\cos\theta(\vecx,t;\vecp_j)
&&\ \textrm{where } 
\alpha_{\vecp_{j+1}}-\alpha_{\vecp_j} = \frac{2\pi}{3}.
\label{base_flow_three} \\
\nonumber
\end{eqnarray}
Note that the base flow (\ref{base_flow_two}) has two Fourier wavevectors
$\vecp_1$ and $\vecp_2$ (and their conjugates) and the base flow
(\ref{base_flow_three}) has three wavevectors $\vecp_1$, $\vecp_2$
and $\vecp_3$ (and their conjugates). 
The choice of these base flows is motivated by the work 
by Sivashinsky and Yakhot \cite{SY} of the large-scale instability
of the following base flows in the 2D Navier Stokes equations
\begin{eqnarray*}
&&\Phi_1^{0} = \cos{x} \\
\\
&&\Phi_2^{0} = \cos{x}\cos{y} = \frac{1}{2}(\cos(x+y)+\cos(x-y)) \\
\\
&&\Phi_3^{0} = \sin{2x}-2\sin{x}\sin{\sqrt{3}y}
=\sin(2x) +\cos(x+\sqrt{3}y) -\cos(x-\sqrt{3}y).
\end{eqnarray*}
Similar to (\ref{base_flow_two}) and
(\ref{base_flow_three}),
$\Phi_2^{0}$ and $\Phi_3^{0}$ have two and three Fourier wavevectors
(and their conjugates), respectively.
Using Multiple Scales
analysis, Sivashinsky and Yakhot showed that base flows $\Phi_1^{0}$ and
$\Phi_2^{0}$ have the critical Reynolds number
$\sqrt{2}$.\footnote{The definition
of the Reynolds number for $\Phi_2^{0}$ in Sivashinsky and Yakhot
is different from our definition. In their definition, the critical
Reynolds number for $\Phi_2^{0}$ is $2\sqrt{2}$.}\label{fn_SY}
An interesting result is that for the base flow $\Phi_3^{0}$, the
linear growth rate at $\veck$ with $k\ll 1$ is independent of the
direction of $\veck$ (i.e., is isotropic)
and is negative for all Reynolds numbers. Thus, there is no
direct instability to large-scales for the flow $\Phi_3^{0}$.
In this section, the same results is obtained for the flow 
(\ref{base_flow_three}) with $Rh=\infty$.

The main result of this section is a generalization of \cite{SY}
to the $\beta$-plane equation for the base flows (\ref{base_flow_two})
and (\ref{base_flow_three}). Sivashinsky and Yakhot obtain the
large-scale linear growth rate using Multiple Scales analysis
with the following scaling:
\begin{eqnarray*} 
\partial_x = \partial_x + \eps\partial_X,
\quad \partial_y = \partial_y + \eps\partial_Y,
\quad \partial_t = \eps^2\partial_\tau.
\end{eqnarray*}
Since the base flows $\Phi_2^{0}$ and $\Phi_3^{0}$ 
depend on both $x$ and $y$,
the analysis becomes rather complicated. Later, Waleffe \cite{WAL}
obtained the same result easily by direct application of formula
(\ref{growth_rate}) for $m=1$ with $Rh=\infty$. We follow Waleffe's
idea to find the large-scale growth rate for the base flows
(\ref{base_flow_two}) and (\ref{base_flow_three}), and
verify the results by Multiple Scales analysis in Appendix
\ref{sec_app_MS}.

Let (\ref{base_flow_two}) be the base flow
which contains two Fourier wavevectors 
(and their conjugates). 
Without loss of generality, we can assume that 
$0\le \alpha_{\vecp_1} \le \pi/4$.
For each wavevector $\veck$,
let $\alpha_{j} = \alpha_\veck - \alpha_{\vecp_j}$ for $j=1,2$.
Since $\alpha_{\vecp_2}-\alpha_{\vecp_1} = \pi/2$,
we have 
$\alpha_2 = \alpha_1 -\pi/2$. By adding terms directly
from (\ref{growth_rate}) with $\vecp=\vecp_j,\ j=1,2$, the growth
rate is given by
\begin{eqnarray}
&&\D{
\Re(\sigma(\veck;\alpha_{\vecp_1},\alpha_{\vecp_2},Re,Rh))
}
\nonumber\\
\nonumber\\
&=&\D{
-\frac{k^2}{Re} \left[
1-\frac{Re^2}{2}\frac{\sin^2\alpha_1}{(1+Re^2\omega^2)^2}
\{1-8\cos^2\alpha_1+Re^2\omega^2\}
\right.}
\nonumber\\
\nonumber\\
&&
\D{
\left.
-\frac{Re^2}{2}\frac{\sin^2\alpha_2}{(1+Re^2\omega^2)^2}
\{1-8\cos^2\alpha_2+Re^2\omega^2\}
\right]
}
+O(k^3)
\nonumber\\
\nonumber\\
&=&
\D{
-\frac{k^2}{Re} \left[
1-Re^2 \frac{1+Re^2\omega^2-4\sin^2(2\alpha_1)}{2(1+Re^2\omega^2)^2}
\right]
}+O(k^3).\label{growth_rate_two}
\end{eqnarray}
Note that we only consider the growth rate up to the order of $O(k^2)$.
This is Waleffe's method \cite{WAL}.
In particular, for $Rh=\infty$ ($\omega = 0$), (\ref{growth_rate_two}) becomes
\begin{eqnarray}
&&\Re(\sigma(\veck;\alpha_{\vecp_1},\alpha_{\vecp_2},Re,Rh=\infty))
\nonumber\\
\nonumber\\
&&\quad 
= -\frac{k^2}{Re}\left[
1-Re^2\frac{1-4\sin^2(2\alpha_1)}{2}\right] + O(k^4).
\label{growth_rate_two_2d}
\end{eqnarray}
The higher order term is $O(k^4)$ since the third order term in
(\ref{growth_rate}) is zero for $Rh=\infty$. This expression 
(\ref{growth_rate_two_2d}) for large-scale growth rate was obtained by
Sivashinsky and Yakhot \cite{SY} and by Waleffe \cite{WAL}. 

Similar to
Section \ref{sec_cre_smallB}, for base flow (\ref{base_flow_two})
we define the marginal Reynolds number 
$Re^{0}(\veck;\alpha_{\vecp_1},\alpha_{\vecp_2},Rh)$ by the smallest
nonnegative Reynolds number 
which makes the leading order term of the growth rate
(\ref{growth_rate_two}) equal to zero by solving
\begin{eqnarray}
Re^{2} = \frac{2(1+Re^2\omega^2)^2}{1+Re^2\omega^2 - 4\sin^{2}2\alpha_1}
\quad\textrm{and}\quad
1+Re^2\omega^2-4\sin^22\alpha_1 > 0.
\label{def_cre_k0_two}
\end{eqnarray}
If there exists no Reynolds number satisfying (\ref{def_cre_k0_two}),
the marginal Reynolds number is defined to be infinity.
$Re^{0}(\veck;\alpha_{\vecp_1}, \alpha_{\vecp_1}, Rh)$ is understood
as the critical Reynolds number in the limit $k\to 0$. 
For $Rh=\infty$ ($\omega=0$), 
from (\ref{def_cre_k0_two}),
the marginal Reynolds number is given by
\begin{eqnarray*}
 Re^{0}(\veck;\alpha_{\vecp_1},\alpha_{\vecp_2},Rh=\infty)
= \sqrt{\frac{2}{1-4\sin^2 2\alpha_1}},
\quad \textrm{if} 
\quad
1-4\sin^2 2\alpha_1 > 0,
\end{eqnarray*}
and its minimum is $\sqrt{2}$ when $\alpha_1 = 0$ or $\pi/2$. Thus,
if $Rh=\infty$, the critical Reynolds number is $\sqrt{2}$
for the 2D Navier Stokes equations with base flow
(\ref{base_flow_two}). This is the same result of Sivashinsky and Yakhot
\cite{SY} and Waleffe \cite{WAL} 
\footnote{ See the footnote of \ref{fn_SY}. }.
As in Section \ref{sec_cre_smallB}, we see that
$Re^{0}(\veck;\alpha_{\vecp_1},\alpha_{\vecp_2},Rh)$ is always larger
than or equal to $\sqrt{2}$, since we have
\begin{eqnarray*}
\left[ Re^{0}(\veck;\alpha_{\vecp_1},\alpha_{\vecp_2},Rh) \right]^2 -2
= \frac{2Re^2\omega^2 + 2Re^4\omega^4 + 4\sin^2 2\alpha_1}
{1+Re^2\omega^2-4\sin^2 2\alpha_1} \ge 0,
\end{eqnarray*}
if $(1+Re^2\omega^2-4\sin^2 2\alpha_1) > 0$. 
For $\alpha_1=0$ or $\pi/2$,
we also have
\begin{eqnarray}
\lim_{Rh\to \infty} Re^{0}(\veck;\alpha_{\vecp_1},\alpha_{\vecp_2},Rh) 
=\sqrt{2},
\label{cre_cont_two}
\end{eqnarray}
recovering the critical Reynolds number $\sqrt{2}$ for the 2D Navier Stokes
equations in the limit $Rh\to \infty$.

Similar to Section \ref{sec_re_k0}, for $0\le \alpha_{\vecp_1} \le \pi/4$
we find that the critical Reynolds number in the limit $k\to 0$ is
given by
\begin{eqnarray}
\lim_{k_0\rightarrow 0} \min_{|\veck|<k_0}
&&Re^{0}(\veck;\alpha_{\vecp_1},\alpha_{\vecp_2},Rh)
\nonumber \\
\nonumber \\
&&
\quad 
= \left\{
\begin{array}{ll}
\D{
\sqrt{\frac{2}{1-4\sin^2(2\alpha_{\vecp_1})}}}
& \quad 0 \le \alpha_{\vecp_1} \le
\frac{1}{2}\sin^{-1}(\frac{1}{2\sqrt{2}})
\\
\\
4\sqrt{2}\ |\sin(2\alpha_{\vecp_1})|
& \quad \frac{1}{2}\sin^{-1}(\frac{1}{2\sqrt{2}}) \le 
\alpha_{\vecp_1} \le \frac{\pi}{4}.
\end{array}
\right.
\label{cre_k0_two}
\end{eqnarray}
We omit the derivation since it is much like that of
(\ref{cre_k0}) for $m=1$. Similar to (\ref{cre_k0}), (\ref{cre_k0_two})
does not depend on $Rh$.

Now, consider the base flow (\ref{base_flow_three}) which contains
three Fourier wavevectors (and their conjugates). For each wavevector,
let $\alpha_{j} =  \alpha_\veck - \alpha_{\vecp_j}$ for $j=1,2,3$.
Since $\alpha_{2} = \alpha_{1} - 2\pi/3$ and
$\alpha_{3} = \alpha_{1} - 4\pi/3$, we have two simple trigonometric 
identities:
\begin{eqnarray*}
&&\sum_{j=1}^{3} \sin^2{\alpha_j} = 3/2 \\
\\
&&\sum_{j=1}^{3} \sin^2{\alpha_j}\cos^2{\alpha_j} = 3/8.
\end{eqnarray*}
Using these identities, by adding terms from (\ref{growth_rate})
for $\alpha_\vecp=\alpha_{\vecp_j},\ j=1,2,3$, the growth rate is
given by
\begin{eqnarray}
&&\Re(\sigma(\veck;\alpha_{\vecp_1},\alpha_{\vecp_2},\alpha_{\vecp_3},Re,Rh))
\nonumber\\
\nonumber\\
&=&
-\frac{k^2}{Re} \left[
1-\frac{Re^2}{2 (1+Re^2\omega^2)^2}\left(
(1+Re^2\omega^2)
\sum_{j=1}^{3}\sin^2{\alpha_j}
\right.
\right.
\nonumber\\
\nonumber\\
&&
\quad 
\left.
\left.
-8\sum_{j=1}^{3}\sin^2{\alpha_j}\cos^2{\alpha_j}
\right)
\right]
+O(k^3)\nonumber\\
&=&
-\frac{k^2}{Re}\left[
1-\frac{3Re^2}{4}\frac{Re^2\omega^2-1}{(1+Re^2\omega^2)^2}\right]
+O(k^3).\label{growth_rate_three}
\end{eqnarray}
In particular, for $Rh=\infty$ it becomes
\begin{eqnarray*}
\Re(\sigma(\veck;\alpha_{\vecp_1},\alpha_{\vecp_2},\alpha_{\vecp_3},
Re,Rh=\infty))=
-\frac{k^2}{Re}\left(1+\frac{3}{4}Re^2\right)+O(k^4)
\end{eqnarray*}
and the leading order expression is isotropic 
in $\veck$ and always negative.
Thus, there is no direct instability to large scales for all
Reynolds numbers. This is the result by Sivashinsky and Yakhot
for the ``isotropic'' base flow $\Phi_3^{0}$ and also obtained 
by Waleffe \cite{WAL}.

An interesting observation is that for $Rh<\infty$, 
the large-scale growth rate (\ref{growth_rate_three})
does not depend on the particular orientation of $\vecp_j$.
However, (\ref{growth_rate_three}) is not isotropic in $\veck$ 
because the $\beta$-plane equation itself is not isotropic.
The growth rate (\ref{growth_rate_three}) depends on $\veck$
only through $\omega(\veck)$, whereas growth rates (\ref{growth_rate})
and (\ref{growth_rate_two}) depend on $\veck$ not only through
$\omega(\veck)$ but also through $\alpha$ or $\alpha_1$, respectively.

As in Section \ref{sec_cre_smallB} and above for (\ref{base_flow_two}), 
we define the marginal Reynolds number 
$Re^{0}(\veck;\alpha_{\vecp_1},\alpha_{\vecp_2},\alpha_{\vecp_3},Rh)$
by the smallest nonnegative Reynolds number satisfying
\begin{eqnarray}
Re^2 = \frac{4}{3}\frac{(1+Re^2\omega^2)^2}{Re^2\omega^2 - 1},
\textrm{ and } Re^2\omega^2 - 1 > 0.
\label{def_cre_k0_three}
\end{eqnarray}
When $Rh = \infty$ ($\omega=0$), there is no nonnegative
Reynolds number satisfying (\ref{def_cre_k0_three}). Thus, the marginal
Reynolds number is infinite and there is no direct large-scale instability
\cite{SY}, \cite{WAL}.
Analogous to (\ref{cre_k0}) and (\ref{cre_k0_two}),
one can show that 
\begin{eqnarray}
\min_{\veck\ne \veczero} Re^{0}(\veck;\alpha_{\vecp_1},\alpha_{\vecp_2},
\alpha_{\vecp_3},Rh) = \sqrt{32/3}
\label{cre_k0_three}
\end{eqnarray}
for $Rh < \infty$.
We omit the derivation of (\ref{cre_k0_three}) since it is similar to that of
(\ref{cre_k0}) for $m=1$. 
The derivation uses the fact that
\begin{eqnarray*}
\min_{\veck\ne \veczero} 
Re^{0}(\veck;\alpha_{\vecp_1},\alpha_{\vecp_2},\alpha_{\vecp_3},Rh) =
\min_{|\veck| < k_0 }
Re^{0}(\veck;\alpha_{\vecp_1},\alpha_{\vecp_2},\alpha_{\vecp_3},Rh)
\end{eqnarray*}
for any positive number $k_0$, since
(\ref{def_cre_k0_three}) does not depend on $\alpha_1$.
Another ingredient in the derivation 
of (\ref{cre_k0_three}) is that the function 
\begin{eqnarray}
f(x) = \frac{3}{4}\frac{x-1}{(1+x^2)}\quad \textrm{for}\quad x\ge 0
\label{f_prop_three}
\end{eqnarray}
has maximum value $3/32$ at $x=3$. 
We conclude that there is a direct instability to large scales
($k\ll 1$) with the ``isotropic'' base flow (\ref{base_flow_three})
for $Re > \sqrt{32/3}$ and $Rh < \infty$. This is different from
the 2D Navier Stokes equations, where no direct large-scale instability
is observed for any Reynolds number with base flow
(\ref{base_flow_three}) \cite{SY}.

\subsection{Numerical study of the growth rate for various $Rh$}

In this section, we consider the linear growth rate at
wavevector $\veck$ with $k<1$ for large-scale instability 
with $m \ge 2$ in (\ref{psi_general}).
As discussed in Section \ref{sec_intro}, it is not straightforward
to solve (\ref{ILS_m}) numerically for $m\ge 2$ since the technique
of the continued fractions used for $m=1$ can not be generalized.
However, in Section \ref{sec_CN} we also observe that for $m=1$ 
the severe truncation (\ref{TCF_one}) with $n = -1,0,1$ is
a qualitatively good approximation for small $k$.
With this background, we truncate 
(\ref{ILS_m}) to
a finite dimensional eigenvalue problem with 
\begin{eqnarray*}
\vecn = \veczero,\pm{\hat{\vece}}_j,\quad\textrm{ for }
j=1,\cdots,m.
\end{eqnarray*}
In terms of triads, this truncation involves only $2m$-triads:
\begin{eqnarray*}
(\veck,\vecp_j,-(\veck+\vecp_j)),\quad
(\veck,-\vecp_j,-(\veck-\vecp_j)),
\quad j=1,\cdots,m.
\end{eqnarray*}
For example, for $m=2$ the corresponding eigenvalue problem with
the truncation (\ref{ILS_m}) is given by
\begin{eqnarray*}
\left[
\begin{array}{ccccc}
-c(0,-1) & 0 & \frac{b_2(0,0)}{q^2(0,-1)} & 0 & 0 \\
\\
0 & -c(-1,0) & \frac{b_1(0,0)}{q^2(-1,0)} & 0 & 0 \\
\\
-\frac{b_2(0,-1)}{q^2(0,0)} & -\frac{b_1(-1,0)}{q^2(0,0)} &
-c(0,0) & \frac{b_{1}(1,0)}{q^2(0,0)} & \frac{b_2(0,1)}{q^2(0,0)^2} \\
\\
0 & 0 & -\frac{b_1(0,0)}{q^2(1,0)} & -c(1,0) & 0 \\
\\
0 & 0 & -\frac{b_2(0,0)}{q^2(0,1)} & 0 & -c(0,1)
\end{array}
\right]
\vecPsi
=\sigma
\vecPsi
\end{eqnarray*}
where
\begin{eqnarray*}
\left\{
\begin{array}{l}
c(\vecn) = \frac{1}{Re}q^2(\vecn) +
i\omega_\vecn,\\
\\
b_{j}(\vecn) = \D{-\frac{1}{2}}(\vecp_j\times\vecq(\vecn))(q^2(\vecn)-1),\\
\\
\vecPsi =
(\psi_{(-1,0)},\psi_{(0,-1)},\psi_{(0,0)},\psi_{(1,0)},\psi_{(0,1)})^{T}.\\
\end{array}
\right.
\end{eqnarray*}
Moreover, we consider for each $m$, the ``isotropic'' base flow with
\begin{eqnarray*}
\alpha_{\vecp_j} = j\frac{\pi}{m},\quad j=1,\cdots,m.
\end{eqnarray*}
For large $m$, this base flow mimics deterministic forcing over
a shell of radius one in Fourier space. For each wavevector
$\veck$ with $Re$ and $Rh$ fixed, we find 
the eigenvalue of the truncated eigenvalue problem with the largest
real part, denoted by $\sigma(\veck;m,Re,Rh)$. By the 
``isotropic'' choice of $\alpha_{\vecp_j}$, $\Re(\sigma(\veck;m,Re,Rh))$
is symmetric about both the $x$-axis and the $y$-axis if $m$ is even.

\begin{figure}
\begin{center}
\begin{tabular}{ccc}
{\small (a) $m=30$, $Rh=\infty$}
&
{\small (b) $m=30$, $Rh=2$}
&
{\small (c) $m=30$, $Rh=1/2$}
\\
\includegraphics[width=4cm]{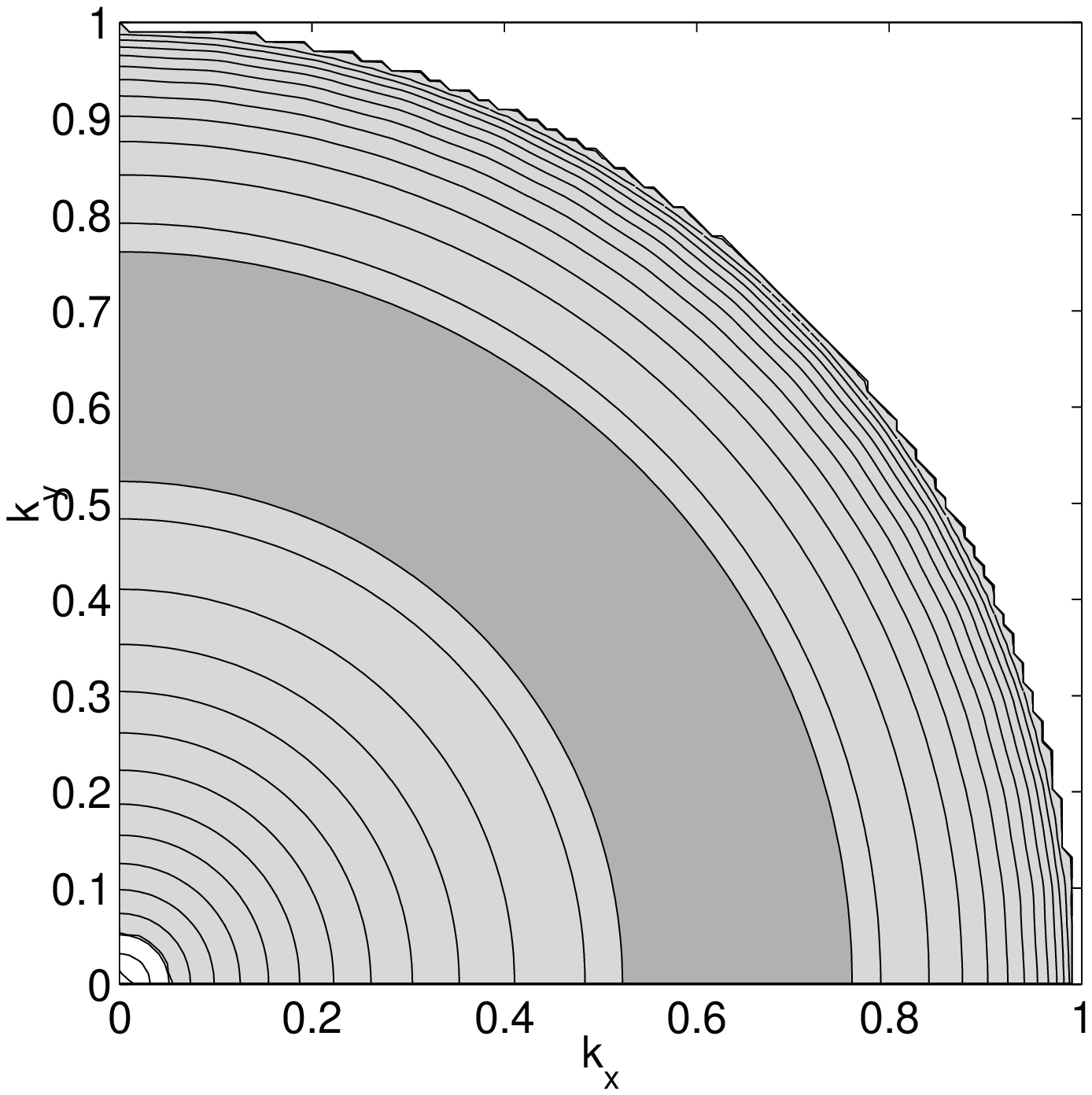}
&
\includegraphics[width=4cm]{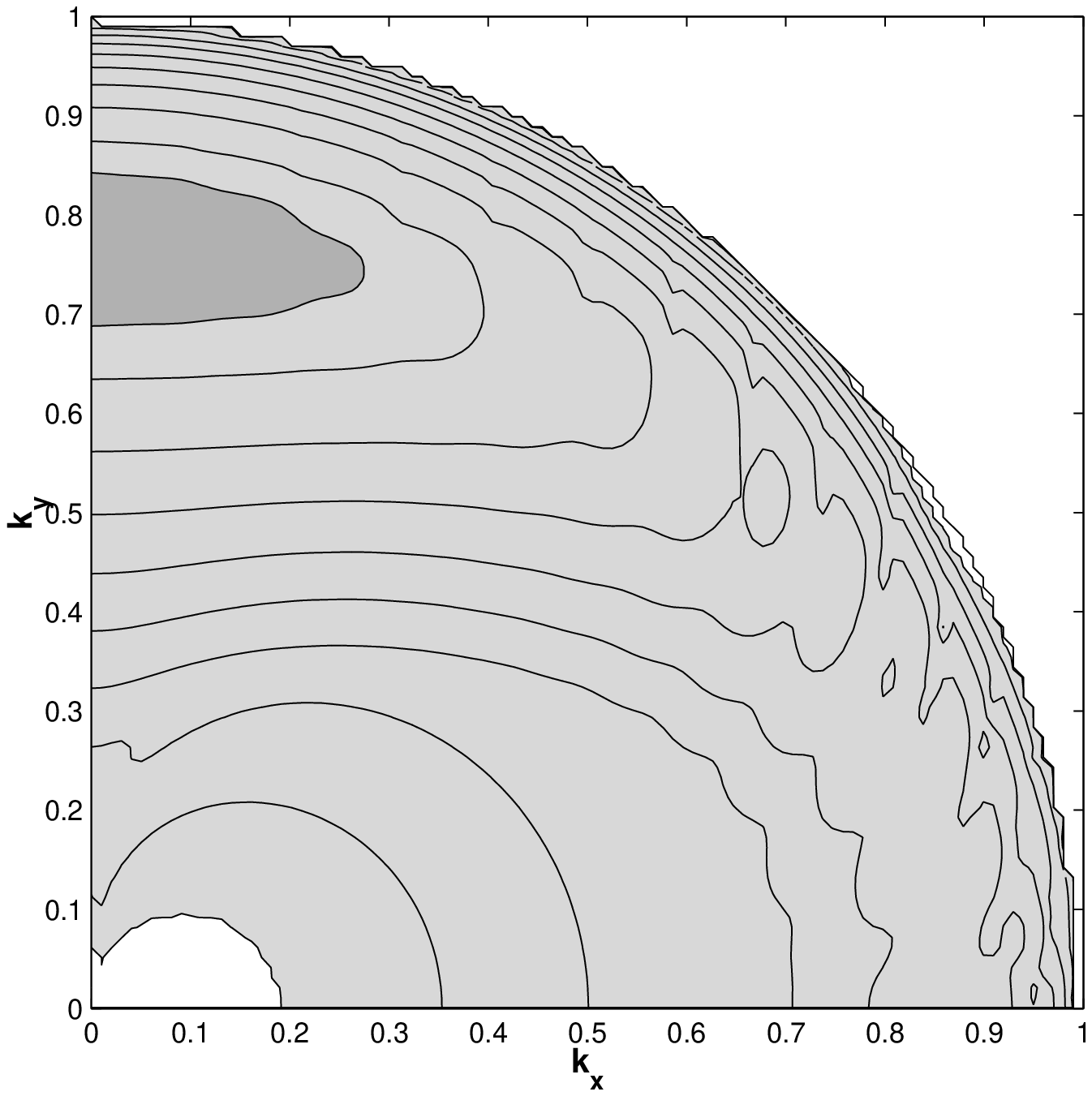}
&
\includegraphics[width=4cm]{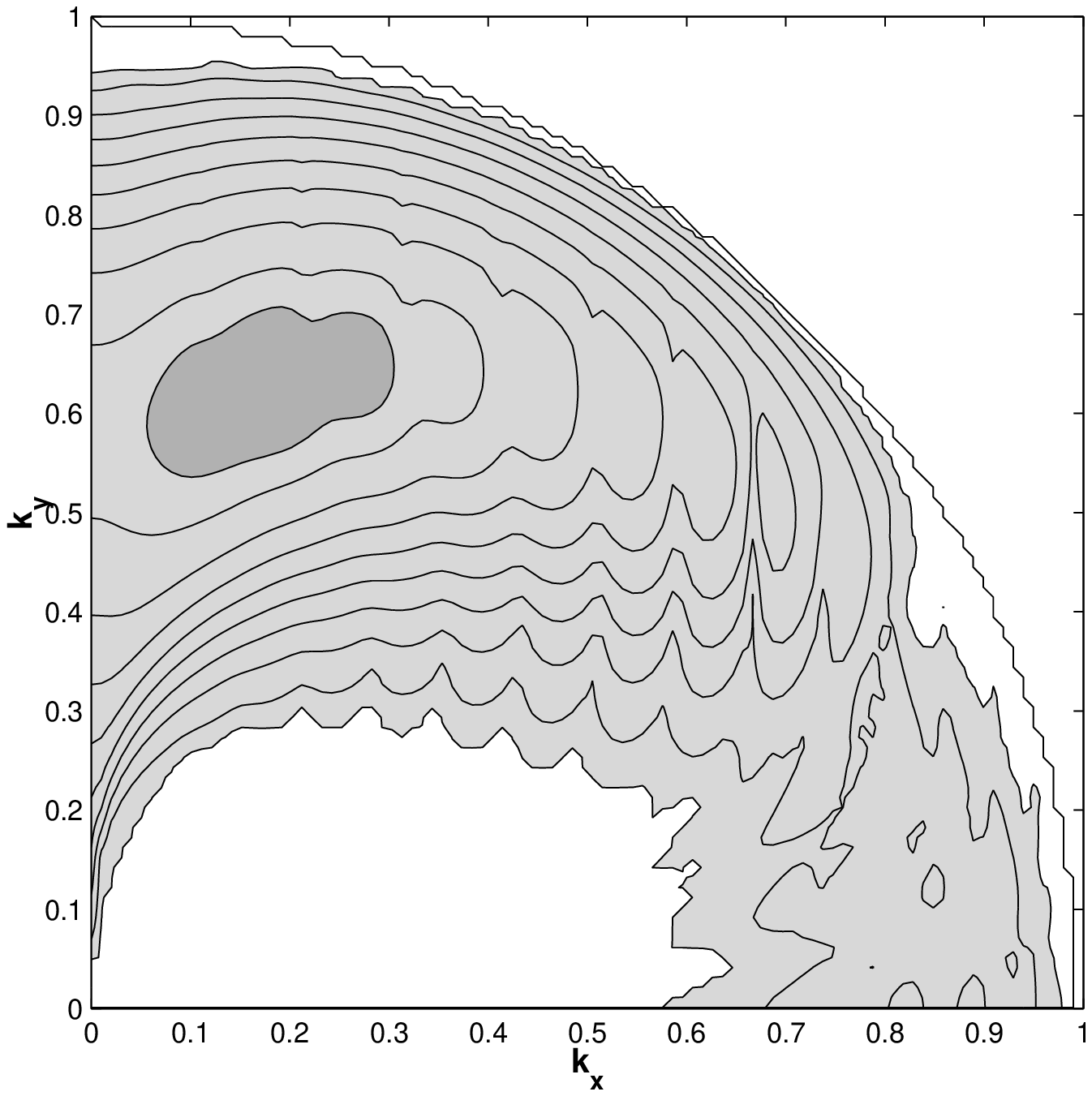}
\end{tabular}
\caption{$\Re(\sigma(\veck;m,Re,Rh))$ at $m = 30$ and $Re = 10$. 
$\Re(\sigma(\veck;m,Re,Rh))$ is computed from (\ref{ILS_m}) with
the severe truncation.}
\label{pic_many_force}
\end{center}
\end{figure}

\begin{figure}
\begin{center}
\begin{tabular}{cc}
(a) $m=30$, $Re = 1000$, $Rh=2$
&
(b) $m=30$, $Re = 1000$, $Rh=1/2$
\\
\includegraphics[width=5cm]{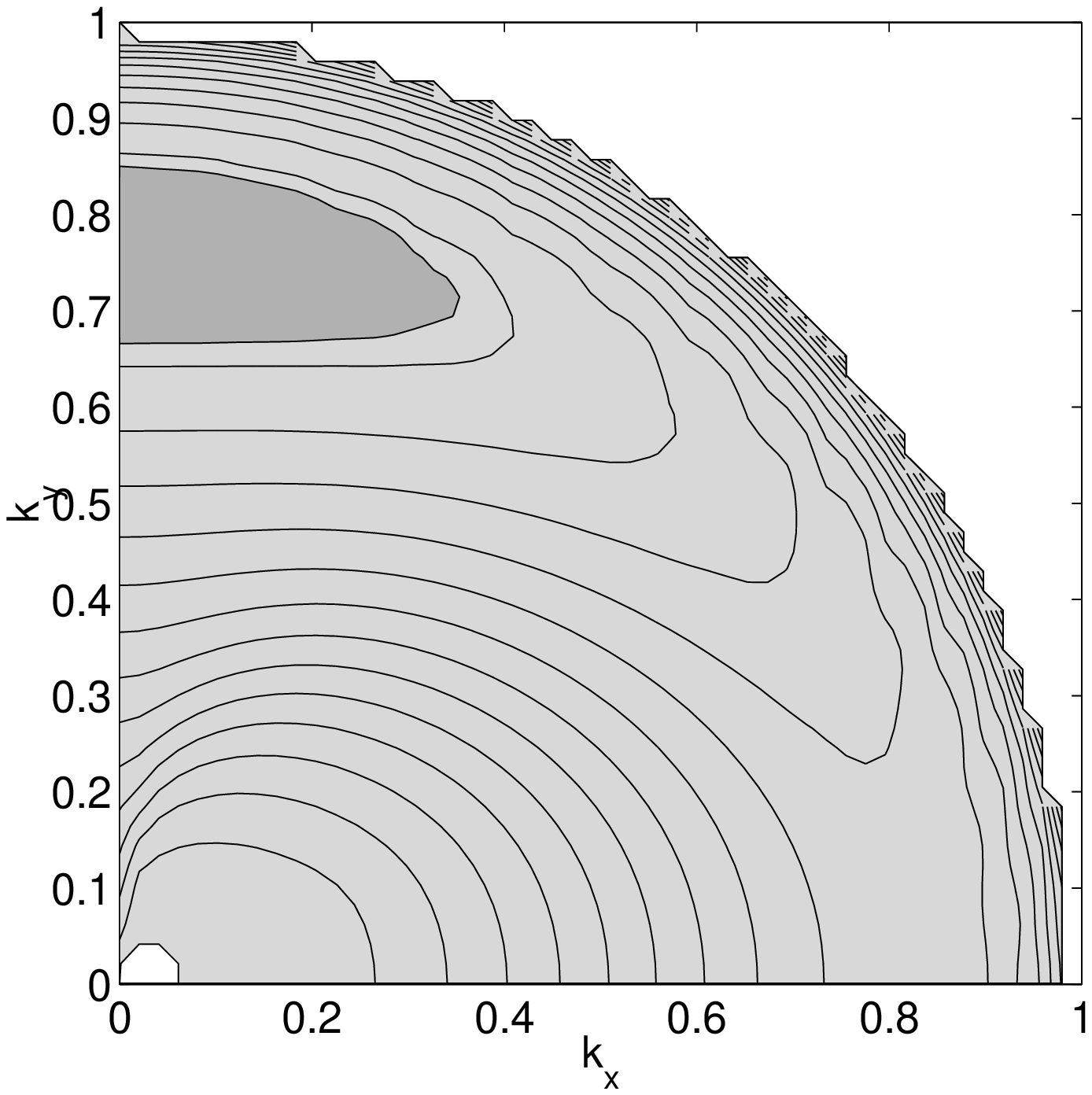}
&
\includegraphics[width=5cm]{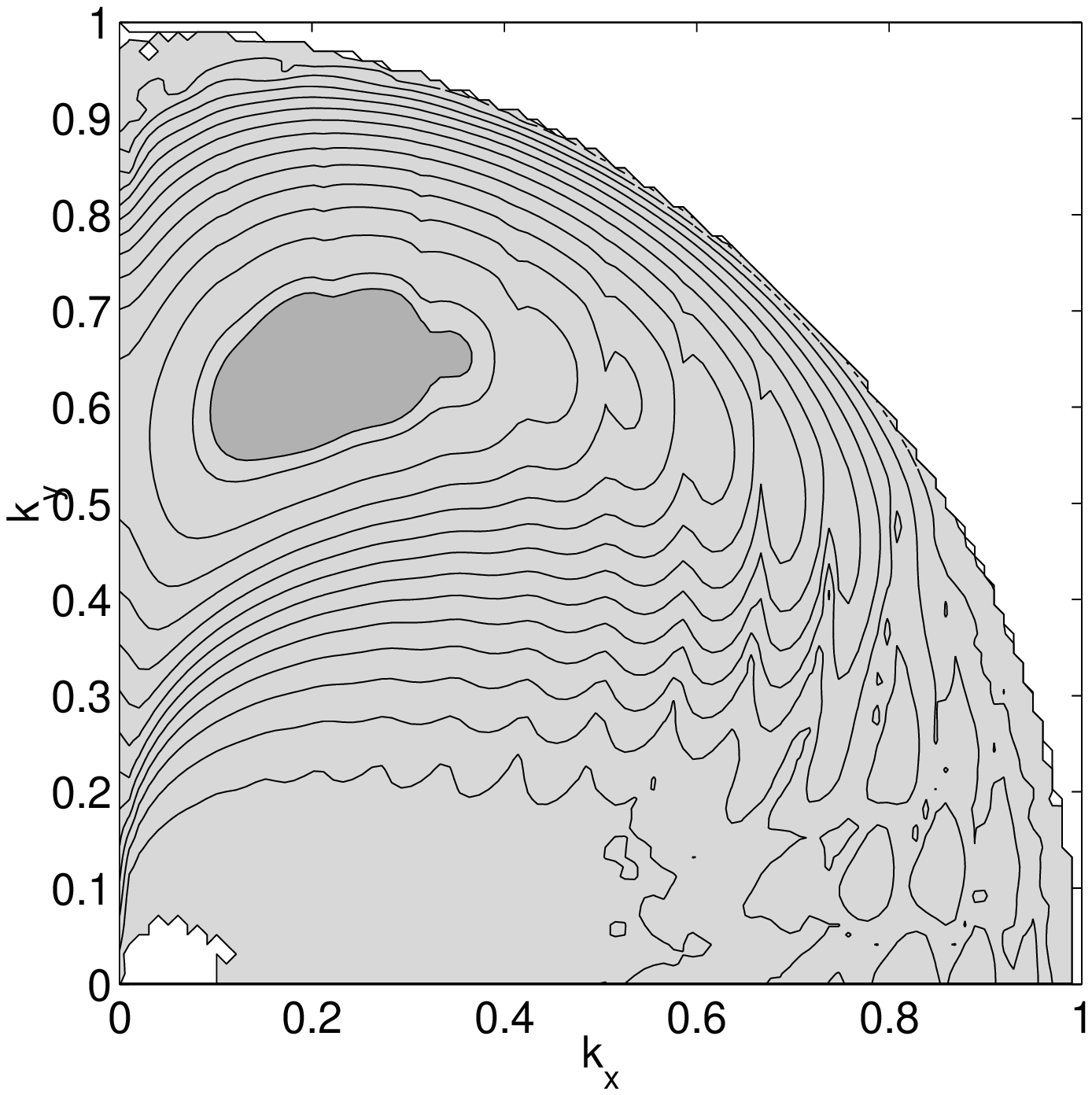}
\\
(c) $m=40$, $Re = 10$, $Rh=2$
&
(d) $m=40$, $Re = 10$, $Rh=1/2$
\\
\includegraphics[width=5cm]{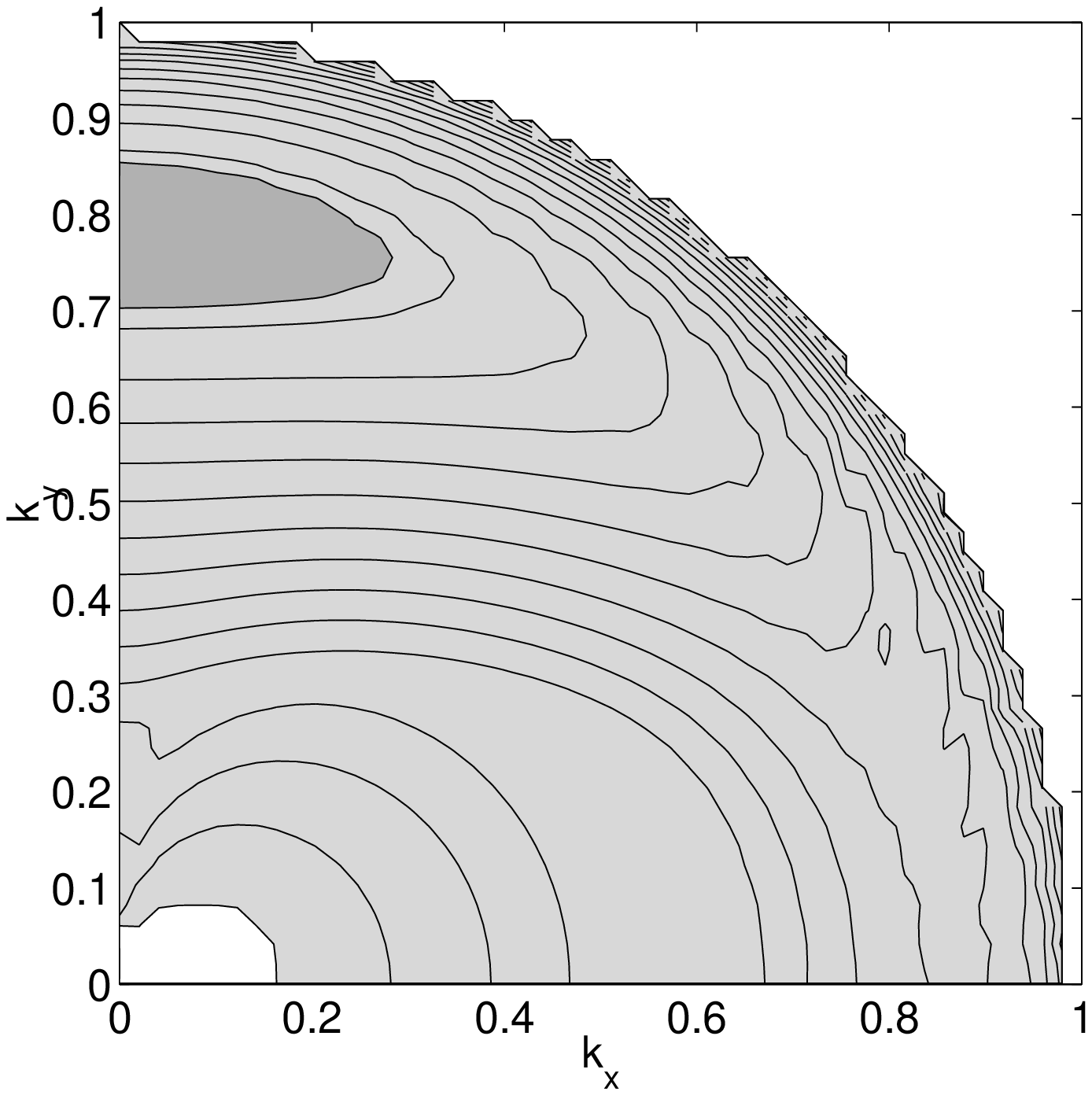}
&
\includegraphics[width=5cm]{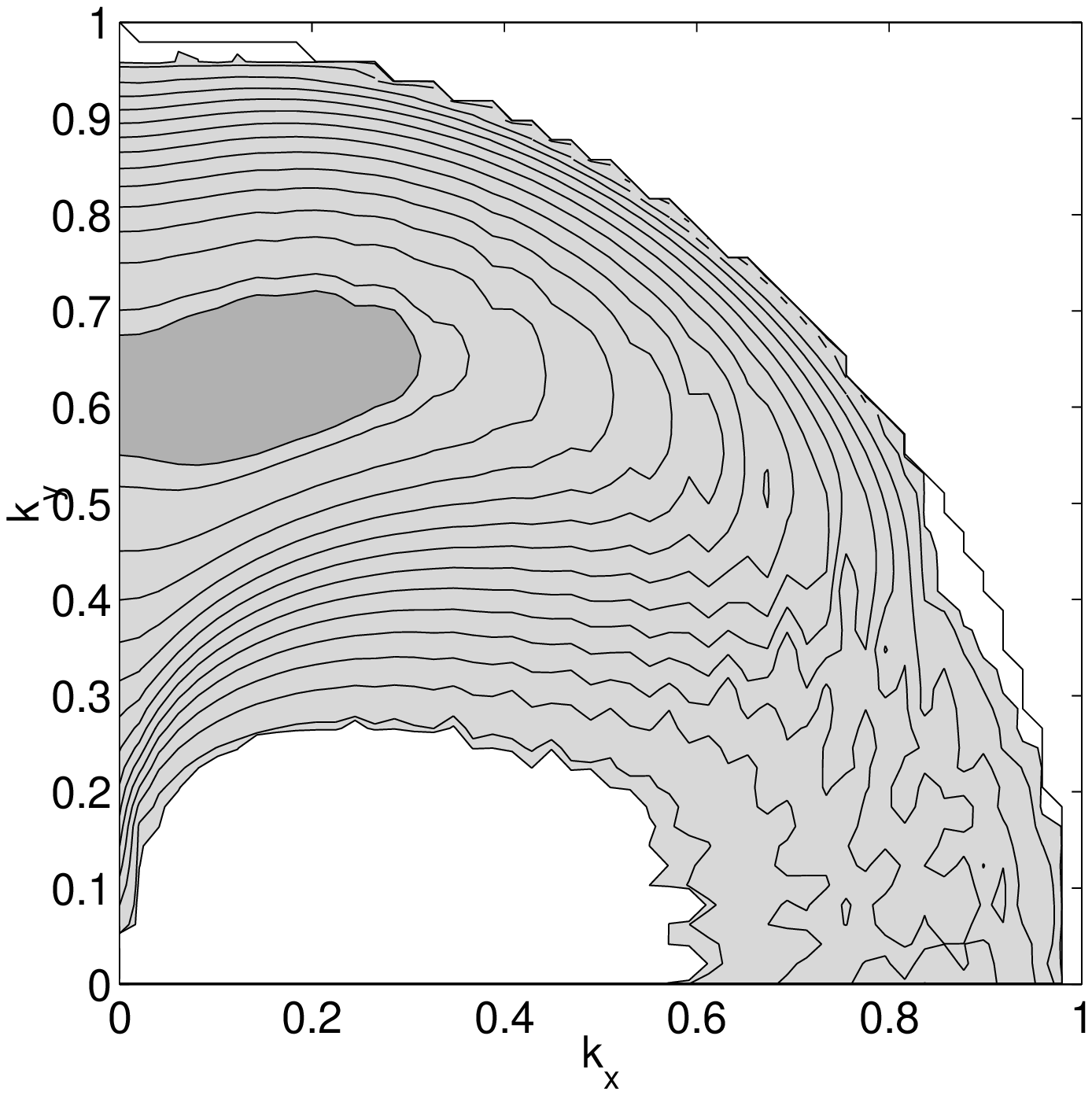}
\end{tabular}
\caption{$\Re(\sigma(\veck;m,Re,Rh))$ at $m = 30$ with $Re = 1000$
and $m=40$ with $Re = 10$. 
}
\label{pic_many_force_ind}
\end{center}
\end{figure}

Each contour plot is for wavevectors $\veck$
with $0<k_x<1$, $0<k_y<1$ and $|\veck|<1$ to study the large-scale
instability. We use a grid of size
$100\times 100$ over $0<k_x<1$ and $0<k_y<1$.
Figure~\ref{pic_many_force} shows the growth rate
$\Re(\sigma(\veck;m,Re,Rh))$ for $m = 30$ and $Re = 10$.
The growth rate for $Rh=\infty$ is invariant under the rotation of
$\pi/m$ since the Navier Stokes equations are isotropic. Also, note that
there is no direct instability into large scales, as consistent with the
result in Section \ref{sec_many_f_cre} for the ``isotropic''
flow (\ref{base_flow_three}) with $m=3$.
For Rhines number $2 \le Rh \le \infty$, the largest growth rate
is observed to be exactly along the line $k_x=0$. However, for 
$Rh = 1/2$, the largest growth rate shifts slightly away from 
$k_x=0$. Again, this transition can
be understood in terms of resonant triad interactions, which cannot
transfer energy into wavevectors with $k_x=0$.
A similar result transition is observed in Section \ref{sec_inf}. 
The $O(1)$ Rhines number for the
transition seems to be independent of $Re$ and $m$,
as demonstrated in Figure~\ref{pic_many_force_ind} for $Re = 1000$ with 
$m = 30$, and $Re = 10$ with $m=40$.

\section{Summary}\label{sec_disc}

The growth rate of linear instability at large scales is
obtained for Rossby wave base flows maintained by deterministic forcing of
one or several Fourier modes.  For forcing of a single mode, we
use the technique of continued fractions following \cite{MS}, \cite{G},
\cite{FR1}, \cite{FR2} and \cite{LI}, and asymptotic analysis.
We show that the mechanism for linear instability changes
from inflectional to triad resonance at an $O(1)$ transition
Rhines number, independent of the Reynolds number.
The critical Reynolds number for instability
has also been obtained in various limits.  In the limits $k\rightarrow 
0$ and $Rh\rightarrow \infty$, we recover the classical result
$Re^c = \sqrt{2}$ found by Meshalkin and Sinai \cite{MS} for the 2D
Navier Stokes equations (see also \cite{SY}, \cite{FLV}, 
\cite{MY2} and \cite{WAL}).  Furthermore, we generalize \cite{MS}
to find $Re^c$ given by equation (\ref{cre_k0}) for $k
\rightarrow 0$ and finite Rhines number $Rh$. 
For $Rh \rightarrow 0$,
we find $Re^c = 0$ for Rossby wave base flows of all orientations
except zonal and meridional ({\it i.e.}, for
$0 < \alpha_\vecp < \pi/2$, where
$\alpha_\vecp$ measures the angle between the forcing wavevector
$\vecp$ and the zonal direction $\hat{\bf x}$, and $|\vecp| = 1$).
For $Rh \rightarrow 0$, a zonal base flow is stable and a
meridional base flow has $Re^c = \sqrt{2}$.
It is clear that, for small Rhines numbers, resonant triad interactions
reduce the critical Reynolds number below $\sqrt{2}$ for
$0<\alpha_\vecp<\pi/2$.

In order to study more isotropic forcings, we consider
base flows consisting of $m > 1$ Rossby waves, thereby
generalizing \cite{SY} to the $\beta$-plane with $Rh < \infty$.   
For $m=3$, we find
instability for $Re > \sqrt{32/3}$, whereas the analogous
base flow was found to be stable for $Rh = \infty$ in \cite{SY}.
For large $m$, such a base flow mimics deterministic forcing
over a shell of radius one in Fourier space.  Numerical
computations of growth rates for $m=30$ and $m=40$ show that the
most unstable mode is purely zonal with $k_x=0$ for $2 \le Rh \ll \infty$
(Figure~\ref{pic_many_force} and \ref{pic_many_force_ind}).
This is because one of the forced Rossby waves is meridional with
$\alpha_\vecp =0$, and the inflectional instability is strongest for 
meridional base flows, as shown in Figure~\ref{pic_inf_line}
and \ref{pic_inf_plane}, and discussed in Section 3.3.  
As the Rhines number is decreased
from $Rh = 2$ to $Rh = 1/2$, the most unstable mode $\veck$ shifts
to a nearly zonal flow with $k_x$ not exactly zero (Figure~\ref{pic_many_force}
and \ref{pic_many_force_ind}).  
This transition in the most unstable wavevector
can be understood as the change from inflectional instability to resonant
triad instability at $Rh = O(1)$, and reflects the result that
resonant triads cannot transfer energy directly to zonal flows
with $k_x=0$ \cite{CH} (see also \cite{LG}, \cite{N}, \cite{S}, 
\cite{SW1}, \cite{SW2} and Section 3).  Waleffe \cite{WAL2} 
showed analytically that resonant triad interactions can transfer
energy to nearly zonal flows with $k_x$ near zero.  In Figure~
11 and 12, however, one also sees that wavevectors with $k_x$
exactly equal to zero remain unstable even for $Rh = 1/2$,
though they are not the most unstable wavevectors.  This is
because for $Rh = O(1)$, the mechanism for instability is
a combination of both the inflectional and triad resonance
instabilities.  The latter observation may be relevant to 
recent simulations of $\beta$-plane flow driven by 
isotropic, stochastic forcing in a wavenumber shell \cite{CH}, \cite{SW1},
\cite{ML}.  With energy input rate $\epsilon_f$
and peak wavenumber $k_f$ of the stochastic force, the Rhines
number for these simulations is appropriately defined as 
$Rh = (\epsilon_f k_f^2)^{1/3}/\beta$, measuring the relative
strength of the forcing and $\beta$ terms.  In \cite{CH}, this
definition of $Rh$ leads to three values $Rh = \infty$,
$Rh = 5$ and $Rh = 1$.  In the latter two cases,
energy is transferred anisotropically to large-scale
zonal flows with $k_x=0$.  Even though our Rhines
number is defined differently, it is nevertheless 
intriguing that linear instability mechanisms transfer energy to 
zonal flows with $k_x=0$ for our $Rh = 2$.  Indeed,
for $Rh = 2$, the most unstable modes are purely zonal flows.  
Thus it is possible that linear instability plays a role in the
population of zonal flows observed in numerical simulations of
the (nonlinear) $\beta$-plane system at moderate Rhines numbers
$Rh > 1$.  Further study of 
stability of the stochastically forced system and 
of nonlinear transfer mechanisms 
is necessary to fully understand numerical simulations 
at both large and small values of the Rhines number.
Finally we note that,
for the atmosphere and the oceans, Rhines number $O(1)$ is
associated with the most energetic eddies \cite{G}.
From the peak wavenumber of the atmospheric spectrum,
Gill \cite{G} estimates that the Rhines number 
of the atmosphere at $40^0$N and $500$ mb is approximately $0.9$.

{\bf Acknowledgements} The authors thank Fabian Waleffe for 
the suggestion to use the method of continued fractions and for many
thoughtful discussions. The support
of NSF is gratefully acknowledged, under grant DMS-0071937.

\appendix
\section{Appendix}\label{sec_app}

\subsection{On the continued fraction}\label{app_CF}
We briefly discuss the convergence of the continued
fractions in equation (\ref{CF}). 
To this end, for complex numbers $\xi_n$ and $\eta_n$, consider 
the continued fraction
\begin{eqnarray}
\displaystyle{
K = \frac{\xi_1}{\eta_1+\frac{\xi_2}{\eta_2+\frac{\xi_3}{\eta_3+\cdots}}}
}.
\label{app_CF_th}
\end{eqnarray}
We have the following classical theorem on the continued fraction by 
\'Sleszy\'nski and Pringsheim \cite{LW}.

\noindent
{\bf Theorem} {\it The continued fraction (\ref{app_CF_th})
converges if for all $n$
\begin{eqnarray*} 
|\eta_n| \ge |\xi_n| + 1.
\end{eqnarray*}
Under the same condition, we have 
\begin{eqnarray*}
|K|\le 1.
\end{eqnarray*}
}

In equation (\ref{CF}), we have two continued fractions with 
$\xi_n = 1$ and $\eta_n = a_{\pm n}$. Since we consider the
case $Re < \infty$, for fixed $\sigma$, $\veck$, $\alpha_\vecp$, $Re$ 
and $Rh$
we have 
\begin{eqnarray*}
\lim_{n\to \infty} |\eta_n| 
= \lim_{n\to \infty} |a_{\pm n}| = \infty.
\end{eqnarray*}
Thus, by the theorem above, both continued fractions converge.
Next, we show (\ref{small_k_prop1}) for $k\ll 1$. 
Since we have
\begin{eqnarray*}
\frac{1}{|a_n|} \le C Re\, k \le \frac{1}{2}
\quad 
\textrm{for } k\le \frac{1}{2CRe}\textrm{ and }n = \pm 2, \pm 3, \ldots,
\end{eqnarray*}
from (\ref{small_k_prop0}), by the theorem, we have
\begin{eqnarray}
\left|\frac{1}{a_{\pm 3}+\frac{1}{a_{\pm 4} +\frac{1}{a_{\pm 5}+\cdots}}}
\right| \le
1.
\label{app_CF_der}
\end{eqnarray}
From (\ref{app_CF_der}), one can derive (\ref{small_k_prop1}) using
an asymptotic expansion in $k$ and (\ref{small_k_prop2}) is a consequence
of (\ref{small_k_prop1}).

\subsection{Critical Reynolds numbers for $\alpha_\vecp=0$ and
$\alpha_\vecp=\pi/2$ in 
the limit $Rh\rightarrow 0$}\label{sec_app_Rh0}

In this appendix, we consider  (\ref{cre_Rh0}) for $\alpha_\vecp=0$
and $\alpha_\vecp = \pi/2$. Since (\ref{cre_Rh0}) can be shown
in a similar way for both cases, we show 
(\ref{cre_Rh0}) only in the case $\alpha_\vecp = 0$ and omit the other
case.

Consider (\ref{cre_Rh0}) for the case $\alpha_\vecp = 0$.
Since we showed that 
\begin{eqnarray*}
\lim_{Rh\to 0}Re^c(\alpha_\vecp=0,Rh) \le \sqrt{2}.
\end{eqnarray*}
in Section \ref{sec_0_90},
it is enough to show that
\begin{eqnarray}
\D{
\lim_{Rh\to 0} Re^c(\alpha_\vecp=0,Rh)
\ge \sqrt{2}.
}
\label{app_cre_Rh0_ineq}
\end{eqnarray}
For this end, we assume that for some $\veck^0$ we have
\begin{eqnarray}
\liminf_{Rh\to 0, \veck\to \veck^0}
Re^c(\veck;\alpha_\vecp=0,Rh) 
=
\lim_{Rh\to 0}Re^{c}(\alpha_\vecp=0,Rh)
\label{app_cre_seq}
\end{eqnarray}
By the symmetries of $\Re(\sigma(\veck;\alpha_\vecp=0,Re,Rh))$
discussed in Section \ref{sec_intro}, we can assume that
$ 0 \le k_x  \le 1/2$ and $ k_y \ge 0$. 
We show (\ref{app_cre_seq}) separately in each of two cases:
$\veck^0 = \veczero$ and $\veck^0 \ne \veczero$.

First, consider the case $\veck^0 \ne \veczero$. 
If $\veck^0$ is not resonant to $\pm \vecp$, then we have
$|\omega_{\pm 1}|\to \infty$ in the $\veck\to \veck^0$ and $Rh\to 0$ limits.
Thus, (\ref{CF}) is reduced to
\begin{eqnarray*}
a_0 = 0
\end{eqnarray*}
which gives that 
\begin{eqnarray*}
\liminf_{Rh\to 0, \veck\to\veck^0}
\sigma(\veck;\alpha_\vecp=0,Re,Rh)=-(k^0)^2/Re < 0.
\end{eqnarray*}
This in turn implies that
\begin{eqnarray*}
\liminf_{Rh\to 0,\veck\to \veck^0}
Re^c(\veck;\alpha_\vecp=0,Rh) = \infty.
\end{eqnarray*}
If $\veck^0$ is resonant to $-\vecp=(-1,0)$, then $\veck^0$ is not resonant
to $\vecp=(1,0)$. Thus, in the $Rh\to 0$ and $\veck\to \veck^0$ limits,
we have $\omega_{-1}\to \omega_{-1}^0$ for
some $\omega_{-1}^0$, 
and $|\omega_{1}|\to \infty$.
\footnote{Since $0\le k_x^0 \le 1/2$, $\veck^{0}$ can not be resonant
to $\vecp=(1,0)$.}
Then, (\ref{CF}) is reduced
to 
\begin{eqnarray*}
a_0 + \frac{1}{a_{-1}} = 0.
\end{eqnarray*}
From this we can find the critical Reynolds number
\begin{eqnarray}
&&\liminf_{Rh\to 0,\veck\to\veck^0}
\left[Re^{c}(\veck;\alpha_\vecp=0,Rh)\right]^2
\nonumber \\
\nonumber \\
&&
\quad \ge
\frac{4k^0(1-2k^{0}\cos\alpha_{\veck^0}+(k^0)^2 )^2}
{\sin^2\alpha_{\veck^0}(-2\cos\alpha_{\veck^0}+k^0)(1-(k^0)^2)}
\label{app_ge4}
\end{eqnarray}
by (\ref{cre_triad}). We claim that the right-hand side of
(\ref{app_ge4}) is larger than or equal to $4$. This can be shown
by closer analysis of (\ref{reson_cond}) for $\alpha_\vecp = 0$
and we omit the proof.
This completes the demonstration of (\ref{app_cre_Rh0_ineq}) 
when $\veck^0 \ne \veczero$.

Now, we consider the case $\veck^0 = \veczero$. 
Consider the following four cases of the limits $\veck\to\veczero$ and 
$Rh\to 0$. Here, $\eps$ and $\delta$ are real constants with 
for $0<\eps<\delta<1$.
\begin{enumerate}
\item { $k_x \le 2k^{2+\eps}$ },
\item { $k_x \ge k^{2+\eps}/2$ and $k/Rh \le 2k^{\delta}$ },
\item { $k_x \ge k^{2+\eps}/2$, $k/Rh \ge k^{\delta}$ and
$k_x\le k_y/\sqrt{3}$ },
\item { $k_x\ge k_y$ }.
\end{enumerate}
Note that these four cases are sufficient to cover all limits
$\veck \to \veczero$ and $Rh\rightarrow 0$ in (\ref{app_cre_seq}).
Of course, this is a formal argument to deal with the limits $\veck\to\veczero$
and $Rh\to 0$.
To be mathematically rigorous, 
these limits have to be understood in terms of sequences. 
However, using sequences complicates the notation.

Since we are interested in $k\ll 1$, we can use (\ref{TCF}) and set
$\Re(\sigma(\veck;\alpha_\vecp,Re,Rh))=0$ to get
\begin{eqnarray}
i\rho + k^2/Re^c
&=&
\frac{k\sin^2{\alpha}(k+2\cos{\alpha})}
{4q^2(1)(i\rho+q^2(1)/Re^{c} + i\omega_{1})}
\nonumber\\
\nonumber\\
&& 
+ \frac{k\sin^2{\alpha}(k-2\cos{\alpha})}
{4q^2(-1)(i\rho+q^2(-1)/Re^{c} + i\omega_{-1})}
+O(k^4)\label{appendix_cre_equation}
\end{eqnarray}
where $\rho=\Im(\sigma(\veck;\alpha_\vecp,Re,Rh))$
is the imaginary part of $\sigma(\veck;\alpha_\vecp,Re,Rh)$, and
$q(\pm 1)=|\vecq(\pm 1)| = |\veck\pm\vecp|$. Here, 
$Re^c=Re^c(\veck;\alpha_\vecp=0,Rh)$.
From (\ref{appendix_cre_equation}), one can see that
$\rho(\veck)=O(k)$  for $k\ll 1$ and all $Rh$.

Now, we consider each case of the $\veck\to\veczero$ and $Rh\to 0$ limits.
\begin{enumerate}
\item { $k_x \le 2k^{2+\eps}$.}

Since $\cos{\alpha} = \D{\frac{k_x}{k}\le \frac{2k^{2+\eps}}{k} =
2k^{1+\eps}}$,
the real part of (\ref{appendix_cre_equation}) becomes 
\begin{eqnarray*}
k^2/Re^{c} 
&=& 
\D{
k^2\Re\left(
\frac{\sin^{2}{\alpha}}{4q^2(1)(i\rho+q^2(1)/Re^{c}+i\omega_{1})}\right.
}
\\
\\
&& 
\left.
+
\frac{\sin^{2}{\alpha}}{4q^2(-1)(i\rho+q^2(-1)/Re^{c}+i\omega_{-1})}
\right)
+O(k^{2+\eps})
\\
\\
&\le& 
\D{
k^2
\left(
\frac{\sin^2\alpha}{4q(1)^4/Re^c}
+\frac{\sin^2\alpha}{4q(-1)^4/Re^c}
\right)
}
+O(k^{2+\eps}).
\end{eqnarray*}
Thus, in the $\veck\to\veczero$ and $Rh\to 0$ limits, we have
\begin{eqnarray*}
\liminf_{Rh\to 0, \veck\rightarrow \veczero}
Re^c(\veck;\alpha_\vecp=0,Rh)\ge \sqrt{2}.
\end{eqnarray*}

\item { $k_x\ge k^{2+\eps}/2$ and $k/Rh \le 2k^{\delta}$.}

Using an asymptotic expansion of $\omega_{\pm 1}$ in $k_x$ and $k_y$
for $k\ll 1$, we can show that
\begin{eqnarray}
\omega_{\pm 1}
&=&
-\frac{1}{Rh}\left(\frac{k_x}{k^2}+k_x
\pm (k^2 - 2k_x^2) + O(k^3) \right)
\label{sec_asym1}\\
\nonumber\\
&=&
-\frac{1}{Rh}\left(\frac{k_x}{k^2}+k_x\right)
+O(k^{2+\delta})\label{sec_asym2}.
\end{eqnarray}
We set $\gamma(\veck) = k_x/k^2 + k_x$.
Then, similar to the derivation of (\ref{growth_rate}), we can show that
\begin{eqnarray}
\Re(\sigma(\veck;\alpha_\vecp,Re,Rh))
=-
\frac{1}{Re}k^2
\left[
1-\frac{1}{2}Re^2\frac{\sin^2\alpha}{(1+\gamma^2Re^2)^2}
\times
\right.
\nonumber\\
\nonumber\\
\left.
\left\{
1-8\cos^{2}\alpha+\gamma^2Re^2 \right\}
\right]
+O(k^{2+\delta}).
\label{appendix_real_sigma}
\end{eqnarray}
This is the same as (\ref{growth_rate}) at leading order. In
the $\veck\to\veczero$ and $Rh\to 0$ limits, the critical Reynolds number 
is obtained by setting the leading order term equal to zero. 
Similar to (\ref{re0gesqrt2}), we have 
\begin{eqnarray*}
\liminf_{Rh \to 0, \veck\rightarrow \veczero}
Re^c(\veck;\alpha_\vecp=0,Rh)\ge \sqrt{2}.
\end{eqnarray*}

\item { $k_x\ge k^{2+\eps}/2$, $k/Rh\ge k^{\delta}/2$ and 
$k_x\le k_y/\sqrt{3}$ }.

Since $\D{\frac{k_x}{k^2}\ge \frac{1}{2}\frac{k^{2+\eps}}{k^2}=
\frac{k^{\eps}}{2}}$,
we have
\begin{eqnarray*}
|\omega_{\pm 1}(\veck)|
= \D{\left|\frac{1}{Rh}(\frac{k_x}{k^2}+O(k))
\right|\ge \frac{C_1}{Rh}k^{\eps}}
\end{eqnarray*}
where $C_1$ is a constant independent of $Rh$.
Combining this with $k/Rh\ge(1/2)k^{\delta}$, we have
\begin{eqnarray*}
|\omega_{\pm 1}(\veck)|\ge\frac{C_1}{Rh}k^{\eps}
\ge (C_1/2)k^{-1+\delta+\eps}.
\end{eqnarray*}

Note that since $k_x\le k_y/\sqrt{3}$, we have $k^2-2k_x^2\ge k^2/2$.
Using this and (\ref{sec_asym1}), one can show that
\begin{eqnarray*}
|\omega_1|\ge |\omega_{-1}| \quad \textrm{for } k\ll 1.
\end{eqnarray*}

Since we have $|\rho(\veck)|=O(k)$ and
$|\omega_{\pm 1}|\ge (C_1/2)k^{-1+\delta+\eps}$,
(\ref{appendix_cre_equation}) becomes
\begin{eqnarray}
&&i\rho+k^2/Re^c
=
\D{
\frac{k\sin^2\alpha(k+2\cos{\alpha})}
{4q^2(1)(q^2(1)/Re^c+i\omega_{1})
\D{\left(1+\frac{i\rho}{q^2(1)/Re^c+i\omega_{1}}\right)}
}
}
\nonumber\\
\nonumber\\
& &+\ 
\D{
\frac{k\sin^2\alpha(k-2\cos{\alpha})}
{4q^2(-1)(q^2(-1)/Re^c+i\omega_{-1})
\D{\left(1+\frac{i\rho}{q^2(-1)/Re^c+i\omega_{-1}}\right)}
}
}
+O(k^4)
\nonumber\\
\nonumber\\
&&=
\frac{k\sin^2{\alpha}\cos\alpha}{2q^2(1)(q^2(1)/Re^c+i\omega_{1})}
\nonumber\\
\nonumber\\
&&
-\frac{k\sin^2{\alpha}\cos\alpha}{2q^2(-1)(q^2(-1)/Re^c+\omega_{-1})}
+O(k^{3-(\delta+\eps)}).
\label{app_case3}
\end{eqnarray}
We note $q(1)\ge q(-1)$ and $|\omega_{1}|\ge|\omega_{-1}|$. Taking
the real part from each side of (\ref{app_case3}), 
in the $k\rightarrow 0$ limit we have
\begin{eqnarray*}
\limsup_{\veck\rightarrow \veczero,\ Rh\rightarrow 0} 1/Re^c \le 0.
\end{eqnarray*}
This in turn implies
\begin{eqnarray*}
\liminf_{Rh\to 0, \veck\rightarrow \veczero}
Re^c(\veck;\alpha_\vecp=0,Rh) \ge \sqrt{2}.
\end{eqnarray*}

\item { $k_x\ge k_y/2$}

One can see that
\begin{eqnarray*}
\D{|\omega_{\pm 1}|\ge \frac{C_2}{Rh}k^{-1}}
\quad\textrm{for some constant }C_2
\end{eqnarray*}
Then, from (\ref{appendix_cre_equation}), taking the real parts,
\begin{eqnarray*}
k^2/Re^c \le C_3 Rh\, k^2 + O(k^3) 
\quad\textrm{for some constant }C_3
\end{eqnarray*}
In the $\veck\to \veczero$ and $Rh\rightarrow 0$ limits,
\begin{eqnarray*}
\liminf_{Rh\to 0, \veck\rightarrow \veczero}
Re^c(\veck;\alpha_\vecp=0,Rh) \ge \sqrt{2}.
\end{eqnarray*}
\end{enumerate}
This completes (\ref{app_cre_Rh0_ineq}) when $\veck^0 = \veczero$
and shows (\ref{cre_Rh0}) for $\alpha_\vecp = 0$.

\subsection{Multiple scale analysis}\label{sec_app_MS}

In this section, we consider the base flows (\ref{base_flow_MY}),
(\ref{base_flow_two}) and (\ref{base_flow_three}) and find the
large scale growth rate of perturbations using Multiple Scales analysis.
These base flows are spatially periodic. For Multiple Scales analysis,
we rotate these base flows to be periodic in both $x$ and $y$.
Being periodic in the coordinate directions simplifies Multiple Scales
analysis. 
For (\ref{base_flow_two}) and (\ref{base_flow_three}), 
let $\vecp = \vecp_1$.
With the rotation of $\alpha_\vecp$,
the $\beta$-plane equation becomes
(\ref{beta_eqn_rot}) and the base flows (\ref{base_flow_MY}),
(\ref{base_flow_two}) and (\ref{base_flow_three}) becomes
\begin{eqnarray}
\tilde{\Psi}^{0}_{M} = -\cos(\tilde{\vecp}\cdot\vecx - 
c \omega(\tilde{\vecp};\alpha_\vecp)t) 
&&\quad\tilde{\vecp} = (1,0) 
\label{base_flow_rot_one}
\\
\tilde{\Psi}^{0}_{2} = 
-\sum_{j=1}^{2}\cos(\tilde{\vecp}_j\cdot\vecx 
- \omega(\tilde{\vecp}_j;\alpha_\vecp)t)
&&\quad\tilde{\vecp}_1 = (1,0),\ \tilde{\vecp}_2 = (0,1)
\label{base_flow_rot_two}
\\
\tilde{\Psi}^{0}_{3} = 
-\sum_{j=1}^{3}\cos(\tilde{\vecp}_j\cdot\vecx 
- \omega(\tilde{\vecp}_j;\alpha_\vecp)t)
&&\quad \tilde{\vecp}_j =
(\cos((j-1)\frac{2\pi}{3}),
\nonumber\\
&&\qquad\qquad \sin((j-1)\frac{2\pi}{3})).
\label{base_flow_rot_three}
\end{eqnarray}
These are the base flows studied by Sivashinsky and Yakhot \cite{SY} 
for the large scale instability in the two dimensional Navier-Stokes
equation ($Rh = \infty$).  Also, 
Manfroi and Young \cite{MY} considered the base flow
(\ref{base_flow_rot_one}) with $c=1$.
Note that (\ref{base_flow_rot_one}), (\ref{base_flow_rot_two})
and (\ref{base_flow_rot_three})
satisfy
\begin{eqnarray*}
\nabla^2\Psi^{0} + \Psi^{0} = 0,
\end{eqnarray*}
and with $c=1$ they satisfy
\begin{eqnarray}
(\nabla^2\Psi^{0})_t + (1/Rh_x)\Psi_x^{0} - (1/Rh_y)\Psi_y^{0} = 0.
\label{app_Rossby_cond}
\end{eqnarray}
Recall that $(1/Rh_x,1/Rh_y) = (1/Rh)(\cos\alpha_\vecp,\sin\alpha_\vecp)$.
A small perturbation $\psi=\Psi-\Psi^{0}$ satisfies
\begin{eqnarray}
&&(\nabla^2\psi)_t + \Psi_y^{0}(\nabla^2\psi+\psi)_x
- \Psi_x^{0}(\nabla^2\psi + \psi)_y  \nonumber
\\
&& +(1/Rh_x)\psi_x - (1/Rh_y)\psi_y = (1/Re)\nabla^4\psi
\label{app_pert_eqn}
\end{eqnarray}
Following Sivashinsky and Yakhot \cite{SY} and
Manfroi and Young \cite{MY}, we introduce the scales
\begin{eqnarray*}
\begin{array}{lll}
\partial_t \to \partial_t + \eps\partial_T + \eps^2\partial_\tau,
& \partial_x\to \partial_x + \eps\partial_X,
& \partial_y \to \partial_y + \eps\partial_Y, \\
Rh_x = \eps^{-1}Rh_x^1
&
Rh_y = \eps^{-1}Rh_y^1.
\end{array}
\end{eqnarray*}
The choice of these scales is discussed in Section \ref{sec_comp_my}.
With these scales, (\ref{app_pert_eqn}) becomes
\begin{eqnarray}
& & (\partial_t + \eps\partial_T + \eps^2\partial_\tau)
\tilde{\nabla}^2\psi 
+\Psi_y^{0}[ (\tilde{\nabla}^2\psi + \psi)_x
+ \eps(\tilde{\nabla}^2\psi + \psi)_X]
\nonumber\\
\nonumber\\
& & -\Psi_x^{0}[ (\tilde{\nabla}^2\psi+\psi)_x
+ \eps(\tilde{\nabla}^2\psi + \psi)_Y]
+\eps\beta_x^1(\psi_x+\eps\psi_X)
\nonumber\\
\nonumber\\
& &
-\eps\beta_y^1(\psi_y+\eps\psi_Y)
= (1/Re)\tilde{\nabla}^4\psi\label{app_expansion}
\end{eqnarray}
\begin{eqnarray*}
\tilde{\nabla}^2\psi 
&=&
(\psi_xx + \psi_yy)+2\eps(\psi_{xX}+\psi_{yY})
+\eps^2(\psi_{XX}+\psi_{YY})\\
\\
\tilde{\nabla}^4\psi
&=&
(\psi_{xxxx}+2\psi_{xxyy}+\psi_{yyyy})
+4\eps(\psi_{xxxX}+\psi_{xxyY}+\psi_{xyyX}+\psi_{yyyY})\\
\\
&&+2\eps^2(3\psi_{xxXX}+3\psi_{yyYY}+\psi_{xxYY}+\psi_{yyXX}+
4\psi_{xyXY})\\
\\
&& +4\eps^3(\psi_{xXXX}+\psi_{xXYY}+\psi_{yXXY}+\psi_{yYYY})
\\
\\
&&+\eps^4(\psi_{XXXX}+2\psi_{XXYY}+\psi_{YYYY}).
\end{eqnarray*}
Integrating (\ref{app_expansion}) in $x$ and $y$ 
over the periodic domain of
$\Psi^{0}$, we obtain
\begin{eqnarray}
&&(\partial_t + \eps\partial_T + \eps^2\partial_\tau)
\int\int(\psi_{XX}+\psi_{YY})dxdy
\nonumber\\
\nonumber\\
&&+\partial_X\int\int\Psi^{0}_y[2(\psi_{xX}+\psi_{yY})+
\eps(\psi_{XX}+\psi_{YY})]dxdy 
\nonumber\\
\nonumber\\
&&-\partial_Y\int\int\Psi^{0}_x[2(\psi_{xX}+\psi_{yY})+
\eps(\psi_{XX}+\psi_{YY})]dxdy
\nonumber\\
\nonumber\\
&&+\int\int (\beta_x^1\psi_{X}-\beta_y^1\psi_Y) dxdy 
\nonumber\\
\nonumber\\
&&=\frac{\eps^2}{Re}\int\int(\psi_{XXXX}+2\psi_{XXYY}+\psi_{YYYY})dxdy.
\label{app_average}
\end{eqnarray}
Note that from (\ref{app_expansion}) and (\ref{app_average}),
$\Psi^{0}$ depends on $x,y,Rh^{1}$ and $T$.
For the details of derivations of (\ref{app_expansion}) and
(\ref{app_average}), we refer Sivashinsky and Yakhot \cite{SY}
and Manfroi and Young \cite{MY} (see also \cite{S}, 
\cite{FLV} and \cite{MY2}).

We seek the solution of (\ref{app_expansion}) of the form
\begin{eqnarray*}
\psi=\psi^{0} + \eps\psi^{1}+\eps^{2}\psi^2+\cdots.
\end{eqnarray*}
and for simplicity we assume that the base flow satisfies
(\ref{app_Rossby_cond}). This assumption is not true for
the base flow (\ref{base_flow_rot_one}) if $c \ne 1$. 

From the zeroth ($O(1)$) approximations of (\ref{app_expansion}) and
(\ref{app_average}), we assume that
\begin{eqnarray*}
\psi^{0} = A(T,\tau)e^{i(\vecK\cdot\vecX - \Omega(\vecK;\alpha_\vecp)t)}
\end{eqnarray*}
where the dispersion relation is given by
\begin{eqnarray*}
\Omega(\vecK;\alpha_\vecp) = \frac{-Rh_x^1 K_x + Rh_y^1 K_y}{K^2}.
\end{eqnarray*}
In the first ($O(\eps)$)
approximation of (\ref{app_expansion}), we find
\begin{eqnarray*}
\psi^{1} = 
\frac{1}{Re^{-1}-i\Omega(\vecK;\alpha_\vecp)}
(\Psi_y^{0}\psi_X^0 - \Psi_x^{0}\psi_Y^0).
\end{eqnarray*}
From the first ($O(\eps)$)
approximation of (\ref{app_average}), we have
$A_T = 0$ and thus $A = A(\tau)$ in $\psi^{0}$.
From the second ($O(\eps^2)$)
approximation, neglecting small terms of the
order $O(Re^2)$, we find
\begin{eqnarray*}
\psi^{2} = 
\frac{(4Re^{-1}-2i\Omega(\vecK;\alpha_\vecp))}
{Re^{-1}-i\Omega(\vecK;\alpha_\vecp)}
(\partial_{xX}+\partial_{yY})
(\Psi^{0}_y\psi_X^{0} - \Psi_x^{0}\psi_Y^{0})
\end{eqnarray*}
Neglecting terms of the order $O(Re^2)$ in the second ($O(\eps^2)$)
approximation
is discussed in Sivashinsky and Yakhot\cite{SY}. Then, the second
approximation in the integral relation (\ref{app_average}) gives
\begin{eqnarray}
\begin{array}{l}
\partial_\tau\int\int\psi_{XX}^0 + \psi_{YY}^0 dxdy \\
\\
\quad = -\partial_{X}\int\int\Psi_y^{0}
[2(\psi_{xX}^2+\psi_{yY}^2)+(\psi_{XX}^1+\psi_{YY}^1)]dxdy\\
\\
\quad + \partial_{Y}\int\int\Psi_x^{0}
[2(\psi_{xX}^2+\psi_{yY}^2)+(\psi_{XX}^1+\psi_{YY}^1)]dxdy\\
\\
\quad + Re^{-1}\int\int\psi_{XXXX}+2\psi_{XXYY}+\psi_{YYYY}
\end{array}
\label{app_growth_rate}
\end{eqnarray}
Inserting $\psi_{0}$, $\psi_{1}$ and $\psi_{2}$, it reduces to
\begin{eqnarray*}
A_{\tau} = \sigma A \quad \textrm{ for some } \sigma.
\end{eqnarray*}
and the linear growth rate is obtained by taking the real part
of $\sigma$.
We refer to Sivashinsky and Yakhot \cite{SY} for the details.

In the case of the base flow (\ref{base_flow_rot_two}), 
(\ref{app_growth_rate}) gives the growth rate
\begin{eqnarray*}
\Re(\sigma(\vecK))
= -\frac{K^2}{Re}\left[
1-Re^2\frac{1+Re^2\Omega^2(\vecK;\alpha_\vecp) -
16K_x^2K_y^2/K^4}{2(1+Re^2\Omega^2(\vecK;\alpha_\vecp))^2}
\right]
\end{eqnarray*}
which is identical to (\ref{growth_rate_two}) with the appropriate
rescaling and rotation.

In the case of the base flow (\ref{base_flow_rot_three}),
(\ref{app_growth_rate}) gives the growth rate
\begin{eqnarray*}
\Re(\sigma(\vecK))
= -\frac{K^2}{Re}\left[
1-\frac{3Re^2}{4}\frac{Re^2 \Omega^2(\vecK;\alpha_\vecp) -1}{(1+
Re^2
\Omega^2(\vecK;\alpha_\vecp))^2}
\right]
\end{eqnarray*}
which is identical to (\ref{growth_rate_three}) after an appropriate
rescaling.

In the case of the base flow (\ref{base_flow_rot_one}), since it
does not satisfy (\ref{app_Rossby_cond}) for $c\ne 1$, 
we can not use (\ref{app_growth_rate}) to find the growth rate. 
Multiple Scales analysis for (\ref{base_flow_rot_one}) can be used
in a similar way as above and the result is (\ref{MY_growth_rate}).
We omit this and refer to Manfroi and Young \cite{MY} for details.


\begin{thebibliography}{10}
\expandafter\ifx\csname url\endcsname\relax
  \def\url#1{\texttt{#1}}\fi
\expandafter\ifx\csname urlprefix\endcsname\relax\def\urlprefix{URL }\fi

\bibitem{CH}
A.~Chekhlov, S.~Orszag, B.~Galperin, I.~Staroselsky, The effect of small-scale
  forcing on large-scale structures in two-dimensional flows., Physica D 98
  (1996) 321--334.

\bibitem{SW1}
L.~Smith, F.~Waleffe, Transfer of energy to two-dimensional large scales in
  forced, rotating three-dimensional turbulence, Phys. Fluid 11 (1999)
  1608--1622.

\bibitem{ML}
P.~Marcus, T.~Kundu, C.~Lee, Vortex dynamics and zonal flows, Physics of
  Plasmas 7 (2000) 1630--1640.

\bibitem{HGS}
H.~Huang, B.~Galperin, S.~Sukoriansky, Anisotropic spectra in two-dimensional
  turbulence on the surface of a rotating sphere, Physics of Fluids 13 (2001)
  225--240.

\bibitem{SW2}
L.~Smith, F.~Waleffe, Generation of slow, large scales in forced rotating,
  stratified turbulence, J. Fluid Mech. 451 (2002) 145--168.

\bibitem{S}
L.~Smith, Numerical study of two-dimensional stratified turbulence, Advances in
  Wave 0 and Turbulence (ed. P.A. Milewski, L.M. Smith, F. Waleffe
  and E.G. Tabak). Amer. Math Soc. Providence, RI .

\bibitem{LG}
M.~Longuet-Higgins, A.~Gill, Resonant interactions between planetary waves,
  Proc. Roy. Soc. A 229 (1967) 120.

\bibitem{N}
A.~Newell, Rossby wave packet interactions, J. Fluid Mech. 35 (1969) 255--271.

\bibitem{MS}
L.~Meshalkin, Y.~Sinai, Investigation of the stability of a stationary solution
  of a system of equations for the plane movement of an incompressible viscous
  liquid, Appl. Math. Mech. 25 (1961) 1140--1143.

\bibitem{FR1}
S.~Friedlander, L.~Howard, Instability in parallel flows revisited, Studies in
  Applied Mathematics 101 (1998) 1--21.

\bibitem{FR2}
L.~Belenkaya, S.~Friedlander, V.~Yudovich, The unstable spectrum of oscillating
  shear flows, SIAM J. Appl. Math. 59 (1999) 1701--1715.

\bibitem{LI}
Y.~Li, On 2d euler equations. i. on the energy-casimir stabilities and the
  spectra for linearized 2d euler equations, Journal of Mathematical Physics 41
  (2000) 728--758.

\bibitem{SI}
G.~Sivashinsky, Weak turbulence in periodic flows, Physica D 17 (1985)
  243--255.

\bibitem{FLV}
B.~Frisch, B.~Legras, B.~Villone, Large-scale kolmogorov flow on the beta-plane
  and resonant wave interactions, Physica D 94 (1996) 36--56.

\bibitem{MY2}
A.~Manfroi, W.~Young, Slow evolution of zonal jets on the beta plane, J. Atmos.
  Sci. 56 (1999) 784--800.

\bibitem{SY}
G.~Sivashinsky, V.~Yakhot, Negative viscosity effect in large-scale flows,
  Phys. Fluids 28 (1985) 1040--1042.

\bibitem{NP}
A.~Novikov, G.~Papanicolaou, Eddy viscosity of cellular flows, J. Fluid Mech.
  446 (2001) 173--198.

\bibitem{L}
E.~Lorenz, Barotropic instability of rossby wave motion, J. Atmos. Sci. 29
  (1972) 258--269.

\bibitem{G}
A.~Gill, The stability of planetary waves on an infinite beta-plane, Geophys.
  Fluid Dyn. 6 (1974) 29--47.

\bibitem{MY}
A.~Manfroi, W.~Young, Stability of $\beta$-plane kolmogorov flow, Physica D 162
  (2002) 208--232.

\bibitem{WAL}
F.~Waleffe, Two-dimensional flows, Unpublished manuscript .

\bibitem{WAL2}
F.~Waleffe, The nature of triad interactions in homogeneous turbulence, Phys.
  Fluids A 4 (1992) 350--363.

\bibitem{WAL3}
F.~Waleffe, Inertial transfers in the helical decomposition, Phys. Fluids A 5
  (1993) 677--685.

\bibitem{KUN}
T.~Kundu, Fluid Mechanics, Academic Press, 1990.

\bibitem{KUO}
H.~Kuo, Dynamic instability of two-dimensional nondivergent flow in a
  barotropic atmosphere, J. Meteor. 6 (1949) 105--122.

\bibitem{JAP}
K.~Gotoh, M.~Yamada, The instability of rhombic cell flows, Fluid Dynamics
  Research 1 (1986) 165--176.

\bibitem{LW}
L.~Lorentzen, H.~Waadeland, Continued Fractions with Applications,
  North-Holland, Amsterdam, Amsterdam, The Netherlands, 1992.

\end{thebibliography}
\end{document}